
 \input harvmac

\def \k {\kappa}

\def \g {\gamma}
\def \td {\tilde} \def\const {{\rm const}}
\def \s {\sigma}
\def\t {\tau}
\def \bt { \bar \tau}
\def \p {\phi}
\def \ha {\half}
\def \ov {\over}
\def \fourth  {{1\ov 4} }
\def \four {{\textstyle {1\ov 4}}}
\def \eit {{\textstyle {1\ov 8}}}

\def \a {\alpha}
\def \lr { \lref}
\def\ep{\epsilon}

\def\vp {\varphi}
\def \bd {\bar \del}\def \B {\bar B}
\def \r {\rho}
\def\const {{\rm const}}\def\bd {\bar \del} \def\m{\mu}\def\n {\nu}\def\l
{\lambda}

\def \b {\beta}

\def\const {{\rm const}} \def \tvp {\tilde \vp}
\def \s {\sigma}
\def\t {\tau}
\def \bt { \bar \tau}
\def \p {\phi}
\def \ha {\half}
\def \ov {\over}
\def \fourth  {{1\ov 4} }
\def \four {\textstyle {1\ov 4}}
\def \a {\alpha}

\def \lr { \lref}
\def\ep{\epsilon}

\def\vp {\varphi}
\def \bd {\bar \del}\def \B {\bar B}
\def \r {\rho}
\def\const {{\rm const}}\def\bd {\bar \del} \def\m{\mu}\def\n {\nu}\def\l
{\lambda}
 \def \sms{{$\s$-models\ }}
\def \four {{\textstyle {1\ov 4}}}

 \def \sm {$\s$-model\ }
\def   \td {\tilde }
\def \h {\chi}
\def \X {{\cal X}}
\def \b {\beta}

 \def \sm {$\s$-model\ }

\def \lr { \lref}
\def \A {{\cal A}}
\def \B {{\cal B}}
\gdef \jnl#1, #2, #3, 1#4#5#6{ { #1~}{ #2} (1#4#5#6) #3}
\def \tvp {{\tilde \vp}}

\lr \burg{C.P. Burgess, \np B294 (1987) 427; V.V. Nesterenko, \ijmp A4 (1989)
2627.}
\lr \tsenul {A.A. Tseytlin, \np B390 (1993) 153. }
\lr \susskind { L. Susskind, ``Some speculations about black hole entropy in
string theory", RU-93-44 (1993), hep-th/9309135. }

\lr \seng{A. Sen, ``Black hole solutions in heterotic string theory on a
torus",
TIFR-TH-94-47, hep-th/9411187.}

\lr \sts{ A.S. Schwarz and A.A. Tseytlin, \np B399 (1993) 691. }

\lr\hrt {G.T. Horowitz and A.A. Tseytlin,  \pr D50 (1994) 5204. }

\lr \gtwo {E. Del Giudice, P. Di Vecchia and S. Fubini, Ann. Phys. 70
(1972) 378; K. A. Friedman and C. Rosenzweig, Nuovo Cimento 10A (1972) 53;
S. Matsuda and T. Saido, Phys. Lett. B43 (1973) 123; M. Ademollo {\it et al},
Nuovo Cimento A21 (1974) 77;
S. Ferrara, M. Porrati and V.L. Teledgi, Phys. Rev. D46
(1992) 3529.}

\lr \plane { D. Amati and C. Klim\v c\'\i k,
\jnl \pl, B219, 443, 1989;
 G. Horowitz and A. Steif,  \jnl \prl, 64, 260, 1990; \jnl \pr,
D42, 1950, 1990;
 G. Horowitz, in: {\it
 Strings '90}, eds. R Arnowitt et. al.
 (World Scientific, Singapore, 1991).}

\lr \tsett {A.A. Tseytlin, \pl B317 (1993) 559.}
\lr \wit { E. Witten,  \prd D44 (1991) 314. }
\lref \dvv  { R. Dijkgraaf, H. Verlinde and E. Verlinde, \np B371 (1992) 269. }

\lr \gibma { G.W.  Gibbons and  K. Maeda, \np B298 (1988) 741;
G.W.  Gibbons, in: {\it Fields and Geometry}, Proceedings of the 22-nd Karpacz
Winter School of Theoretical Physics, ed. A. Jadczyk (World Scientific,
Singapore,  1986).}

\lr \gibwi { G.W.  Gibbons and  D.L.  Wiltshire, \np B287 (1987) 717.}

\lref \tsnul { A. Tseytlin, \jnl \np, B390, 153, 1993.}

\lr\duval  { C. Duval, Z. Horvath and P.A. Horvathy, \jnl \pl,  B313, 10,
1993.}

\lr\gro{D.J.  Gross, J.A. Harvey,  E. Martinec and R. Rohm, \np B256 (1985)
253; \np B267 (1985) 75.}

\lr \incer { E.J. Ferrer, E.S. Fradkin and V.de la Incera, \pl B248 (1990)
281.}
\lr \green {M.B. Green, J.H.  Schwarz and E.  Witten, {\it Superstring Theory}
(Cambridge U.P., 1987).}

\lr \quev {C. Burgess and F. Quevedo,  \np B 421 (1994) 373. }
  \lr \nahm { W. Nahm, \np B124 (1977) 121. }
 \lr \kkkk {  E. Kiritsis and C. Kounnas,  ``Curved four-dimensional spacetime
as infrared regulator in superstring theories", hep-th/9410212. }
\lr \sen{A. Sen, \pr D32 (1985) 2102; \prl 55 (1985) 1846.}
\lr \hulw { C. M. Hull and E.  Witten,  \jnl \pl, B160, 398, 1985. }

\lr\hult { C. Hull and P. Townsend, \jnl \pl, B178, 187, 1986. }
\lr \gps {S.  Giddings, J. Polchinski and A. Strominger, \jnl  \pr,  D48,
 5784, 1993. }

\lref \horts {G.T. Horowitz and A.A. Tseytlin, ``A new class of exact solutions
in string theory", Imperial/TP/93-94/54, UCSBTH-94-31, hep-th/9409021.   }

\lr\horrt{G.T. Horowitz and A.A. Tseytlin, \prl 73 (1994) 3351.}

\lr \attick {J.J. Attick  and E. Witten, \np B310 (1988) 291. }
\lr\fund{A. Dabholkar, G. Gibbons, J. Harvey and F. Ruiz Ruiz, \jnl \np, B340,
33, 1990;
D. Garfinkle, \jnl \pr, D46, 4286, 1992; A. Sen, \jnl \np, B388, 457, 1992;
D. Waldram, \jnl \pr, D47, 2528, 1993. }

\lr \duality { L. Brink, M. Green and J. Schwarz, \np B198 (1982) 474;
K. Kikkawa and M. Yamasaki, \pl B149 (1984) 357;
N. Sakai and I. Senda, { Progr. Theor. Phys. }  75 (1984) 692. }

\lr \canon {A. Giveon, E. Rabinovici and G. Veneziano, \np B332 (1989)167;
K. Meissner and G. Veneziano, \pl B267 (1991) 33;
E. \' Alvarez, L. \' Alvarez-Gaum\' e and Y. Lozano, \pl B336 (1994) 183. }

\lr\vene{G. Veneziano, \pl B265 (1991) 287;
K. Meissner and G. Veneziano, \pl B267 (1991) 33; \mpl A6 (1991)3397.}

\lr \kkl {E. Kiritsis, C. Kounnas and D. L\" ust, \pl B331 (1994)321.}
\lr \kk {E. Kiritsis and  C. Kounnas,  \pl B320(1994)361;
E. Kiritsis, C. Kounnas and D. L\" ust, \pl B331 (1994)321.}
\lref \kallosh { E. Bergshoeff, R. Kallosh and T. Ort\' \i n, \jnl \pr,  D47,
5444,
1993; E. Bergshoeff, I. Entrop and R. Kallosh,
\jnl \pr, D49, 6663, 1994.   }

\lref \duff{M. Duff, B. Nilsson and C. Pope, \pl { B163} (1985) 343;
R.~Nepomechie, \pr { D33} (1986) 3670. }

\lr\frats { E.S. Fradkin and A.A. Tseytlin , \pl B163 (1985) 123. }
\lr\tsey{  A.A. Tseytlin, \pl B202 (1988) 81.  }
\lr\tset { A.A. Tseytlin, \np B350 (1991) 395.}
\lr\metsa {R.R. Metsaev and A.A. Tseytlin, \np B298 (1988) 109. }
\lr\abo { A. Abouelsaood, C. Callan, C. Nappi and S. Yost, \np B280 (1987) 599;
V.V. Nesterenko, \ijmp A4 (189) 2627.
}

\lr\mono{T. Banks, M. Dine, H. Dijkstra and W. Fischler, \pl B212 (1988) 45.}
\lr\rb {I. Robinson, Bull. Acad. Polon. Sci. 7 (1959) 351;
B. Bertotti, \pr 116 (1959) 1331. }

\lr\kirr{E. Kiritsis, \np B405 (1993) 109.}

 \lr \napwi {C. Nappi and E. Witten, \prl 71 (1993) 3751.}
\lr\tset{A.A. Tseytlin, \pl B317 (1993) 559.}
\lr \ruts { J.G.  Russo and A.A. Tseytlin, ``Constant magnetic field in closed
string theory: an exactly solvable model", CERN-TH.7494/94,
Imperial/TP/94-95/3, hep-th/9411099. }
\lr \givki{A. Giveon and E. Kiritsis,  \np B411 (1994) 487.}
\lr\sft{  K. Sfetsos, \pl B324 (1994) 335; \pr D50 (1994) 2784.}
\lr\sfts{K. Sfetsos and  A.A. Tseytlin, \np B427 (1994) 325.}
\lr\kum{A. Kumar, \pl B293 (1992) 49; S. Hassan and A. Sen,
\np B405 (1993) 143; E. Kiritsis, \np B405 (1993) 109. }

\lr \tsee{A.A. Tseytlin, ``Exact string solutions and duality", to appear in:
{\it Proceedings of the 2nd Journ\' ee Cosmologie}, ed. H. de Vega  and N. S\'
anchez (World
Scientific), hep-th/9407099. }

\lr \alv {
E. \'Alvarez, L. \'Alvarez-Gaum\'e, J.L.F.
Barb\'on and Y. Lozano, \np B415 (1994) 71. }
\lr\rabi{A. Giveon, M. Porrati and E. Rabinovici, Phys. Rep. 244
(1994) 77.}
\lr\ghs { D. Garfinkle, G.T. Horowitz and A. Strominger,  \pr  D43 (1991) 3140;
and {D45} (1992) 3888(E).}

\lr\busch{T.H. Buscher, Phys. Lett. B194 (1987) 51; B201 (1988) 466.}

\lr \rocver{M.  Ro\v cek  and E. Verlinde, \np B373 (1992) 630.  }

\lr \dabh {A. Dabholkar, \np B439 (1995) 650.}
\lr \stro{D.A. Lowe and A.  Strominger, ``Strings near a Rindler or black hole
horizon", UCSBTH-94-42, hep-th/9410215.}
\lr \orb {L. Dixon, J. Harvey, C. Vafa and E. Witten, \np B274 (1986) 285.}

\lr\hrt {G.T. Horowitz and A.A. Tseytlin, \pr D50 (1994) 5204. }
\lr\bag{ J.A. Bagger, C.G.  Callan and J.A.  Harvey, \np B278 (1986) 550. }

\lr\ant{I. Antoniadis, C. Bachas and A. Sagnotti, \pl B235 (1990) 255.}
\lr\bak{C. Bachas and E. Kiritsis, \pl B325 (1994) 103. }
\lr\rutsn{J. Russo and A.A. Tseytlin, to appear }

\lr \kltspl { C. Klim\v c\'\i k and A.A. Tseytlin, \pl B323 (1994) 305.}

\lr \gibma { G.W.  Gibbons and  K. Maeda, \np B298 (1988) 741.}
\lr \gibb{
G.W.  Gibbons, in: {\it Fields and Geometry}, Proceedings of the 22-nd Karpacz
Winter School of Theoretical Physics, ed. A. Jadczyk (World Scientific,
Singapore,  1986).}

\lr \gps {S.  Giddings, J. Polchinski and A. Strominger,   \pr D48 (1993)
 5784. }

\lr\horts{ G.T. Horowitz and   A.A. Tseytlin, \pr  D51 (1995) 2896.}

\lr \hoho {J. Horne and G.T. Horowitz, \np B368 (1992) 444. }
\lr \sftse { A.A. Tseytlin, \pl B317 (1993) 563; K. Sfetsos and A.A.  Tseytlin,
 \pr D49 (1994) 2933.}

\lr\khu{R. Khuri, \pl B259 (1991) 261; \np B387 (1992) 315.}

\lr \nels {W. Nelson, \pr D49 (1994) 5302.}
\lr \kalor { R. Kallosh and T. Ort\'\i n, ``Exact $SU(2)\times U(1)$ stringy
black holes", SU-ITP-94-27, hep-th/9409060. }

\lr \busc {T.H. Buscher, \pl  B194 (1987) 59; \pl B201 (1988) 466;
 M. Ro\v cek and E. Verlinde, \np B373 (1992) 630. }

\lr \los { D.A. Lowe and A. Strominger, \prl 73 (1994) 1468.}

\lr\bachas {C. Bachas and M. Porrati , \pl B296 (1992) 77.}
\lr\ferrara{S. Ferrara and M. Porrati, \mpl A8 (1993) 2497.}
\lr \sfets { K. Sfetsos and  A.A. Tseytlin, \np B427 (1994) 325.  }
\lr\rs {J.G. Russo and L. Susskind, \np B437 (1995) 611.}
\lr\thermal {J.G. Russo, \pl B335 (1994) 168.}
\lr \napwi{ C. Nappi and E. Witten,  \prl 71 (1993) 3751.}

\lr\tsetl { A.A. Tseytlin, \pl B208 (1988) 221.}
\lr\polch{ J. Polchinski, Commun. Math. Phys. 104 (1986) 37.}
\lr\obrien{ O'Brien -Tan , McClain-Roth }
\lr \ginsp { W. Lerche, A. Schellekens and N. Warner, Phys. Repts. 177 (1989)
1; P. Ginsparg, in: {\it Fields, Strings and Critical Phenomena, } ed. by E.
Brezin and J. Zinn-Justin (Elsevier Science Publ., 1989).  }
\lr\salam{A. Salam and J. Strathdee, \np B90 (1975) 203.}
\lr\nielsen{ N.K. Nielsen and P. Olesen, \np B144 (1978) 376;
J. Ambjorn and P. Olesen, \np B315 (1989) 606 and \np B330 (1990) 193.}

\lr\anto{I. Antoniadis and N. Obers, \jnl \np, B423, 639, 1994}
\lr \nonsemi { D. Olive, E. Rabinovici and A. Schwimmer,  \jnl \pl, B321, 361,
1994.}
\lr\sfee{
 K. Sfetsos,  \jnl \pl, B324, 335, 1994; \pr D50 (1994) 2784.}
\lr\nons{A.A.  Kehagias and P.A.A.  Meesen, \pl B331 (1994) 77;
J.M. Figueroa-O'Farril and S. Stanciu, \pl B327 (1994) 40;
A. Kehagias, ``All WZW models in $D\leq 5$", hep-th/9406136. }

\lr \busc {T.H. Buscher, \pl  B194 (1987) 59; \pl B201 (1988) 466.}
 \lr\rocver{ M. Ro\v cek and E. Verlinde, \np B373 (1992) 630. }
\lr \bsf{I. Bars and K. Sfetsos, \pr D46 (1992) 4510.}
\lr\sfees{K. Sfetsos, \ijmp A9 (1994) 4759.}
\lr \sftse { A.A. Tseytlin, \pl B317 (1993) 563.}
\lr\sfew{K. Sfetsos and A.A.  Tseytlin,  \pr D49 (1994) 2933.}

\lr\givroc{A. Giveon and M. Ro\v{c}ek,  \np B380 (1992) 128.}

\def\np {  Nucl. Phys. }
\def \pl { Phys. Lett. }
\def \mpl { Mod. Phys. Lett. }
\def \prl { Phys. Rev. Lett. }
\def \pr  { Phys. Rev. }

\def \ijmp {Int. J. Mod. Phys. }
\def \prd  { Phys. Rev. }

\baselineskip8pt
\Title{\vbox
{\baselineskip 6pt{\hbox{CERN-TH/95-20 }}{\hbox
{Imperial/TP/94-95/17}}{\hbox{hep-th/9502038}} {\hbox{revised}}} }
{\vbox{\centerline { Exactly solvable    string models    }
\centerline { of curved   space-time backgrounds   }
}}

\vskip -20 true pt

\centerline  { {J.G. Russo\footnote {$^*$} {e-mail address:
 jrusso@vxcern.cern.ch
} }}

 \smallskip \smallskip

\centerline{\it  Theory Division, CERN}
\smallskip

\centerline{\it  CH-1211  Geneva 23, Switzerland}

\medskip
\centerline {and}
\medskip
\centerline{   A.A. Tseytlin\footnote{$^{\star}$}{\baselineskip8pt
e-mail address: tseytlin@ic.ac.uk}\footnote{$^{\dagger}$}{\baselineskip8pt
On leave  from Lebedev  Physics
Institute, Moscow, Russia.} }

\smallskip\smallskip
\centerline {\it  Theoretical Physics Group, Blackett Laboratory}
\smallskip

\centerline {\it  Imperial College,  London SW7 2BZ, U.K. }
\bigskip
\centerline {\bf Abstract}
\medskip
\baselineskip8pt
\noindent
We consider a new 3-parameter class of exact 4-dimensional solutions
in closed string theory and solve the corresponding string model,
determining the physical spectrum and  the partition function.
The background fields (4-metric, antisymmetric tensor, two
Kaluza-Klein vector fields, dilaton and modulus) generically
describe axially symmetric stationary rotating (electro)magnetic
flux-tube type universes. Backgrounds of this class include
both the `dilatonic'  ($a=1$) and  `Kaluza-Klein' ($a=\sqrt 3$) Melvin
solutions
and the uniform magnetic field solution, as well as some
singular space-times. Solvability of the string $\s $-model
is related to its connection via duality to a simpler model
which is a ``twisted" product of a flat 2-space and a space
dual to 2-plane. We discuss some  physical properties of this
model (tachyonic instabilities in the spectrum, gyromagnetic ratio,
issue of singularities, etc.).
 It  provides
 one of the  first examples of  a consistent  solvable
 conformal string model with explicit  $D=4$  curved
space-time interpretation.

\Date {February 1995}


\noblackbox

\vfill\eject

\def \F {{\tilde F}}

\lr \tsem { A.A. Tseytlin, \pl B346 (1995) 55. }
\lr \ruts { J.G.  Russo and A.A. Tseytlin, ``Constant magnetic field in closed
string theory: an exactly solvable model", CERN-TH.7494/94,
Imperial/TP/94-95/3, hep-th/9411099. }

\def \sms {$\s$-models\  }
\lr \mahar {J. Maharana and J.H. Schwarz, \np B390 (1993) 3. }
\lr \mel { M.A. Melvin, \pl 8 (1964) 65; W.B. Bonner, Proc. Phys. Soc. London
A67 (1954) 225. }

\lref \cec { S. Cecotti, S. Ferrara and L. Girardello, \np B308 (1988) 436;
A. Giveon, E. Rabinovici and G. Veneziano, Nucl. Phys. B322 (1989) 167;
A. Shapere and F. Wilczek, \np B320 (1989) 669; A. Giveon, N. Malkin and E.
Rabinovici,
Phys. Lett. B238 (1990) 57.}

\lr \gibb{
G.W.  Gibbons, in: {\it Fields and Geometry}, Proceedings of the 22-nd Karpacz
Winter School of Theoretical Physics, ed. A. Jadczyk (World Scientific,
Singapore,  1986).}
\lr\gaun {
 D. Garfinkle and A. Strominger,  \pl  B256  (1991) 146;
  H.F. Dowker, J.P. Gauntlett, S.B. Giddings and G.T. Horowitz, \prd D50 (1994)
2662;  S.W. Hawking, G.T. Horowitz and  S.F. Ross, ``Entropy, area and black
hole pairs", NI-94-012, gr-qc/9409013.}
\lr \gaunn{H.F. Dowker, J.P. Gauntlett, D.A. Kastor and J. Traschen,
\prd D49 (1994) 2909.
 }

\newsec{Introduction}

 Conformal  $\s $-models describing the propagation of closed strings in
curved space-times  are, unfortunately, so complicated that
the spectrum of the physical  string excitations is  known  only in  a few
special cases.
In this paper we will introduce a new  class of exact
conformal  string models  representing non-trivial space-time backgrounds,
for which  the corresponding CFT's can be described  in terms of free
oscillators.
 The resulting world-sheet Hamiltonian is quartic
in the  free
creation/annihilation operators  and  is  diagonal in Fock space.
This makes it possible to calculate the
spectrum and the partition function.
 The reason for  solvability of these models is  in their formal connection
through  world-sheet (or target space)
duality to a flat theory.  Because of non-trivial
boundary conditions the duality does not leave us within flat CFT but  leads to
a new  conformal theory.

The corresponding string  backgrounds  interpolate  continuously  between very
different configurations, e.g.   between
 a homogeneous ``rotating" universe
with  a uniform magnetic field  \refs{\horts,\ruts}
\eqn\hts{ ds^2_4 = - (  dt + \ha \b  \r^2 d\vp)^2  + d\r^2 + \r^2 d\vp^2 +
dx_3^2 \ , }
$$ \A = -\B=  \ha \b   \r^2 d\vp\   , \ \ \
 B=  \ha  \b  \r^2 d\vp\wedge dt\ , \ \ \ \p=\p_0=\const \ , \ \ \s=0 \ , $$
the  dilatonic  $``a=1"$  Melvin-type \mel\  static  magnetic flux-tube
universe \refs{\gibma, \tsem}
\eqn\melv{ ds^2_4=  - dt^2  + d\r^2 +  F^2(\r)  \r^2 d\vp^2 + dx_3^2 \ , }
$$ \A= -\B =  \b  F(\r)   \r^2 d\vp\ , \ \ \ B=0\ , $$ $$
 e^{2(\p-\p_0)}=  F \ , \ \ \
  \   \s=0\ , \ \ \ F\inv = 1 + \b^2 \r^2 \  ,    $$
the ``$a=\sqrt 3$"  (Kaluza-Klein) Melvin solution \gibma
\eqn\melvi{ ds^2_4=  - dt^2  + d\r^2 +  \td F(\r)  \r^2 d\vp^2 + dx_3^2 \ , }
$$ \A=  q_+  \td F(\r)   \r^2 d\vp\ , \ \ \  \ \B=0\ , \ \ B=0\ , $$
 $$
 \p=\p_0   \ , \ \ \ \ e^{2\s} = {\td F}\inv = 1 +  q_+^2 \r^2 \  ,    $$
and  singular ``rotating" space-times  with no gauge fields.
Here
$\A$ and $\B$ are   the  abelian vector and  axial vector
  1-forms, $B$ is the antisymmetric tensor 2-form,  $\p$ is the dilaton
and $\s$ is the modulus scalar corresponding to the  compact ($x^5\equiv y\in
(0,2\pi R))$
 Kaluza-Klein  dimension.

Explicitly, our
  3-parameter ($\a,\b,q_+$ or $a_+,c_+,c_-$)  class of  $D=4$
 axially symmetric exact
string solutions is represented by
the following  background fields:
\eqn\qqqs{  ds^2_4 =
 - f_1 (\r) dt^2
+ f_2 (\r)  d\vp dt
+ f_3(\r)  d\vp^2 +  d\r^2 +  dx_3^2 \ , \   }
$$f_1 = 1 + \four a_+^2 c_-^2 \r^4 F(\r) \F(\r) \  ,  \ \ \
\  f_2= c_- [1+ \four(c_+^2 -a_+^2 -c_-^2)\r^2] F(\r) \F(\r) \r^2 , $$ $$
\ f_3= (1-\four c_-^2\r^2) F(\r) \F(\r) \r^2\   , $$
\eqn\ttts{\A  = \ha  \F(\r)  \r^2[ (a_+  + c_+) d \vp + a_+c_-dt] \ , \ \ \
    }
$$\B = \ha  F(\r) {\r^2 } [ (a_+  - c_+) d \vp + a_+c_- dt] \ ,  \ \ \
 B=
  - \ha   c_- F(\r)  \r^2 d\vp\wedge dt  \ , $$
\eqn\sssu{ e^{2(\p-\p_0) } =  F (\r) \ , \ \ \  \ \ \  e^{2\s } =   {F(\r) \ov
\F(\r)}\ ,
  }
\eqn\dees{
F (\r) \equiv {1\ov 1 + {\r^2 /\rho^{2}_0} } \ , \ \ \
\F (\r) \equiv {1\ov 1 +{\r^2/\td \rho^{2}_0}}\ ,  }
$$
 \rho^{-2}_0\equiv  \four [(a_+ - c_+)^2 - c_-^2] = \a\b\ , $$
$$
\td \rho^{-2}_0 \equiv  \four [(a_+ +  c_+)^2 - c_-^2] = q_+ (q_+ +\b-\a) \  ,
$$ $$ a_+=q_+-\a\ , \ \ \  \  c_+ = q_+ + \b \ , \ \ \ \  c_-= \a-\b \ .$$
The previously known  magnetic solutions  \hts, \melv\  and \melvi\
are obtained  in the special cases:  $a_+=0, \ c_+= -c_-=\b$;\
$\ \ a_+=0, \ c_+=2\b, \ c_-=0$ and $a_+=c_+=q_+, \ c_-=0$.

In general,  the metric is  stationary  and describes a rotating
electro-magnetic flux tube universe.   Asymptotically the space-time is  that
of   a
  product  of a flat space and a rotating cylinder with  radius going  to zero
at large $\r$.
These  models  can be interpreted, in particular,  as describing  the geometry
induced by  the two magnetic fields (in $d\A$ and $d\B$) associated with  the
Kaluza-Klein
gauge groups $U(1)_{\rm v}$ and
 $U(1)_{\rm a}$,
\eqn\para{
{\bf B}_{\rm v}={{{\bf B}_{{\rm v}}}_0\ov \big(1+ {\rho ^2/\td \rho_ 0^{2}}
\big)^2}\ ,\ \ \
{\bf B}_{\rm a}={{{\bf B}_{{\rm a}}}_0\ov \big(1+ {\rho ^2/\rho_ 0^{2}}\big)^2}
\ , }
$$ {{\bf B}_{{\rm v}}}_0 = a_+ + c_+ \ , \ \ \ \ \ {{\bf B}_{{\rm a}}}_0 = a_+-
c_+ \ . $$
For generic values of the parameters   there are  also  non-vanishing
electric fields.
When  $\rho^2_0$ or  $\td \rho^2_0$  are
negative,  the  curvature  has singularities  at $\rho ^2=-\rho^2_0$ or
$\rho^2=-\td \rho^2_{0}$.

The important  special cases correspond to
\eqn\speci{\a\b q_+ (q_+ +\b-\a)=0 \ , }
 i.e.  to  $ \rho^2_0=\infty$
or $\td \rho^2_0=\infty$, when  at least one  of the magnetic fields
 uniformly extends to infinity.  For these values of the parameters
the corresponding  CFT simplifies; in particular,   the  quantum
 Hamiltonian  becomes quadratic in oscillators
and the  partition function  takes  the form  of  a  modular integral and  a
sum over winding sectors of a left-right
factorized expression.

These  string solutions    represent   typical   flux tube   type uniform
electromagnetic backgrounds in  closed string theory.
Such backgrounds
 are interesting for several reasons.
Strings in magnetic fields  are expected to undergo phase transitions with   a
possible symmetry restoration in a  way analogous  to
 gauge theories \nielsen. This is suggested by the emergence of tachyons in the
spectrum for critical values of the magnetic field  where the partition
function
develops new divergences \ruts.
Similar electromagnetic backgrounds   are also related to a description  of
pair creation of charged  black holes \refs{\gibb,\gaun,\gaunn}.
Knowing  the solution of the string model (i.e. of the conformal field theory)
corresponding, e.g., to  the Melvin solution
of the low-energy effective field theory,  is a necessary step towards
the analysis  of such processes at the level of  string theory.

The present construction  illustrates, in particular,    how one  can find
complicated  exact string solutions without actually solving the
equations of   low-energy effective  field theory but instead
 looking    directly  for    conformal 2d $\s$-models with the  desired
properties.
The  world-sheet
duality transformations  relating  complicated \sms to simpler ones
provide a useful  tool.
 For example, starting with  a dimension $D\geq d+n$ flat model
with  a number $d$ of periodic coordinates and making  formal
$O(d+n,d+n;R)$  world-sheet duality transformations (see e.g.
\refs{\cec,\vene,\rabi})
 with {\it  continuous } parameters,
one obtains new {\it inequivalent}  conformal theories (with  $O(d,d;Z)$
dualities as symmetries),  corresponding to
 complicated  space-time backgrounds which solve
the  low-energy effective equations.

The contents  of this  paper, which is a  sequel to \refs{\ruts,\tsem},
 is the following.

We shall start in Section 2.1  with a   construction of  our string model
by  applying  coordinate shifts and  simple  duality
(or, equivalently, a special $O(3,3;R)$ duality transformation)
to a   (dual to) flat  model.  This $O(3,3;R)$ duality
leads to a non-trivial CFT since two of our coordinates are compact.
We shall also consider a  simple duality transformation in the
Kaluza-Klein coordinate which will act on our class of    models relating
members with  certain different  values of the parameters.
Next, in  Section 2.2 we shall  show that the resulting \sm
is locally  (ignoring topology)
 equivalent  to a particular  model with two null Killing symmetries
considered in \horts\ and,   as such,  is conformal to all orders in $\a'$.
In Section 2.3 we shall discuss some special cases,  namely
plane-wave type models (corresponding, e.g., to the solution \hts)
and the  model behind the Melvin solution \tsem.
Some generalizations,  involving,  in particular,
the replacement of the (dual to) 2-plane by a $D=2$ black hole space
and applying a similar ($O(3,3;R)$) duality transformation, will be briefly
considered in Section 2.4.

Section 3  will be devoted to a
space-time/low energy  field theory   interpretation of
our  conformal string model.
Rearranging the  \sm action in a Kaluza-Klein way
(i.e. separating terms with compact  $x^5$-coordinate)
we shall determine the background fields
\qqqs--\dees\  which solve the equations of the corresponding $D=4$ effective
field theory.
We shall then discuss in turn
various special cases, emphasizing, in particular,
that differently looking backgrounds which  originate from \sms  related  by
simple  duality  in  a compact coordinate
represent different  ``sides" of the  same string solution, i.e. of the same
CFT.
We shall also mention (in Section 3.7) some closely related
exact string solutions which may have an interpretation of  (3+1)-dimensional
black holes  in external electromagnetic fields
(but for which  we are unable to  solve the corresponding string model).

 In Section 4  we shall start the discussion of  the  solution of this
string model.  We  shall first explain
 (on a  simple special case)
 how the
theory can be effectively  expressed in terms of free fields
and then proceed with  a computation of the partition function  $Z$
on the torus   defined by  the  string path  integral.
Although the model is not gaussian, we  will  show that all path integrals can
be
evaluated  explicitly and
obtain  the expression  for $Z$ in terms of the
 standard modular integral and sum over winding numbers and  also of
two additional ordinary  integrals. The latter integrals
can be easily computed in the  special cases \speci.
We shall also discuss peculiar  target space
duality invariance properties of $Z$.

In  Section 5 we  shall systematically  describe a   solution of the   string
model using  the canonical
operator  quantization   approach.
We shall  first  (in Section 5.1)
derive the  expression for the  general
solution of the classical
equations of motion  on the cylinder (free string propagation)
 in terms of constant zero-mode parameters and
free oscillators.    We  shall then
canonically quantize the  theory  (Section 5.2)
 using a light-cone type gauge  and derive
the quantum Hamiltonian. The latter,  in general,
will be fourth order in  creation/annihilation operators (becoming quadratic
only
in the special case of \speci) but will be diagonal in Fock space, enabling   a
straightforward determination of the string spectrum.
 In Section 5.3 we  shall  illustrate this construction
by considering  its  point-particle limit in the Melvin model case, i.e. derive
the expression for the
zero level  scalar (tachyon)  spectrum directly from the Klein-Gordon operator
in the  Melvin background \melv.
 In Section 5.4 we shall   show that
the operator approach leads to the same expression
 for the partition function that was earlier obtained
 in the path integral approach.

 Some physical properties of our class of string  models will be
considered in Section 6. In  Section 6.1 we
shall discuss  the string spectrum
in the  special case  when $\a=\b$ (which includes the  Melvin model).
Such models turn out to have tachyonic instabilities
in the charged state sector of the spectrum.
Implications for the value of the gyromagnetic factor in closed string theory
will be mentioned in Section 6.2.  In Section 6.3 we shall look at the
string models  with $\a\b q_+(q_+  + \b -\a ) <0$ corresponding to
singular space-time backgrounds and point out the existence of  new tachyonic
states
related to  the presence  of the quartic term in the string Hamiltonian.
 We will find  that in these cases
  even the state which is the counterpart of the usual graviton becomes
tachyonic,  what  reflects  a strong instability of these backgrounds.
We shall also  comment on the relation of  the  corresponding  backgrounds  to
$D=3$ black string and black hole  geometries.

Section 7 contains  some  concluding remarks.
In the Appendix we give the expression for the  curvature scalar for the metric
\qqqs.

\newsec{String  model}
We shall start with a construction of  our   class
of exact conformal  $D=4$ string  solutions directly at the string \sm level,
using   duality transformations  to relate  simple \sms  to a more complicated
one. This will provide  a clue  to  why the resulting   string  models,
describing complicated curved space-time backgrounds  are actually  exactly
solvable.

\subsec{  Duality transformations}
We shall consider  axially-symmetric $D=4$ string backgrounds
which  are direct products of  a non-trivial three-dimensional $(t,x_1,x_2)$
part and a   line  ($x_3$-direction).
We shall also introduce  a  coupling of a closed string to an external
 gauge field  background  by using a  stringy  version of the Kaluza-Klein
approach,
i.e.   by adding  an extra  compact internal direction $x^5=y$ (with period
$2\pi R$).
We shall often use the  ``light-cone" coordinates
 \eqn\uv { u= y-t\ , \ \ \ \ v= y+t\ , \  \ \ \ \ y \in (0,2\pi R ) \ . }
 Let us start with  an auxiliary   \sm  describing  a string in $D=4$
space-time
 which is a direct product of a $D=2$ space-time cylinder $(t,y)$  and
a space dual to
 a 2-plane\foot{The construction that follows can be repeated in the case
when
the dual plane is replaced  by a plane itself (with a  trivial dilaton).
The resulting  backgrounds   will also represent exact string solutions.
Their form   can be obtained from the  expressions
that follow
 by replacing $\r $ by $ 1/\r$  in the functions (but not in  derivatives or
differentials). This class of   backgrounds will generically be singular at
$\r=0$ (and will not contain  solutions \hts,\melv,\melvi).
Moreover,  in contrast to the model (2.3) discussed below,
the corresponding string model will {\it not}
be  exactly solvable  by our methods (for example, the partition function will
not be explicitly computable).
One may also consider a  generalization  of (2.2) (or of a similar model with
the 2-plane part)
by adding an arbitrary constant in front of the $\r^{\mp  2} \del \td \vp \bd
\td \vp$ term. The  resulting models will have conical singularities and do
not seem to be well defined (for  a discussion in  a particular case see
\tsem).}
 \eqn\lagrq{ I={1\over \pi\alpha '}\int d^2 \s\big[  \del u \bd v
+ \del \r \bd \r + \r^{-2} \del \tvp\bd  \tvp  + {\cal R} (\p_0 -  \ha \ln \r^2
) \big]\  , }
where ${\cal R}\equiv  \fourth  \a' \sqrt g R^{(2)}$ and  $\tvp$  should have
period $2\pi\a'$ to preserve equivalence of the ``dual plane" model to the
flat 2-plane CFT \rocver.  A more general  model
is obtained from \lagrq\ by  making a coordinate shift and adding    constant
antisymmetric  tensor terms
 \eqn\lag{ I={1\over \pi\alpha '}\int d^2 \s \big[
( \del u + \a \del \tvp)  (\bd v + \beta  \bd \tvp)
+ \del \r \bd \r + \r^{-2} \del \tilde \vp \bd \tilde \vp
}
 $$ + \ q_1 (\del u \bd \tvp - \bd u \del \tvp)
+ q_2 (\del v \bd \tvp - \bd v \del \tvp)
+ {\cal R} (\p_0 - \ha \ln \r^2 ) ]\ .   $$
Here $\a,\b, q_i$ ($i=1,2)$ are constant free parameters of dimension
$cm^{-1}$, i.e. the dimensionless parameters of our model are $\sqrt{\a'}\a, \
\sqrt{\a'}\b,
\ \sqrt{\a'}q_i$ and  $r=R/\sqrt{\a'}$.
Ignoring  target space topology, the  two models \lagrq\ and \lag\  are of
course ``locally-equivalent"; in particular,
\lag\  also  solves the  conformal invariance equations.
The corresponding  conformal field theories will  not, however,  be equivalent
because of the compactness of
$y= \ha (u+v)$ and $\tvp$ (if $R\not=\infty$ the
redefined coordinates $u + \a \td \vp $ and $v + \b \td \vp$
will be  well-defined, i.e. periodic,
 only for special values of $\a'\a/ R$ and $\a'\b/ R$).

The string models we are going to discuss can be
 obtained from \lag\ by making
the duality transformation in  the $\tvp$   direction. Gauging the
corresponding isometry and introducing the zero gauge field constraint with the
Lagrange multiplier $\vp$  \refs{\busch, \rocver}
we find the dual action (we add also the  flat $x_3$-direction):
$$  I={1\over \pi\alpha '}\int d^2 \s\big[  \del u \bd v
+  \del \r \bd \r $$
\eqn\laggd{ + \  F (\r)  \r^2
[  \del \vp  + (q_1 + \b ) \del u + q_2 \del v ]
  [ \bd \vp + q_1\bd u + (q_2 -\a)\bd v] }  $$
  +  \ \del x_3 \bd x_3  +   {\cal R} (\p_0 +  \ha \ln F ) \big]\ , \ \  \ \ \
\  F\equiv (1 + \a\b \r^2)\inv  \ ,     $$
or, equivalently (dropping the
 total derivative term $\del y \bd t - \del t \bd y$)
$$ I={1\over \pi\alpha '}\int d^2 \s\big[ - \del t \bd t
+\del y \bd y +  \del \r \bd \r $$
\eqn\lagg{
 + \  F (\r)  \r^2
[  \del \vp  +  c_+  \del y  + c_- \del t ]
  [ \bd \vp + a_+ \bd y  +   a_- \bd t]   }   $$
  +  \  \del x_3 \bd x_3 +   {\cal R} (\p_0 +  \ha \ln F ) \big]\ ,  $$ $$
  \  F\inv =   1 + \four [(a_+-c_+)^2 - (a_- - c_-)^2] \r^2  \ .     $$
We have introduced the following linear combinations of the parameters which
will be convenient to use below along with  $\a,\b,q_i$
$$  a_+\equiv  q_+-\a\ ,  \ \ \  c_+\equiv  q_+ + \b  \ ,  \ \ \
a_-\equiv  -q_- -\a\ ,
\ \ \ \   \ c_-\equiv -q_- -\b\ , \ \ \ q_\pm \equiv q_1 \pm q_2 \ ,    $$
\eqn\dee{ \a= \ha (c_+ -a_+ + c_- - a_-) \ ,
 \ \  \ \ \b = \ha (c_+ -a_+ - c_- + a_-) \ ,  } $$
 \ \  q_+ = \ha (c_+ + a_+ + c_- - a_-) \ ,
\ \  \ \  q_- = \ha (-c_+ + a_+ - c_- - a_-) \ .
$$
The physical meaning of these parameters will become clear in the next section.
The angle $\vp$ in \lagg\  has  period $2\pi$ so that, in
the  trivial
case of $\a,\b,q_i=0$,  we get  a flat 5-space with  coordinates $t,\ y,\
x_1=\r \cos \vp, \ x_2 = \r \sin \vp, \ x_3$.

The  \sm  \laggd\ can be interpreted as a particular  $O(3,3;R)$  ($t,y,\td
\vp$)  duality transformation
(making  shifts of $t$ and $y$ by  $\td  \vp$, adding torsion terms, and
performing  the simple duality in $\td \vp$)
of the direct product model \lagrq\  $R_t \times S^1_y \times $ (dual
2-plane)$_{\r,\vp}$ or, since the latter  itself is a duality rotation of a
plane,  directly
of the  flat model  $R_t \times S^1_y \times  R^2$.
Since $y$ and $\td \vp$ are compact,  this $O(3,3;R)$   duality transformation
with  continuous values of the parameters
is $not$ a symmetry of one  conformal theory (cf. \givroc)  but maps a trivial
flat-space
CFT into a non-trivial one we  shall discuss below.  Members of the  resulting
class of conformal theories   parametrized by $\a,\b,q_\pm $  will be
invariant  only  under some  special $ O(2,2;Z)$ duality transformations.

Eq. \laggd\  or \lagg\ is the  action of the  string model investigated
in this paper (another form of the action is (2.13)).
It  contains four parameters but one of them, $q_- =q_1-q_2$,
can be removed by a coordinate transformation.  In fact,   the dependence on
$q_i$ can be  formally
eliminated by  introducing the new coordinate
\eqn\new{ \vp' \equiv  \vp + q_1 u + q_2 v  = \vp + q_+y -
 q_- t\ . }
However, this transformation  does not,   in general,  preserve  the
equivalence of the corresponding conformal models because of the different
periods of the compact coordinates $\vp$ and  $y=\ha(u+v)$
(the equivalence is preserved only in the special cases when
$q_+R=m= \pm 1, \pm 2, ...$).
At the same time, assuming    $t$ is  non-compact,
 we can, in fact,  shift $\vp \to \vp +   \l t$, eliminating  the dependence on
$q_-$  by  choosing  $\l=q_-$.
As a result, the  models we shall discuss  below will essentially
 depend
only on the {\it three} parameters $\a,\b$ and $  q_+$ (as well as  on $R$).

Different choices of $q_-$ correspond to different definitions of the
coordinate $\vp$  (or different  choices of  ``frames" or ``gauges").
For example, one may set  $q_-=0 $, i.e. (see \dee)
 \eqn\gggq{ a_+ =  q_+ -\a\ , \  \ \ a_-=  -\a\ ,\ \
\  c_+= q_+ + \b\    , \ \ \     \ c_-= -\b\ , }
or  $q_-=-\a $, i.e.
\eqn\gau{ a_-=0\ , \ \ \  c_-=\a-\b\ , \ \ \  a_+ =  q_+-\a, \ \
\  c_+= q_+ + \b  \ .   }
This second gauge  may be called    a ``chiral gauge"   since when $a_-=0$
there is no $\bd y$ coupling  term in  \lagg.
Below it  will be convenient  to present
some of the results   in  a  general  form without  specifying  a particular
gauge.

 Since the models \lag\ and \lagg\ are related by the  simple $\vp$ -
duality transformation  they should represent the same string solution, i.e.
the  corresponding CFT's  should be  completely equivalent  \refs{\rocver} (see
also \refs{\alv, \rabi}). We  will explicitly  demonstrate  this   in Sections
4,5
 by  solving the theory (computing the quantum
Hamiltonian and the  partition function)
   starting directly  with
  the  action  \lagg\   and    with  the simpler dual  action  \lag.

Let us now perform the duality transformation  of  the \sm
\lagg\ in the  compact $y$-direction.
We  get\foot{If one  adds to  \lagg\
the total derivative term  $C (\del y \bd t - \del t \bd y), \ C=\const$,
one gets  $\td I$  with $\td y$ replaced by ${\td y}' = \td y - C t$.
This leads to a redefinition of parameters $a_-, \ c_-$. }
 $$ \td  I={1\over \pi\alpha '}\int d^2 \s\big[ - \del t \bd t
+\del \td y \bd \td y +  \del \r \bd \r   $$
\eqn\ldu{
 + \   {\td F }(\r)  \r^2
[  \del \vp  +  c_+  \del \td y  + c_- \del t ]
  [ \bd \vp -  a_+ \bd \td y  +   a_- \bd t]   }
$$
  +  \  \del x_3 \bd x_3 +   {\cal R} (\p_0 + \ha \ln  \td F ) \big]\  ,  $$
$$ {\td F\inv  (\r) } \equiv  F\inv (\r)  [ 1 + a_+c_+ F(\r) \r^2]
 =   1 + \four [(a_+ + c_+)^2 - (a_- - c_-)^2] \r^2 \ .  $$
This action is  related  to \lagg\ by  any of the two transformations
\eqn\ded{ y\to \td y\  , \ \ \ \ \ a_+ \to - a_+\ ,  \   \  \ \ \  or \  \ \  \
\ \
 y \to -\td y\ , \ \ \ \ \ c_+ \to - c_+ \ .     }
The same  conclusion is reached by making  the $y$-duality directly
in  our starting model \lag\ (the duality transformations in $\tvp$ and $y$
directions are independent).
Note that the \sms with $a_+=0 $ or  $ c_+=0$ are ``self-dual", i.e.
preserve their form under $y$-duality.

Again, the  two $y$-dual models \lagg\ and \ldu\    should represent the same
conformal theory (provided the period of the dual coordinate $\td y$  is  taken
to be
 $2\pi \a'/R$):
the CFT (Hamiltonian, spectrum, etc.)
 will
be invariant under  replacing   $R$ by
$\a'/R$, interchanging the winding  $w$ and momentum  $m$  quantum numbers
 in the $y$-direction
$and$ transforming $a_+ \to - a_+$ (or $w\leftrightarrow -m, \ \ c_+ \to -
c_+$).
The additional  transformation of a parameter $a_+$ (or $c_+$)
 is an interesting  feature of this  model.
 Naively, it  may seem   that  the model has  just $one$
   non-trivial compact direction ($y$),  so that   the CFT should be invariant
just under the transformation $R\to \a'/R$ and interchanging  winding   and
momentum numbers
($a_+ \to - a_+$  should then  be an additional
 symmetry of CFT).
As we shall see in Sections 4,5  this is not the case; for example, the
partition function  $Z(R, a_+, c_+, q_+)$ is  $not$ invariant  just under
$R\to \a'/R$
(as happened in the  ``self-dual" $a_+=0, q_i=0$  model of \ruts)
but satisfies
\eqn\duytr{ Z(R, a_+, c_+, q_+)=Z(\a'/R, -a_+, c_+, q_+)=Z(\a'/R, a_+, -c_+,
q_+) \ . }
This  may be compared to the case of 2-torus model, where  one
has 4 parameters transforming under $O(2,2;Z)$ duality.
 The  need to transform
an extra parameter ($a_+ $ or $c_+$) under the $y$-duality
is  related to the  presence  in \lagg\  (or \lag) of the couplings  of $y$ to
  another angular variable $\vp$ (or $\td \vp$).

\subsec{Conformal invariance }
Using the combination $\vp'$ \new\ the  model  \laggd\  can be
represented  also  as
 \eqn\lagge{ I={1\over \pi\alpha '}\int d^2 \s\big[
 F(\r) (\del u  -2 \a  A) (\bd v   + 2 \b \bar A  )
+ \del \r \bd \r +
   \r^2 \del \vp'  \bd \vp'   } $$
+\  \del x_3 \bd x_3 +   {\cal R} (\p_0 +  \ha \ln F ) \big]\ ,     $$
\eqn\aaa{  A=  \ha \r^2 \del \vp' \ , \ \ \  \ \ \ \bar A = \ha \r^2 \bd \vp' \
. }
This form of the action is useful
in order to demonstrate its exact conformal invariance.\foot{The
 argument about  the duality  to  the simple  model \lag, which is itself
formally (ignoring topology)  related
to a flat space by a combination of a coordinate transformation and duality,
guarantees only that the leading-order $\b$-function equations are satisfied,
i.e.  is not by itself  sufficient in order to  prove that there exists a local
scheme in which \laggd\ is  conformal to all orders in $\a'$.}
In order to check the \sm
conformal invariance conditions  (which are local and covariant target space
equations)
 one  may ignore the difference between $\vp$ and $\vp'$, or simply set
$q_i=0$.
Then  the Lagrangian in \lagg\ becomes  ($I=\int d^2 \s L{/\pi\alpha '}$)
\eqn\laggo{ L(q_i=0) \  =
 F(x) [\del u  -2 \a  A_i (x)  \del x^i][\bd v   + 2 \b A_i (x)
\bd x^i ] } $$
+\ \del x_a \bd x^a +   {\cal R} (\p_0 +  \ha \ln F ) \ , $$
$$ A_i  = - \ha \ep_{ij} x^j\ , \ \ \ \ \ \ F\inv = 1 + \a\b x^i x_i \ ,  \ \ \
x^a=(x^i,x^3) \ ,  $$
i.e. it corresponds to a special case of a  generalized ``$F$-model"
considered  in Section 5 of ref. \horts
\eqn\twvec{ L = F(x) \big[ \del u + 2B_a (x)  \del x^a ] [   \bd v
+ 2\bar B_a (x) \bd x^a\big] +
       \del x_a \bd x^a +  {\cal R} \phi(x) \ .    }
When  the vectors  $B_a$ and $\bar B_a$ have constant field strengths
${\cal F}_{ab}$ and $\bar{\cal  F}_{ab}$
this model was shown    to be conformally invariant
to all loop orders
provided  \horts
\eqn\sonm{ \del^a\del_a  F\inv
+   2  { \cal F}^{ab}  {\bar {\cal  F} }_{ab} =0 \  , \ \ \ \  \ \
\p= \p_0 + \ha \ln F  \ .   }
In the  case of \lagg\
\eqn\son{ {\cal F}_{ij} = -\a \ep_{ij} \ , \ \ \ \bar{\cal  F}_{ij} = \b
\ep_{ij}\ ,
\ \ \   F\inv = 1 + \a\b x^i x_i\ ,  \ \ \ i=1,2 \ , }
so that  the condition \sonm\ is  indeed satisfied.

\subsec{Special cases}
The $R=\infty$ limit of our
model \lagg\  can    also be represented  by the
Lagrangian  \laggo\  corresponding to the choice  of $q_i=0$.
 Indeed, when both $y$ and $t$ are non-compact
 the transformation $\vp\to \vp'$  \new\ is
completely legitimate and thus the  resulting CFT
 should  depend only on the two parameters $\a,\b$.\foot{The finite $R$ theory
with $q_i=0$ and the $R=\infty $ theory are of course inequivalent as conformal
field theories.}
If we drop the trivial $x_3$-direction,   \laggo\ describes an
interesting  exact $D=4$ string background  (see Section 3.6) on  which, as on
flat space, the
string theory  can be explicitly   solved as discussed in Sections 4 and 5.

The  simplicity of the $q_i=0$  model \laggo\ is also  reflected in
  the fact that  the Lagrangian
related to \laggo\  by the   $y$-duality transformation
is quadratic in $x^i$ and thus
represents a straightforward generalization of the model of
\refs{\horts,\ruts}.
In fact, setting $q_i=0$ (${\td F  (\r) }=1$) in  \ldu\   we get a  generalized
``non-chiral"
 plane-wave type model  (see  \refs{\hrt,\horts} and references there)\foot{To
obtain this  dual action
directly from \laggo\ one needs to add to \laggo\ the total derivative
term  $(\del y \bd t - \del t \bd y)$  which was dropped
in going from \laggd\ to \lagg. If one starts directly with \laggo\
one obtains $\td I$  with  $\td y $ replaced by $\td y + t$. }
 \eqn\lduu{\td  I={1\over \pi\alpha '}\int d^2 \s\big(\del \td u  \bd \td v
 +  \a\b \r^2 \del \td u \bd \td u
+ \a \r^2 \del \vp\bd \td u + \b \r^2 \bd \vp\del \td u} $$
 + \
 \del \r \bd \r +  { \r^2} \del \vp \bd \vp
  +   \del x_3 \bd x_3 +   {\cal R} \p_0 \big)\
, \ \  \ \ \td u\equiv  \td y -  t \  , \  \ \td v \equiv  \td y + t \ ,   $$
or, equivalently,
\eqn\lduu{\td L= \del \td u  \bd \td v
 +  \a\b x^ix_i \del \td u \bd \td u
+ \a \ep_{ij} x^i \del x^j \bd \td u + \b \ep_{ij} x^i \bd  x^j \del \td u
 +  \del x_i \bd x^i
  +   \del x_3 \bd x_3 +   {\cal R} \p_0 \ . }
Solving for $\td v$ one finds that $\td u$ satisfies the free equation of
motion
and then the  linear equation for $x^i$  is also easily solvable. Thus
the operator quantization of this model can be carried out essentially in the
same way as  was done for its special case of $\a=0$ in \ruts.
Equivalently, in the path integral approach,  the integral over $\td v$
implies the constraint that $\td u$ is given by the zero (winding)
 mode term  only,
so that, e.g.,   the partition  function on the torus is readily computable.
Since the resulting CFT is  duality-invariant,
in this way we get  also the  partition function of the equivalent $q_i=0$
model \laggo\  (see Section 4.2).

 Let us now consider  some  other special cases of models
\laggd\ or \lagg.  When $\a\b=0$ the dilaton field is constant.
For $q_i=0$ and $  \a=0$ ($\b=0$ gives  a  similar  model with  $\del \to \bd$)
 we  get back to  the  constant magnetic field model
of \refs{\horts,\ruts}
  \eqn\lgg{ L(q_i=0, \a=0) \
= \ \del u \bd v   + \b \r^2 \bd \vp  \del u   +  \del \r \bd \r
+ \r^2 \del \vp \bd \vp +   \del x_3 \bd x_3 +   {\cal R} \p_0  }
$$ = \ \del u \bd v   + \b \ep_{ij} x^i
\bd  x^j \del u   +  \del x_i \bd x^i +   \del x_3 \bd x_3 + {\cal R} \p_0\  .
$$
More generally, let us  choose the gauge
$a_-=0$ \gau\  and consider the   2-parameter subclass of   models defined by
the condition $a_+=0$
(i.e. $q_1=0, \ q_2 =\a$)
  \eqn\lggr{L(a_\pm=0) \  = \ \del u \bd v
+  \del \r \bd \r +
 F(\r)\r^2  (  \del \vp  + \b \del u + \a \del v )  \bd \vp }     $$
  + \
\del x_3 \bd x_3  +  \  {\cal R} [\p_0 +  \ha \ln  F (\r)  ]  \  ,   \ \ \ \
 F\inv = 1 + \a\b \r^2 \  , $$
or, up to total derivative,
\eqn\lggr{ L(a_\pm=0) \ = \ -   \del t \bd t + \del y \bd y +  \del \r \bd \r +
   {\r^2 \ov 1+ \a\b \r^2 }
(\del \vp   +c_+\del y + c_- \del t )  \bd \vp
 }
$$ +  \ \del x_3 \bd x_3   +  {\cal R} [\p_0 -  \ha \ln (1+ \a\b \r^2)] \  . $$
These models with $a_+=0$ (or $c_+=0$) are ``chiral" (or ``heterotic") in the
sense that the  background gauge field only couples to the left or to the right
sector.
The special case of   \lggr\  with $ c_-=0$ (i.e. with $\a=\b$)
 is the string model  \tsem\
  corresponding  to the dilatonic Melvin solution
 \gibma\  which  describes  a magnetic flux-tube  background  \melv\
 $$L(a_\pm=c_-=0) \   = \ - \del t \bd t +  \del \r \bd \r +  {F(\r) \r^2}
\del \vp   \bd \vp  +
\del y \bd y +
  {2\b F(\r)  \r^2  }
\del y   \bd \vp      $$
\eqn\lggm{  +
\del x_3 \bd x_3  +  \  {\cal R} [\p_0 +  \ha \ln F(\r) ]  \  ,  \ \ \ F\inv =
1+ \b^2 \r^2 \ ,
\ \  \  \b=\a =\ha c_+ \ .  }
In the non-compact case $R=\infty$
the  non-trivial 3-dimensional $(y,\r,\vp)$-part of this model can be
considered
\tsem\ as a special singular limit of the $SL(2,R)\times R/R$ ``charged black
string"
coset  model \hoho.\foot{Correspondingly, the  $(\r,\vp)$-part of the Melvin
background \melv\ can be interpreted as a  limit of the Euclidean version of
the charged $D=2$ black hole. A.T. is grateful to A. Strominger for this
remark.}

Changing the parameter $\a$ in eq. \lggr\ from $0$  to $\b$ we thus interpolate
between the  constant magnetic field   model \lgg\  and the flux-tube  Melvin
model
\lggm.  The parameter $1/\sqrt{\a \b }$   represents an effective
width of the flux tube inside   which the magnetic field is approximately
constant. As  \lgg\ and    \lggm\   all   the
members  of the  2-parameter class of models  \lggr\  are  invariant with
respect to the duality transformation in the Kaluza-Klein $y$-direction (cf.
\ded).

\subsec{Generalizations}
The idea of  constructing  non-trivial  $D=4$ string backgrounds
by starting with simple \sms\ such as \lag\
 and applying  duality transformations in  angular coordinates
can be generalized  in several ways.
The model \lag\  admits the following generalization
which preserves the number of its isometries:  if we shift $v \to v + \k u,
\ \ \k=\const$  in \lag\
then the new model  will not be equivalent to the $\k=0$ one
as long as $y$ is compact.
The corresponding dual  \sm action \laggd\ will take the form
$$  I={1\over \pi\alpha '}\int d^2 \s\big[  \del u \bd v  +  \k \del u \bd u
+  \del \r \bd \r $$
\eqn\laggde{ + \  F (\r)  \r^2
[  \del \vp  + (q_1' + \b' ) \del u + q_2 \del v ]
  [ \bd \vp + q_1'\bd u + (q_2 -\a)\bd v] }  $$
  +  \ \del x_3 \bd x_3  +   {\cal R} (\p_0 +  \ha \ln F ) \big] \ .    $$
$\k$ will be a new   parameter of the corresponding CFT. For example, in
computing the partition function on the torus $u$ and $v'=v + \k u = (1+\k )y +
(1-\k )t$
will now have different radii of their  winding zero mode parts, namely,  $R$
and $R'= (1+\k)R$. As a result, $Z$ will depend on $R$ $and$ $R'$  (as well as
on $\a,\b,q_+$).
In the rest of  this paper we shall  assume for simplicity that $\k=0$.

\lr\give{A. Giveon, \mpl A6 (1991) 2843.}
One  may  also  replace the ``dual 2-plane" part of \lag\
by the \sm representing  the (dual to the) Euclidean  $D=2$ black hole  (i.e.
the
gauged WZW model $SL(2,R)/U(1)$ \refs{\wit,\dvv,\give}),
\eqn\lagx{ I={1\over \pi\alpha '}\int d^2 \s \big[
( \del u + \a \del \tvp)  (\bd v + \beta  \bd \tvp)
+ \del \r \bd \r +  f^2(\r) \del \tilde \vp \bd \tilde \vp
}
 $$ + \ q_1 (\del u \bd \tvp - \bd u \del \tvp)
+ q_2 (\del v \bd \tvp - \bd v \del \tvp)
+ {\cal R} (\p_0  - \ln g(\r) ) \big]\ ,    $$
with
\eqn\fii{ f(\r) = b  \coth b\r \  , \ \ \  \ g(\r) = b\inv  \sinh b \r \ ,  }
 or
\eqn\see {  f(\r) = b\inv   \tanh b\r  \ , \ \ \ \   g(\r) =  \cosh b \r \ . }
The resulting \sm is conformal to all orders in the ``leading-order" scheme
\tsett.
The constant $b$  ($\a' b^2 = 1/k$)  is fixed by the condition  of the
vanishing of the total central charge, $ 2 + 3k/(k-2) -1  + N -26=0$,
where $N$ is a number of extra free bosonic dimensions. The value of $b$ can be
made
continuous by introducing a linear dilaton coupling in $t$ direction.
The action \lagx\ with \fii\ reduces to \lag\ in the limit $b\to 0$ (i.e. $N\to
22$).
The \sm dual to  \lagx,\fii\  is given by the following  generalization of
\lagg\
$$  I={1\over \pi\alpha '}\int d^2 \s\big[  \del u \bd v
+  \del \r \bd \r $$
\eqn\lggdx{ + \  b^{-2}   {\rm tanh}^2b\r\  F (\r)
[  \del \vp  + (q_1 + \b ) \del u + q_2 \del v ]
  [ \bd \vp + q_1\bd u + (q_2 -\a)\bd v] }  $$
  +  \ \del x_3 \bd x_3  +   {\cal R} (\p_0 +  \ha \ln F'(\r) ) \big] \ ,
$$
\eqn\uuu{ F\inv = 1 + {\a\b \ov b^2} {\rm \tanh}^2 b\r \ ,
\ \ \ \   {F'} =  F  (\r) {\rm \cosh}^{-2} b\r   \ . }
The parameter $b$ ``regularizes" the large $\r$  form  of the
 models \lagx\ and \lggdx\ (cf. \lag, \laggd):  the  target space becomes  a
product of a Minkowski
3-space $(t, \r, x_3)$  and  a  ``twisted"  2-torus $(y, \vp)$.
The  ``mixing" of the Kaluza-Klein and angular $\vp$ coordinates
 makes the corresponding conformal theory and its space-time interpretation
quite  non-trivial.

The  \sm  \lggdx\ can be   described as a particular  $O(3,3;R)$  duality
transformation
(shifting of $t$ and $y$ by  $ \td \vp$, adding torsion, and  performing
duality in $\td \vp$)
of the direct product  model $R_t \times S^1_y \times$ $[SL(2,R)/U(1)
$WZW$]_{\r,\vp}\times R_{x_3}$.  As was already mentioned in Section 2,  for
general values of the parameters such  $O(3,3;R)$  duality transformation  is
not a symmetry of the  original CFT, i.e. it
 gives a new conformal theory.

For special  values of the parameters the 3-dimensional  $(y, \r, \vp)$
part of  this   model is  equivalent
to the  $[SL(2,R)_{k} \times U(1)_{k'}]/U(1) $  gauged WZW model
(or, for  non-compact $y$ and $\vp$, to the  ``charged black string" model of
\hoho).
For $b\not=0$ the   conformal theory corresponding
to \lagx\ or  \lggdx\  is much more complicated than the  $b= 0$  one
 discussed in the main part   of the present  paper.\foot{Another
generalization which will not be explored here is to consider
 the coordinate  $t$ in \lag\  to be periodic.
It may be of interest  in connection with a finite temperature  description
as well as for  construction of related
(e.g. by   analytic continuation) models where   the role
of time   can be assigned to one of the other  coordinates. This may permit,
in
particular,  to obtain models with   expressions for  the
 electric and magnetic fields  being interchanged.
One more possibility  is to start with
the  Euclidean  $D=4$ model  which is a combination of the two ``dual
2-planes", make coordinate  shifts preserving the number of
 isometries, apply  the duality in two angular coordinates
and then look for  the range of parameters  (or analytic continuations)
for which the resulting \sm has Minkowski signature.}

\newsec{ Space-time/low energy  field theory   interpretation }
To give a space-time interpretation  to
 the string models \lagg, i.e. to determine the corresponding
string background geometry,
   we shall  make  the  Kaluza-Klein--type
rearrangement of terms in the  above \sm actions
    (see, e.g.,   \refs{\duff,\horts})
$$ I_5={1\over \pi\alpha '}\int d^2 \s\big[ ( G_{MN} +B_{MN})(X)
\del X^M \bd  X^N  + {\cal R} \p
(X) \big]$$  $$
= {1\over \pi\alpha '}\int d^2 \s\big[  (\hat G_{\m\n} +B_{\m\n})(x)
\del x^\m \bd x^\n + \  e^{2\s(x) } [\del y+ {\cal A}_\m (x) \del x^\m][\bd y+
{\cal A}_\n (x)
\bd x^\n]  $$
\eqn\lagr{
 + \  {\cal B}_{\m   } (x) (\del x^\m \bd y- \bd x^\m \del y )  + {\cal R} \p
(x) \big]\ , }
where $X^M= (x^\m, x^5), \ x^\m=(t, x^i, x^3), \ x^5\equiv y$
and
\foot{The modulus field $\s $ should  not be confused with the
world-sheet coordinates $\s^\a $.}
\eqn\fgfg{\hat G_{\m\n} \equiv  G_{\m\n} - G_{55}{\cal A}_\m {\cal A}_\n
\ , \ \ \ G_{55}\equiv  e^{2\s}\ , \ \ \ \
 {\cal A} _\m\equiv   G^{55}  G_{\m 5}\ ,  \ \ {\cal B} _\m \equiv
B_{\m 5}\ .  }From the point of view of the
low-energy effective  field theory,
  this decomposition corresponds to starting
 with the $D=5$ bosonic string effective action and
 assuming  that one spatial dimension $x^5$ is compactified on a small
circle. Dropping the massive Kaluza-Klein modes  one
finds the following dimensionally reduced  $D=4$  action (see,  e.g., \mahar)
\eqn\acttp{  S_4 = \int d^4 x \sqrt {\hat G }\  e^{-2\phi + \s}    \ \big[
  \   \hat R \ + 4 (\del_\m \phi )^2 - 4 \del_\m \phi \del^\m \s }
$$  - {1\ov 12} (\hat H_{\m\n\l})^2\  - \fourth e^{2\s} ({  F}_{\m\n}
({\cal A}))^2
-\fourth  e^{- 2\s} (F _{\m\n} ({\cal B}))^2
  + O(\a')   \big]  \  , $$
where,   in addition to \fgfg,  we have  defined
 \eqn\jjj{  F_{\m\n} ({\cal A}) = 2\del_{[\m}
{\cal A} _{\n]}  \ ,   \ \ F _{\m\n}({\cal B}) = 2 \del_{[\m} {\cal B}
_{\n]}  \  , \ \ \
\hat H_{\l\m\n} = 3\del_{[\l} B_{\m\n]} - 3 {\cal A}_{[\l} F_{\m\n]}
({\cal B})
\   .   }
As follows from \lagr, the  \sm duality transformation in $y$
induces the following  target space transformation
\eqn\daa{\A \to \pm \B\ , \  \ \
 \B \to \pm  \A \ , \ \   \
\s \to -\s\   , \  \  \ \p \to \p - \s\ , \  \ \  }
$$ \hat G_{\m\n}  \to \hat G_{\m\n}  \ , \  \ \
B_{\m\n}  \to B_{\m\n} + \A_\m \B_\n - \B_\m \A_\n\ , \ \ \
\hat H_{\m\n\l} \to \hat H_{\m\n\l}  \ , $$
which is obviously  the invariance of \acttp.
Dual backgrounds related by  \daa\  correspond to the same string
solution (CFT).

\subsec{3-parameter class of new string solutions}
Let us define the two functions $F$ and $\td F$ (which already appeared in
\lagg, \ldu)
\eqn\deey{
F (\r) \equiv {(1 + {\r^2 \ov\r _0^2 })\inv  }\   , \   \ \ \ \
\F (\r) \equiv {(1 +  {\r^2\ov \td \r_0^2})\inv  }\ , }
\eqn\deeyy{
\r_0^{-2}=\four [(a_+ - c_+)^2 - (a_--c_-)^2]=\a\b  \  , }
$$
  \td \r_0^{-2}=\four [(a_+ + c_+)^2 - (a_- -c_-)^2]=q_+(q_++\b-\a) \ , $$
$$
 \F (\r)= [F(\r)]_{a_+\to  -a_+}= [F(\r)]_{c_+\to  -c_+}\ .
$$
Note that $\r_0^2 $ and $\td \r_0^2$ can  take  both positive and negative
values.
Starting with  \lagg\ we find the following expressions for the  $D=4$
background fields:

\noindent
 dilaton and modulus scalars
\eqn\scal{ e^{2(\p-\p_0) } =  F (\r)   \ ,  \ \ \
\ \ \   e^{2\s } =     {F(\r) \ov \F(\r)}\    \ ,
}
abelian vector  and axial-vector  potentials
\eqn\vee{\A=
{\cal A}_\m dx^\m  = \ha
\F (\r) \r^2  [ (a_+  + c_+) d \vp + (a_+c_- + c_+a_-)dt] \ , \ \ \ } $$
\B=  {\cal B}_\m dx^\m  = \ha
{F(\r) \r^2 } [ (a_+  - c_+) d \vp + (  a_+c_- - c_+a_-)dt] \ ,  $$
with field strengths
 \eqn\fies{  d{\cal A} =  \F^2 (\r)  \r d\r \wedge  [ (a_+  + c_+) d \vp +
(a_+c_- + c_+a_-)dt]\   ,   }
$$
d \B =  F^2 (\r)  \r d\r \wedge [ (a_+  - c_+) d \vp +  (  a_+c_- - c_+a_-) dt]
\ ,
 $$
the effective  $D=4$ metric
\eqn\qqqs{  ds^2_4\equiv  \hat G_{\m\n}dx^\m dx^\n =
 -dt^2  +      F(\r) \r^2 (d\vp + c_-dt)(d\vp + a_-dt)  } $$
 - \ \four \F(\r )F(\r) \r^4  \big[ (a_+  + c_+) d \vp + (a_+c_- +
c_+a_-)dt\big]^2 +  d\r^2 +  dx_3^2 \ ,
   $$
\eqn\deter{
\det \hat G =   -\r ^2 F(\r )\F (\r ) \ ,
}
and the antisymmetric tensor
\eqn\mmeh{ B=   \ha B_{\m\n} dx^\m \wedge dx^\n
 =  \ha (a_- - c_-)  F(\r)  \r^2 d\vp\wedge dt \ . }
Note that  the invariant antisymmetric tensor field strength
$\hat H_{\m\n\l} $ in \jjj,  is, like $B$ itself,  proportional to $a_- - c_-
=\b-\a$.
The  duality  transformation in the  $y$ direction
\daa\ is indeed induced by  $a_+\to -a_+$ (i.e. $F\to \F$, etc.) in agreement
with   \ldu,\ded.
The  duality invariant `shifted'  $D=4$ dilaton  which determines the
effective coupling in the  $D=4$ action  \acttp\  is  given by
\eqn\dii { \Phi \equiv \p - \ha \s = \p_0   + \four \ln [ F(\r) \F(\r) ]  \ . }
In general,  we get a stationary (``rotating") metric  of the form
\eqn\meet{  ds^2_4= - f_1 (\r) dt^2 + f_2 (\r)  d\vp dt  +  f_3 (\r) d\vp^2  +
 d\r^2  + dx_3^2 \ , }
and  the gauge field strengths  which have   both magnetic and electric
components.  There are  pure magnetic solutions but no purely electric ones,
except in special limits.\foot{It is  possible to   formally exchange
the roles of the electric and magnetic fields by assigning the  role  of time
to the $\vp $ coordinate (e.g. by the rotation $\vp\to i\vp $, or by specifying
to
 the  regions where $\vp $ has  a time-like
signature). }

Using the  expressions for  $F$, $\F $  in \deey\ and  \deeyy\ one finds  that
for generic values
of the parameters the metric becomes degenerate   at $\rho \to \infty $
and the  asymptotic $\r\to 0$ and $\r \to \infty$ forms of the metric are
\eqn\meeta{  (ds^2_4)_{\r \to 0}= -  dt^2 +  (a_- + c_-) \r^2  d\vp dt   +
d\r^2  +  \r^2 d\vp^2 + dx_3^2 \ , }
\eqn\meett{  (ds^2_4)_{\r \to \infty }= -  k_0 ( dt +  k_1 d\vp )^2  + \r^{-2}
k_2 d\vp^2  +   d\r^2   + dx_3^2 \ ,  \ \ k_s= \const \ , }
i.e. the 2-space $(\r, \vp)$
 which looks like a  2-plane near  the origin  $\r=0$
 at large $\r$
becomes a (rotating) cylinder with  radius going to zero  as $\r\to \infty $.
This is similar to the topology of the Melvin solution \gibma.

In  the  special cases when $F=1$ (or $\F=1$)
the determinant  of the metric \deter\ at large $\r$ approaches   a finite
value and,  for $\a\not =\b$, we get instead  a rotating Rindler-type
asymptotic space
\eqn\meetg{  (ds^2_4)_{\r \to \infty }= -\r^2(  dt +  n_1 d\vp )^2   +\rho^{-2}
n_2 d\vp^2   +   d\r^2   + dx_3^2 \ ,  \ \ \ n_s= \const \ .  }
For $\a=\b$ the asymptotics is that of a cylinder,
$(ds^2_4)_{\r \to \infty }= -  dt^2 +   n' d\vp^2   +   d\r^2   + dx_3^2 $.
$F$ or $\td F$ are equal to 1 when either $\a\b =0$ or $q_+(q_++\b-\a )=0$ (see
\dee), i.e.  when either  of the axial or vector field strengths  (cf. \vee) is
 constant (uniformly extends  to infinity).
When both $F=1$ and $\td F=1$  the asymptotic form of the metric is that of the
plane.

For $\a\b <0 $ or $q_+(q_++\b-\a )<0$ the determinant \deter\ has singularities
at $\rho^2=-\r_0 ^2$ or
$\r^2 =-\td \r^2_0 $, respectively. These are, in fact, curvature
singularities, as can be seen from the general expression for the curvature
scalar of metric \qqqs \ which is given in the Appendix.

Excluding the singular cases (and  considering, e.g., the case when the
electric fields vanish, i.e. $a_-=c_-=0$)  the  resulting
geometry  admits  a straightforward
physical interpretation as the one  induced  by the  two
Kaluza-Klein ($U(1)_{\rm v}$ and $U(1)_{\rm a}$)
magnetic fields of strengths ${{\bf B}_{\rm v}}_0=a_++c_+$ and ${{\bf B}_{\rm
a}}_0=a_+ -c_+$
which
are non-vanishing  inside  the  regions
  of characteristic scales $\r_0 $ and $\td \r_0$ (their profile
functions are
$\td F^2 $ and $ F^2$,  see \fies).
As will be discussed in Section 6.3, the  singular configurations  can be
related to  black or naked strings.

We have mentioned in Section 2 that  one  combination
 of the  parameters  $a_-$ and $ c_-$  can be fixed by  a  coordinate
transformation (cf. \new).
 Below we shall use  the ``chiral"  gauge $a_-=0$ \gau\  in which the metric
\qqqs\
 can  be put into  the following    form:
 \eqn\qqqq{  ds^2_4 =
 -[1 + \four a_+^2 c_-^2 \r^4 F(\r) \F(\r) ]dt^2
+ c_- [1+ \four(c_+^2 -a_+^2 -c_-^2)\r^2] F(\r) \F(\r) \r^2 d\vp dt  } $$
+ \ (1-\four c_-^2\r^2) F(\r) \F(\r) \r^2 d\vp^2 +  d\r^2 +  dx_3^2 \ . $$

To clarify the properties of this   3-parameter class of solutions
\scal--\mmeh\ corresponding to the action \acttp\  let us  now consider some
special choices of the free parameters $a_+, c_+,c_-$.

\subsec{Chiral magnetic backgrounds  with constant modulus $\s$ ($a_+=0$) }
The  solutions with $a_+=0$ (or $c_+=0$)
 are precisely the ones which are self-dual, i.e. invariant under \ded, \daa.
 They have  zero modulus,
$ \s=0$, and   are  ``chiral"   (cf.  \lggr):   the two vector
fields $\A$   and $\B$  in \vee\  are equal   up to sign (and have only
magnetic components). In the $a_-=0$ gauge
\eqn\veey{\A= -\B  = \ha c_+
{F(\r) \r^2 } d \vp \  ,   \
\ \ \  d \A = c_+
{F^2(\r) \r } d\r \wedge
 d \vp  \ ,  \ \ \  c_\pm = \a\pm \b \ ,
}
\eqn\ppp{   e^{2(\p-\p_0)}= F = \F =
 {1\ov 1 + \four (c_+^2 - c_-^2) \r^2 }  \ ,  \ \  \ \s=0 \  , \
\ \ \   B=  - \ha  c_-  F(\r)  \r^2 d\vp\wedge dt   \ , }
\eqn\qqqp{  ds^2_4=
 -[dt - \ha  c_-  F(\r)  \r^2 d\vp]^2
 +    d\r^2 +  F^2(\r) \r^2 d\vp^2  +   dx_3^2
   \    .   }
The Melvin solution  \melv\ corresponds to $c_-=0$. Then the metric is static
and
the antisymmetric tensor vanishes.
The solution of {\horts}\ (eq. \hts)   is obtained for  $c_+=c_- = \b $: then
$F=1$ so that the magnetic field strength  and the dilaton become constant
everywhere in space, the  metric is of ``rotating"  type  and $B_{\m\n}\not=0$.

Among other members of this $a_+=0$ subclass there is a
special solution  with
$c_+=0, \ c_-\not=0$ (i.e. with $\a=-\b$).
In this case  both gauge fields  vanish identically,  but the metric, dilaton
and $B_{\mu\nu}$  remain  non-trivial,
\eqn\qeqqp{  ds^2_4=
 -dt^2   +   F(\r)  \r^2 ( d\vp + c_- dt) d\vp
 +    d\r^2   +   dx_3^2
   \     , \ \ \     }
$$  B=  - \ha  c_-  F(\r)  \r^2 d\vp\wedge dt   \ , \ \ \  e^{2(\p-\p_0)}= F=
(1 - \four c_-^2 \r^2 )\inv \ . $$
 Dropping  the trivial $x_3$-direction we get
a $D=3$ string solution  which (like  the $\a=\b$ Melvin model \tsem)
 can be formally (ignoring the issue of different periodicities of coordinates)
identified   with   a special  singular
limit  of the charged black string background of  \hoho.
The corresponding string model (given by  \lggr\ with $c_+=0, c_-\not=0$)
is  an ``antipode"  of  the Melvin model (which has $c_-=0, \ c_+\not=0$)
being  connected  to it
by a  change  of sign of $\a$ and formal replacement of   $y$ by $t$. Thus,
 like  the Melvin model \tsem,   it can be related  (ignoring the
global issues)  to  a  special limit of the $SL(2,R)\times R/R$
gauged WZW model. Equivalently,   if  all  the coordinates  are formally taken
to be non-compact,  this model
is the same as  the $D=3$ \sm  corresponding to  $E^c_2/U(1)$ gauged WZW theory
\sft.

Solutions with $|c_-| > |c_+|$  (i.e. $\a\b<0$ ) correspond to singular
geometries.
 In  fact,
the  derivatives of  $\p$  and $B_{\m\n}$  and the curvature   blow up  on a
$(x_3, \vp$)-cylinder  with radius   $\r^2=4/(c_-^2 - c_+^2)$ where $F\to
\infty$.
Computing the curvature for the metric \qqqp\ in the obvious vierbein basis,
one finds  the following non-vanishing vierbein components
\eqn\rim{
R_{ 0101} = R_{ 0202 }= \fourth   c_-^2 F^2 (\r) \ ,
\ \ \
R_{ 0112 } = R_{ 1201 } = - \fourth  c_-(c_+^2 - c_-^2) \r  F^2 (\r) \ , } $$
 R_{ 1212}= {1\ov 8}  [12c_+^2  -6 c_-^2  - (c_+^2 - c_-^2)^2  \r^4]F^2 (\r)  \
. $$
Note  that the duality-invariant shifted dilaton \dii\ is
also singular at the points where $F=\infty$, i.e. where thus  the effective
string coupling diverges.

\subsec {Static magnetic backgrounds with vanishing antisymmetric tensor field}
According to  \vee\  the subclass of solutions
for which the vector field strengths have only magnetic components
is determined (in the gauge  $a_-=0$) by  the condition
$  c_-a_+ =0.  $
We  have already discussed the case of $a_+=0$ so let us now assume  that
$c_-=0$.
As follows from \lagg,  for $c_-=a_-=0$  (i.e. for $\a=\b$)
  the time direction decouples and  one gets   a static  $D=4$ metric and zero
antisymmetric tensor
(in fact,
$\a=\b$  is the only case  when the metric  \qqqq\  becomes of non-rotating
type).
Indeed, in this case  we get from \scal,\qqqq,\vee \
\eqn\sal{ e^{2(\p-\p_0) } =  F (\r) \ , \
\ \ \   e^{2\s } =  {F(\r)\ov \F(\r)} \ ,}
$$ F\inv = 1 + \four (a_+ - c_+)^2  \r^2  \ , \ \ \
\F\inv  = 1 + \four (a_+  +   c_+)^2  \r^2  \ , $$
\eqn\veme{\A
 = \ha (a_+  + c_+)
{\F(\r)} \r^2  d \vp  \ ,  \ \ \
\B  = \ha (a_+  - c_+)
{F(\r) \r^2 }   d \vp \ , \ \ \  B=0 \ ,  }
\eqn\qqq{  ds^2_4=
 -dt^2  +     d\r^2   + F (\r)  \F (\r)\r^2  d \vp^2 +  dx_3^2 \ . }
The metric \qqq\  is flat near $\r=0$ and, for $\a\not=\b$, it  becomes
effectively
3-dimensional
at large $\r$,
$ (ds^2_4)_{\r\to \infty} =-dt^2  +     d\r^2   + k \r^{-2 }  d \vp^2 +
dx_3^2.$
An interesting simple special case  is  $\a=\b=0, \ q_+\not=0$, i.e.
\eqn\salq{ e^{2(\p-\p_0) } = 1 \ , \
\ \ \   e^{2\s } =  { \F\inv (\r)}   = 1 + q_+^2  \r^2  \ ,  }
\eqn\vemeq{\A
 = q_+  {\F(\r)} \r^2  d \vp  \ ,  \ \ \  \B  = 0  \ , \ \ \   B=0 \ ,  }
\eqn\qqqx{  ds^2_4=
 -dt^2  +     d\r^2   +  \F (\r)\r^2  d \vp^2 +  dx_3^2 \ ,  }
which corresponds to the ``$a=\sqrt 3$" (or ``Kaluza-Klein")  Melvin solution
\gibma.\foot{As was pointed out in \gaunn,
it is possible to obtain the dilatonic Melvin solution
(with arbitrary  value of the coupling parameter $a$) by applying the duality
transformation in  the angular coordinate to the flat space  solution of the
dilaton-Einstein-Maxwell theory.
For $a=1$ and $a=\sqrt 3$ (which are the only cases that can be embedded
into the string effective action \acttp)
this is  the effective field theory analogue  of our construction of the
corresponding string theory in Section 2.}

\subsec{Backgrounds with    uniform magnetic field }
We have seen in  Section 2  that  the string model with  $q_i=0$ \laggo\
and its dual \lduu\  have a particularly simple form.  More generally,
  a simplification occurs
for   $F=1$ or $\td F=1$ when
one of the two parameters $\rho_0 $ or $\td\r _0$ in \deeyy\ is infinite, i.e.
when
$\a\b =0 $ or $q_+(q_+ +\b-\a  )=0$.

Let us first consider the case of $\td F=1$, i.e. $q_+(q_+ +\b-\a  )=0$, or, in
particular,
 $q_+=0$,  corresponding to the case when the magnetic field
associated with the vector  $U(1)_{\rm v}$ is  constant.
To present the expressions for the corresponding background fields  we shall
use  the gauge $q_1-q_2=0$ \gggq\ (as discussed in Section 2, different gauges
are related by the coordinate  transformation  $\vp \to \vp + \l t$).
For $q_i=0$ we  have  $a_+ =  a_-=  -\a\ ,\ \
 c_+= -c_-= \b$ and thus find from  \scal--\mmeh\
that  the modulus is equal
to the  (non-constant part of the) dilaton
\eqn\salr{ e^{2\s } =  e^{2(\p-\p_0) } =  F(\r) = (1 + \a\b  \r^2)\inv  \ ,
\ \ \ \F (\r) =1 \  ,  }
the vector $\A$   has a constant magnetic field strength
\eqn\veme{\A
 = \ha (\b-\a) \r^2  d \vp  \ , \ \  \
\B  = - \ha
{F(\r) \r^2 } [ (\a+\b)   d \vp  -2\a\b dt]  \ ,  }
and the metric and the antisymmetric tensor have  the following simple form
\eqn\qqq{    ds^2_4=  -F(\r)[ dt +  \ha (\a + \b) \r^2 d\vp]^2   +     d\r^2
+  \r^2   d \vp^2 +  dx_3^2 \ ,  }
$$ B
 =    \ha (\b -\a)F(\r)  \r^2 d\vp\wedge dt \ . $$
This class of  backgrounds  is  another generalization  of the  constant
magnetic field solution  \hts\  different from the Melvin-type generalization
\veey--\qqqp: here the deformation  is in the $a_+=-\a$ direction. It induces
nontrivial scalars $\s$ and $\p$,  and
makes $\B$ different from $-\A$  (while  still  preserving   the
``rotational" structure of the metric).

The  curvature  components corresponding to  \qqq\ are
 proportional (as in \rim)  to the  powers of $F$ so that
 for  $\a\b < 0$
 they are singular on a  cylinder  corresponding to the special  value of  $\r
^2=-\td \r_0^2$   where $F=\infty$,
and where   $\p$ and $\Phi$ \dii\ also  blow up.

An interesting special case of  this subclass is obtained for $\a=-\b$.
Then $\A$ has only magnetic component
 while  $\B$ has only the electric one,
\eqn\salry{ e^{2\s } =  e^{2(\p-\p_0) } =  F(\r) = (1 - \b^2  \r^2)\inv  \ ,
\ \ \  B
 =   \b F(\r)  \r^2 d\vp\wedge dt \ ,   }
\eqn\vemey{\A
 = \b \r^2  d \vp  \ , \ \  \
\B  = - \b^2
{F(\r) \r^2 } dt =  -  F(\r) d t  + dt \ ,  }
\eqn\qqqy{    ds^2_4=  -F(\r) dt^2    +     d\r^2   +  \r^2   d \vp^2 +  dx_3^2
\ .  }
The vierbein components of the corresponding Riemann tensor  and curvature
scalar (see the Appendix) are
\eqn\rrr{ R_{01 01 } = \b (2\b^4\r^4 -\b^2 \r^2 -1 ) F^3(\r ) ,
\ \   R_{02 02 } = \b^2 F(\r )  ,  \ \  R=-2\b^2 (2  +  \b^2 \r^2)F^2 (\r) ,  }
 so that $\r^2 = 1/\b^2$ is the singularity.

\def \aa{\td \a}
\def \bb {\td \b}

 The   solutions  with $\a\b=0$  are  $y$-dual \daa\ to
\salr--\qqq. These are the solutions with constant dilaton,
\eqn\salrn{  e^{2(\p-\p_0) } =  F(\r) = 1 \ ,
\ \ \  e^{2\s } = \F\inv (\r) = 1 + \aa\bb \r \ ,      }
\eqn\vemen{\B
 = \ha (\aa-\bb) \r^2  d \vp  \ , \ \  \
\A  =  \ha
{\F(\r) \r^2 } [ (\aa+\bb)   d \vp  -2\aa\bb dt]  \ ,  }
\eqn\qqqn{    ds^2_4=  -\F(\r)[ dt +  \ha (\aa + \bb) \r^2 d\vp]^2   +
d\r^2   +  \r^2   d \vp^2 +  dx_3^2 \ , \ \ \
 }
$$ B
 = \ha (\bb -\aa)  \r^2 d\vp\wedge dt \ ,  \ \ \ \aa \equiv q_+ \ , \ \ \bb
\equiv
q_+ +\b-\a  \ . $$
Indeed, this background  corresponds to the  ``plane-wave-type"
string model \lduu\  which is dual to \laggo.
Thus the two backgrounds \salr--\qqq\  and \salrn--\qqqn\   are  different
``faces" of the same
string solution (they correspond to the same CFT).

The metric \qqq\  or \qqqn\
is that of a homogeneous space only when both  $F$ and $\td F$ are equal to 1,
i.e. when   both magnetic fields are uniform (both $\rho _0$ and $\td\r _0 $
are infinite). This requires that   $\a\b=0 $ and $q_+(q_++\b-\a )=0$,  i.e.
that $c_+=\pm a_+$, leaving only one free parameter (representing the  magnetic
field strength).
 This is  essentially  the solution  \hts\  of \horts\
investigated in \ruts\ with the metric  representing  the space of the
Heisenberg
group.

\subsec{Backgrounds corresponding to the $\vp$-dual model \lag}

In addition to  $y$-dual backgrounds  related by  \daa\
(i.e.  by $a_+\to -a_+$ or $c_+\to -c_+$)
which represent two  different ``sides" of the same string solution as ``seen"
by point-like modes   with linear  or winding Kaluza-Klein momentum  (charge)
 there is another
``face" of  our string solution  (CFT)
which is described by the $\vp$-dual \sm
\lag.
 Representing \lag\ in the form \lagr\
we find  another  $D=4$ solution
of the equations for  the action \acttp\
(we drop the exact pure-gauge   $O(dt)$ and $
 O(d\td\vp\wedge dt) $ parts of the
1-forms $\A,\B$  and  the 2-form $B$)
\eqn\scale{ e^{2(\p - \p_0) } = \r^{-2}  \ , \ \  \ \   e^{2\s } = 1\ , }
\eqn\veee{\A  = -\ha  (a_+ - c_+ ) d \td \vp  \ , \ \ \
\B = \ha  (a_+ + c_+) d \td \vp  \ , \ \ \
\   B=  0 \ , }
\eqn\qqqse{  ds^2_4   =
 -[dt + \ha (c_- - a_-) d\td \vp]^2   +  d\r^2 +  \r^{-2} d\td\vp^2  +  dx_3^2
\ .  }
The metric \qqqse\  can be called  a  ``rotating dual 2-plane".
While $\A$ and $ \B$   are locally trivial and have zero field strengths,
 their configuration  is globally  non-trivial:  $d \td \vp$
is closed but not exact since  $\td \vp$ is periodic
and  $\r=0$ is the  curvature  singularity.\foot{The ``regularized" version
of the dual 2-plane -- the  dual $D=2$ black hole -- has  a ``trumpet" topology
(with the radius of the $\td \vp$-circle being nonvanishing everywhere),
suggesting that $\td \vp$ should be considered as a ``true"  angular
coordinate.}
It is remarkable that the  two such  different  $D=4$ backgrounds as
 \scal--\mmeh\ and \scale--\qqqse\ correspond to the same string solution.


\subsec{$D=4$ solution
 associated with non-compact $R=\infty$ model \laggo}
 When  the coordinate $y$ is non-compact
(i.e. $R =\infty $)
 the model  is characterised by   the
two parameters $\a, \b $.
Dropping the trivial $x_3$ direction we then get  from \laggo\
the following  $D=4$  ($u=y-t,\  v=t+y,\  x_1,x_2)$  exact string solution:
\eqn\rinfi {
ds^2= F(\r ) [du dv+ \r^2 d\vp (\b du-\a dv)+  \r^2 d\vp^2]  +
d\r^2 \ , }
$$
B=-\ha \r^2 F(\r ) d\vp \wedge (\b du +\a  dv )             \ ,\ \ \
e^{2(\phi-\phi_0)} = F (\r) = (1+\a\b\rho ^2)\inv \ .
$$
This is a special case of the  class of solutions   represented by
\twvec,\sonm\
which were found in \horts.
The metric \rinfi\ has two null Killing vectors which  become covariantly
constant
if $\a\b=0$.
In the case  of $\a =0$   this  background corresponds
to  the  $E^c_2$ WZW model of  \napwi\ with \rinfi\  representing (in  proper
coordinates)  the metric of  a monochromatic left-moving plane wave.
Similarly,  for  $\b = 0$ \rinfi\ describes  a right-moving plane  wave.
For $\a\b\neq 0$ the  metric
 is static, as can be seen by  the coordinate transformation
$ u'= 2\b u, \ v'= -2\a v$, i.e.  $y'=\b u-\a v$.
 It may be possible to interpret this background  as    describing  a
``superposition" of two interacting  gravitational  waves moving in  opposite
directions.
  The corresponding curvature  asymptotically (in  transverse space)
goes to zero (and is singular at finite $\r$  if $\a\b <0$).

As in the compact ($R < \infty$) case
the corresponding conformal  field theory
can be  studied   explicitly.
 We shall show in   Sections  4 and  5 that    the    Hamiltonian  of the
$R=\infty$ model  is quadratic in oscillators (but still   non-trivial)
 and the partition function  is equivalent to the partition function of  the
free closed string theory.

\subsec{Generalizations}
Starting with the generalizations   \laggde\ and \lggdx\  of our model \laggd\
we obtain  two different one-parameter extensions of the class of solutions
\scal--\mmeh. The second one \lggdx\  may be  of particular interest  since
 the introduction of the parameter $b$ changes the large $\r$  behaviour of the
background fields (while for small $\r$ the form of the background fields
remains the same as for $b=0$).
In particular, we obtain generalizations of the Melvin  \melv\ and the constant
magnetic field  \hts\ solutions. For example, in the latter case:
\eqn\htse{ ds^2_4 = - (  dt + \ha \b b^{-2} { \rm tanh}^2 b\r \  d\vp)^2  +
d\r^2 + b^{-2} { \rm tanh}^2 b\r \ d\vp^2 + dx_3^2 \ , }
$$ \A = -\B=  \ha \b   b^{-2} { \rm tanh}^2 b\r \ d\vp\   ,  $$ $$
 B=  \ha  \b  b^{-2} { \rm tanh}^2 b\r\ d\vp\wedge dt\ , \ \
\ e^{\p-\p_0} = \cosh^{-1} b\r \ , \ \ \s=0 \ . $$
 The magnetic field is  no longer  uniform
everywhere but  decays  asymptotically with  a characteristic scale $b\inv$,
the space is not homogeneous and  the dilaton is non-constant.

It is possible also to construct new exact string solutions by generalizing the
$q_i=0$ model
\laggo. Indeed, as was pointed out in \horts,
the model \twvec\ is conformal for any $F$ satisfying \sonm.
Thus we may consider solutions of \sonm\  more general than  \son.
In particular, we may add to $F\inv   = 1 + \a\b \r^2$ a solution of the
homogeneous $D=3$  $(\r,\vp,x_3)$ Laplace equation,  obtaining,  e.g.,
 \eqn\geneg{   F\inv (\r)  = 1 + {\m \ln { \r\ov\r_0}}  + \a\b \r^2 \ .  }
Another possibility is
\eqn\gene{   { F}\inv (\r, x_3)  = 1 + {M\ov r}  + \a\b \r^2 \ , \ \ \ \ \ \ \
r^2\equiv  \r^2 + x_3^2 \ . }
Dimensionally reducing the resulting string model along the $y$-direction
we  obtain  again  an  exact solution  similar to \salr--\qqq\   with $F$ now
being given by  \geneg\  or \gene.
These backgrounds  seem to represent (3+1)-dimensional   string and  black-hole
type  configurations  in external electromagnetic fields  (cf. \horts).
 The
corresponding string model, however,  is no longer solvable by our methods.

For example, the $q_i =0$ background \salr,\veme,\qqq\
with $F$ given by \gene\  represents a   generalisation to the case of
$\a\not=0$
of the solution in \horts\   which  was  an extension ($\b\not=0$) of  the
extremal  Kaluza-Klein  ($a=\sqrt 3$) black hole \gibma.\foot{At the same time,
the solutions in our class  with $F$ given by \gene\
do not
include  a generalization of the extremal
$a=1$ dilatonic black hole \refs{\gibma,\ghs} since the model
\lagge\ does not contain  the  term $K\del u \bd u$ (or $K\del v \bd v $) which
is necessary in order to obtain the $a=1$ extremal black hole by dimensional
reduction \refs{\horrt,\horts}.
In fact,  for
$\a\b\not=0$ such a term (or, e.g., its ``gauge-invariant" generalization
$K(\del u -2\a A)(\bd u -2\a \bar A)$)
cannot  be added to \lagge\   whithout
spoiling its conformal invariance  (this can be shown following the discussion
in  \horts).
}

\newsec{Solution of the string model: path integral approach}

Looking at the  world-sheet action of our model \laggd\ or \lagg\
it may seem unlikely  that such a complicated interacting 2d theory
may have explicitly solvable classical equations and computable
path integral. The reason for the   solvability of this model is
that it is $\vp$-dual to a  much simpler theory \lag\ which, in turn, is
locally (ignoring topology) related  (by $\vp$-duality and coordinate
transformation) to a flat free-field model.
This explains, in particular, why the classical equations corresponding to
\laggd\ can be solved in terms of the free fields: the classical solutions of
the two dual \sms  are related in a simple way.
It is the  topology (periodicity of $y$ in  \uv\ and $\vp$) that  makes the
model nontrivial, and  it turns out to be possible to take the boundary
conditions into account in a  rather straightforward way.

Below we shall first  discuss the reduction to free fields
(on a simple example of the  $q_i=0$ model) and then present the computation
of the  general expression for the partition  function in the path integral
approach.

\def \g {\gamma}
\subsec{Reduction to free fields }
To clarify  why this model can be effectively  transformed   (up to zero modes)
into  a free-field one, it is helpful to  consider  first  a particularly
simple  case of the  $q_i=0$ model.\foot {Similar simplifications  occur also
in the models with either $\a ,\ \b $ or $q_+ +\b-\a $ equal to zero.}
As was already noted in Section 2.2, the $q_i=0$ model \laggo\
is $y$-dual to \lduu\ which is quadratic in $x^i$ and in which
integrating out   $\td v$ restricts   $\td u$  to the free classical solution.
As a result, this model is essentially ``gaussian",  like  the constant
magnetic field model \lgg\ solved in \ruts.
 A similar conclusion can be reached by starting directly with  \laggo.
In Minkowski world-sheet notation ($\s_\pm\equiv \tau\pm\s $; in this section
we shall ignore the trivial direction $x_3$)
\eqn\twe{ L (q_i=0) \   =  \ F(x) [ \del_+ u -2\a  A_+ (x)][\del_- v
+ 2\beta  A_- (x) ] + \del_+ x \del_- x^* + \a' {\cal R} \phi(x) \ ,   }
$$ A_\pm = \fourth  i ( x \del_\pm   x^*  -  x^*\del_\pm  x) \ , \ \ \ F=
e^{2(\p-\p_0)} = (1+ \a\b xx^*)\inv\ ,  $$
\eqn\ddq{
 x=x_1+ix_2=\rho e^{i\vp }\ ,\ \ \ \ x^*=x_1-ix_2=\rho e^{-i\vp }\ .}
The classical equations for $u$ and $v$ can be integrated once, giving
\eqn\inti{ F(x) [ \del_+ u -2\a  A_+ (x) ] = h_+(\s_+)\ ,
\ \ \  F(x)[\del_- v
+ 2\beta  A_- (x)  ] = h_-(\s_-) \ , }
where $h_\pm$ are arbitrary functions. Then the  equation for $x$ becomes {\it
linear}
\eqn\xxe{ \del_+\del_-x  +  i\b  h_+ \del_-x  - i\a  h_- \del_+ x  +  \a\b
h_+h_- x=0 \ , }
and is readily solved
\eqn\soo{ x = e^{ i\a g_- - i \b  g_+} X\  ,  \  \ \  \del_\pm  g_\pm  \equiv
h_\pm  \ , \ \ \   X= X_+ + X_- \ ,  \ \ \ X_\pm = X_\pm (\s_\pm) \ , }
where $X$ satisfies the free  wave
 equation, $\del_+\del_- X =0$.\foot{Note that in  the  general case  of $\a\b
q_+(q_+ +\b-\a )\neq 0$ the equation for $x$
will  no longer  be  linear  but will still be solvable, see  Section  5.1.}
 The functions  $h_\pm (\s_\pm) $ can be fixed to be constants by using the
remaining
freedom of  conformal transformations ($\s_\pm \to f_\pm (\s_\pm)$).  This is a
natural light-cone type gauge in this model,  in which
\eqn\lcc{ x = e^{ i\a h_-\s_-  - i \b  h_+\s_+} X \ , \ \ \  \ \  \ h_\pm
=\const \ .  }
As in the special case of the $\a=0$ model \lgg\ discussed in \ruts\ the
 transformation  $x \to X$ in \soo\ or \lcc\ makes the theory effectively a
free one and is   a  key to  its solution.
Using \soo\ one finds that  the equations for $u$ and $v$
\inti\ take  a very simple form (because of the special quadratic form of
$F\inv$ all $XX^*$-terms cancel out)
\eqn\hhh{  \del_+ u = h_+ +  \ha i\a ( X \del_+ X^*  -  X^*\del_+  X) \ , \  \
\ \del_- v = h_-  -   \ha i\b ( X\del_-  X^*  -  X^*\del_- X)  . }
To proceed, one needs to specify the boundary conditions.  Let us first
consider the
case of the cylindrical world sheet.
The closed string periodicity condition $x (\s + \pi,\tau) = x(\s,\tau)$ is
solved if
$X$ satisfies  the ``twisted" boundary condition (see also  \ruts)
 \eqn\ffff{ X(\s + \pi, \tau)= e^{ i\g \pi } X(\s,\tau)\ , \ \ \ \
\g\equiv \b h_+  +   \a h_-  \ ,   }
  implying
\eqn\yplumin{
 X_+ = e^{i\g \s_+} \X_+ \ ,  \ \ \ \ \ \  X_-= e^{-i\g \s_-} \X_- \ , }
where $\X_\pm =\X_\pm (\s_\pm) $ are single-valued free fields
\eqn\fourie{ \X_+  =  i  \sqrt{\a'/ 2 } \sum_{n=-\infty}^\infty \tilde a_n
\exp (-2in \s_+)
  \  , \ \ \ \
\X_-  =  i  \sqrt{\a'/ 2 } \sum_{n=-\infty}^\infty  a_n
\exp (-2in \s_-)
  \ .}
One can then solve \hhh\ expressing $u$ and $v$ in terms
of momentum and winding  $y$-modes  and oscillators in \fourie\ (see Section
5.1).

The  solution    of the
general model \lagg\ or \lagge\ with $q_i\not=0$
 can be  essentially reduced to that of  the  $q_i=0$ case.  Since the only
difference between \lagge\ and the $q_i=0$ model \laggo\
is in the substitution  \new, i.e. $\vp \to \vp' = \vp + q_1 u + q_2 v$,
starting with \lagge\  we get \twe,\inti,\soo,\ etc.,
with $x=\r e^{i\vp} $ replaced by
\eqn\xdx{ x'= e^{i(q_1 u + q_2 v)}  x  \ .  }
If $y$ were non-compact,   $x'$ would
 be single-valued like $x$ and the $q_i\not=0$ theory would be equivalent to
the $q_i=0$ one.  The only subtlety is thus to take into account the winding
mode part  of $y$, which should be treated separately,
while the  single-valued part of $y$ and $q_-t$-term in \new\ can be eliminated
 by the transformation \xdx\ with  $u,v$ replaced by $ u',v'$, which   do not
contain the winding  part of $y$.
A systematic way of doing
this will be discussed below in Section 5.1  using angular coordinates.

\subsec{Path integral computation of the partition function on the torus}
Let us now illustrate the solvability of the model by computing the
partition function on the torus  $Z$ using the  path
integral approach.
It turns out to be possible to compute all the path integrals explicitly
expressing $Z (R, \a,\b,q_+)$ in terms of sums over winding numbers and two
extra  (in addition to modular) ordinary integrals.  The latter  will be
absent  (easily computable)  in the special cases
when $\a\b q_+ (q_+ +\b-\a)=0$.

The  fields $x$ and $t$  are single-valued, i.e.,  on the torus,  $x (\s_1 +n,
\s_2  +m ) = x (\s_1, \s_2) $,
$ \ t (\s_1 +n, \s_2  +m ) = t (\s_1, \s_2)$ ($n,m$ are integers).\foot{ We
shall follow the notation of \ruts\  with the following exceptions:  we use $y$
and not $\phi$ for the compact coordinate,
and $c_+$ instead of $f$ for the (``left") magnetic strength parameter. In
particular,  for the torus
$ds^2 = |d \s_1 + \t d \s_2 |^2  , \ \
 0<\s_\a\leq 1\ , \ \  \t=\t_1 + i \t_2   ,  $
$\
g_{\a\beta}=\left(\matrix {1&\t_1\cr \t_1& |\t|^2 \cr}\right),\   \
\sqrt g g^{\a\beta}= \t_2^{-1} \left(\matrix {|\t|^2&-\t_1\cr -\t_1& 1
\cr}\right).
$
Also, $\del =\ha (\del_2 -\t
\del_1), \ \    \bd = \ha (\del_2 -\bar \t
\del_1)$.}
Since $y=\ha (u + v) $
has  period $2\pi R$  it should
satisfy the  condition
\eqn\foo{y (\s_1 +n, \s_2 +m ) = y (\s_1, \s_2) + 2\pi R (nw + m w') \ , }
 where $w,w'$ are two
integer  winding numbers. Then
\eqn\zerr{ y= y_* + y'  \ ,  \ \  \   y_* =y_0 +
 2\pi R(w \s_1 + w' \s_2)\ , \ \ \
y' = \sum_{n,n'} y_{nn'} e^{2\pi i(n\s_1 +
n'\s_2)} \ ,  }
where $y'$ is the single-valued part of $y$.
The computation in the general case  of $q_+\not=0$ turns out to be a simple
generalization of the   $q_+=0$ case. If one starts with the $q_i=0$ action
\laggo\
and  separates
 the single-valued parts in $u,v$ ($u=y_* + u',\ v=y_* + v'$) one obtains
\eqn\laoy{ I=  {1\ov \pi \a'\t_2}  \int d^2\s \big[
 F(x) (\del u'   +  A_1) (\bd v'  + A_2)
 +  \del x \bd x^*  +  \t_2 {\cal R} (\p_0 +  \ha \ln F ) \big]  ,   }
\eqn\ppp{ A_1\equiv \del y_*
 - \ha i\a    ( x \del  x^*  -  x^*\del  x) \ , } $$
\ A_2 \equiv   \bd y_*  +
   \ha i\b   ( x \bd  x^*  -  x^*\bd  x) \ ,
 \ \ \ F\inv = 1 + \a\b xx^* \ , $$
\eqn\yyt{ \del y_* = \pi R (w' -\t w) \ , \ \ \
\bd y_* = \pi R (w' -\bar \t w)\ . }
For  $A_{1,2}=0$  the  integral over $u',v'$   would
lead to the conclusion that the dilaton term is cancelled out \sts\
and that
the partition function is  thus given by the free-theory  one \polch.
For  $A_{1,2}\not=0$ the integral over $u',v'$
gives also the product of the ``zero-mode" parts
of $A_i$, i.e. the term  $\sim  \langle F\rangle \langle A_1\rangle
\langle  A_2\rangle ,$ \  \
($\langle ... \rangle  \equiv \int d^2\s  ...$)
which is non-gaussian  in $x,x^*$.  To
 be able   to then integrate    over $x,x^*$, it is convenient
to ``split"  this term into   quadratic parts   using  an ordinary
integral over   two auxiliary parameters.
Equivalently, one may
\def \C {\bar C}
 introduce  from the very beginning
 an auxiliary vector field  $(C,\C)$ representing
$e^{-I} $ as\foot{ We   omit the dilaton term  which is  cancelled  out at the
end
after one integrates
over $u,v,C,\C$. Note  also that the measure of integration over $u,v$,  which
originally contained
the $\sqrt {-G} = F(x)$ factor,  becomes trivial after  the introduction of
$(C,\C)$
(the $F$-factor is ``exponentiated").
 }
\eqn\aux{ e^{-I} = \int [dCd\C] \exp \big(- {1\ov \pi \a' \t_2}  \int d^2\s
\big[
 F\inv (x) C \C   } $$
 +\   \C (\del u'   +  A_1) - C (\bd v' + A_2)
 +  \del x \bd x^* \big]\big)  \ . $$
Integrating over $u',v'$ we find that (on the torus)
$C$ and $\C$ are constrained to be  equal to constants. We shall denote  these
constants  as $C_0$ and $  \C_0$ (the integral over $C_0,\C_0$  will contain
the  factor of $\t_2\inv$ in the  measure).
The remaining action  is  then quadratic in $x,x^*$
so that the expression for the partition function takes the  following
form\foot{The modular measure contains the contribution
of the    22 extra free scalar  degrees of freedom  added  to satisfy
the zero central charge condition.
In general, the integrand of the
partition function
is modular invariant (as it should be, being derived  from  a reparametrisation
invariant world-sheet theory):
  the transformations $\t\to \t +1 $ and $\t\to -1/\t$ are  ``undone"
by
the redefinitions of other integration and  summation parameters.}
$$  Z(r, \a, \b, q_+) = \int
 {d^2\t \ \tau_2^{-12} }
 e^{11\pi \t_2/3} |f_0(e^{2\pi i \t})|^{-44}\ { \cal Z } (\t, \bar \t ) \ ,
$$
 \eqn\zee{ { \cal Z} = r \sum_{w,w'=-\infty}^{\infty}
 \int  dC_0d\C_0 \ \t_2\inv \   \exp[ -  {1\ov \pi \a' \t_2 }
 (C_0 \C_0  + \C_0   \del y_* - C_0  \bd y_* )]  \  {\cal Z}_x  \ , }
$$ {\cal Z}_x = \int [dx dx^*] \exp (- I'[x,x^*; C_0,\C_0, w, w', \t, \bar \t]
) \ , $$
\eqn\ffa{ I'=   {1\ov \pi \a' \t_2 }  \int d^2\s \big[
 \a\b  C_0 \C_0   xx^* - \ha i\a    \C_0 ( x \del  x^*  -  x^*\del  x)  }
 $$
 -\  \ha i\b  C_0  ( x \bd  x^*  -  x^*\bd  x)  +  \del x \bd x^* \big]  \ . $$
 The   factor $r= R/\sqrt{\a'}$ in \zee\
 as usual comes from the integral over the   compact constant  mode  $y_0$  (we
drop out  an  infinite integral over $t_0$).
Expanding
 \eqn\yty{ x= x_0+ x'= x_0 + \sum_{(n,n')\not=0} a_{nn'} e^{2\pi i(n\s_1 +
n'\s_2)}\ , }
 one finds that the  gaussian
integral over  $x',x'^*$ leads to the following
 simple  result (cf. ref. \ruts)
\eqn\ggg{
{\cal Z}'_x=
c_0 [{\rm  det'}\Delta_0 ]^{-1} \  Y \inv (\t,\bar \t, \h, \td \h) \ ,   }
\eqn\ddd{
{\rm  det'}\Delta_0 = \t^2_2 \eta^2 \bar \eta^2 \ ,
\ \   \eta =  e^{i\pi \t/12} \prod^\infty_{n=1} (1-e^{2\pi in\t} ) \equiv
e^{i\pi \t/12} {
f_0} (e^{2\pi i \t}) \ , }
\eqn\yy{ Y (\t,\bar \t, \h, \td \h)= \prod_{(n,n')\not=(0,0)}
\big( 1 +   {\h \ov n'- \t n}\big)
\prod_{(n,n')\not=(0,0)}
\big( 1 +   {\td \h \ov n'- \bar \t n}\big)\ , }
\eqn\eee{      \h
 \equiv  {1\ov \pi}  \beta C_0 \ , \  \  \ \ \td \h\equiv  {1\ov \pi}  \a \C_0
\  .  }
The factorized  ``chiral" form of $G$ is spoiled by a diffeomorphism (modular)
invariant regularisation.
 To define explicitly the formal expression \yy\
 consider
 \eqn\yyyi{
U(\t,\bar \t, \chi, \td \h ) \equiv    \prod_{(n,n')\not=(0,0)}
 (n'- \t n + \chi )(n'-\bar \t n + \td \h )
} $$  = \prod_{k\not=0}
( k  + \chi )(k + \td \h)   \prod_{ n\not=0, n'}
  (n'- \t n + \chi ) (n'-\bar \t n + \td \h) \ . $$
Computing first the product over $k$ and $n'$ using
$$\prod_{n=-\infty}^\infty (n + \chi) = \chi  \prod_{n=1
}^\infty (-n^2) ( 1 - {\chi^2\ov n^2}) =
2i \sin \pi \chi\ , $$
one gets the product of $\sin$-functions.
Separating the factor
$$\prod_{n\not=0}  \exp (-i\pi \t
n + i \pi \chi
) \ \exp (i\pi \bar \t n  - i \pi \td \chi
 ) \ , $$
and defining it as
 $$ \exp \big[2\pi \t_2  \sum_{n\not=0} \big(n + i {\chi-\td \h \ov 2
\t_2}\big) \big]
\ , $$
one should compute the sum using the generalised $\zeta$-function
regularisation
\eqn\zere{  \sum_{n=1}^\infty (n + c )= \lim_{s\to
-1}\sum_{n=1}^\infty (n + c )^{-s}
= -{1\over 12}   + {1\ov 2}   c (1-c) \  . }
As a result,
\eqn\yyrr{Y (\t,\bar \t, \h, \td \h)\equiv  {U(\t,\bar \t, \chi, \td \chi )\ov
U(\t,\bar \t, 0, 0)}
=\  \exp[{{\pi  (\chi-\td \chi)^2
\ov 2 \t_2}}]  \ } $$ \times   \big[{ \sin \pi   \chi
 \ov  \pi  \chi}
\prod_{ n=1}^\infty {  (1 - \r\inv
   q^{n })(1 -  \r
   q^{n }) \ov (1 -  q^{n })^{2}}  \big]  \
\big[{ \sin \pi \td  \chi
 \ov  \pi  \td \chi}
\prod_{ n=1}^\infty {  (1 - {\td\r}\inv
  \bar q^{n })(1 -  \td\r
   \bar q^{n }) \ov (1 -  \bar q^{n })^{2}}  \big]  \ , $$
$$
  \r\equiv \exp ({2\pi i
\h }) \ , \ \   \td \r\equiv \exp ({2\pi i
\td \h }) \ ,   \ \ \ q=\exp({ 2\pi i  \t})\ ,
$$
or, finally,
\eqn\ttyy{
Y (\t,\bar \t, \h, \td \h) =
 \exp[{{\pi  (\chi-\td \chi)^2
\ov 2 \t_2}}]  \     {\theta_1(\h| \t)
\ov \h\theta'_1 (0| \t) }  \  {\theta_1(\td \h|\bar  \t)
\ov \td \h\theta'_1 (0|\bar  \t) }  \  . }
The contribution
 of the integral over the constant  parts  $x_0, x_0^*$
is
\eqn\consm { {\cal Z}_{x0} =
 \int  dx_0dx^*_0   \exp[ -  {( \pi \a' \t_2)\inv  }
\a\b  C_0 \C_0  x_0x_0^*] = {\a'\pi^2\t_2\ov \a\b C_0 \C_0 }={\a'\t_2\ov \h\td
\h} \ . }
For $\a\b=0$ this expression gives a divergent factor corresponding to the area
of the
$x_1,x_2$ plane.\foot{For $\a\b\not=0$   the   external
fields
 break down the translational invariance on the $x_1,x_2$-plane (in particular,
 $x_{cl}=x_0$ is no longer a classical solution for the \sm action
\laoy\ when $u_{cl}=v_{cl}=y_*$ and $w,w'\not=0$)  and thus ``regularize"
the  divergent area factor $\int dx_0dx_0^*$  in $Z$.}
If one  defines $Z$  using the factor \consm\   then the free-theory limit
becomes singular. Alternatively, one may
 leave the integral over $x_0,x_0^*$ to the end so that
 the limit $\a\b \to 0$  is regular in the integrand.\foot{In general,  the \sm
partition function on the torus
is \ \  $Z= \int d^D X_0 Z_1 (X_0),$  \ \  $Z_1= \sqrt {-  G(X_0) }  Z'_1
(X_0)$,
where $X_0^M$ are the constant parts of all   \sm coordinates. The \sm
path integral $\int [dX] \exp (-\int G_{MN}(X) \del X^M\bd X^N)$  is defined
using the measure
$ |\delta X|^2= \t_2 \int d^2 \s G_{MN}(X) \delta X^M\delta  X^N$.
  In the present case of \laoy\  (before one introduces $C,\C$)
 the volume $\int d^D X_0 \sqrt {  -G(X_0) }
= 2\pi R \int dt_0  \int dx_0 dx_0^*  (1 + \a\b x_0x_0^*)\inv $
is divergent in the $x_0,x_0^* $ direction: logarithmically if $\a\b \not=0$
and quadratically -- if $\a\b=0$.}
Another possibility to obtain $Z$  with a regular
free-theory limit (equal
to the standard   partition function of a free string
compactified  on a circle)
is to project out the constant mode factor
\consm\ (e.g. by inserting  the $\delta$-functions
$\delta(x_0) \delta(x^*_0)$).
Using the latter prescription we  find\eqn\etr{
 Z(r, \a,\b , q_+)  =c_1   \int [d^2\t]_1 \ W(r, \a,\b,q_+|\t, \bar \t) \ ,
}
\eqn\meas{ [d^2\t]_1 \equiv
 {d^2\t \ \tau_2^{-14} }
 e^{4\pi \t_2} |f_0(e^{2\pi i \t})|^{-48}\  , }
\eqn\wwwe{  W(r, \a,\b,q_+=0)\   = \  r (\a'\a\b\t_2)\inv
\sum_{w,w'=-\infty}^{\infty}
 \int d\h d\td\h\  }
$$
\times\  \exp\big( -  \pi (\a' \a\b\t_2 )\inv
 [ \h\td \h   + \sqrt {\a'} r\b   (w'-  \t w) \td \h
 - \sqrt {\a'} r \a    (w'- \bar \t w)\h  ]\big) $$ $$
\times  \  \exp\big[{-{\pi  (\h - \td \h)^2 \ov 2 \t_2}}\big] \
 {  \h \td \h |\theta'_1(0|\t )|^2 \ov \theta_1(\h|\tau )
\theta_1 (\td\h |\bar \t )\ } \ .
 $$
When  $\a=0$ this expression  reduces to the partition function  for the
``chiral"   $\a=q_+= 0$ theory  \lgg\
 found in \ruts:  for $\a\to 0$
the integral over $\bar C_0= \pi \td \h/\a$  produces  the  $\delta$-function
constraint  $\h=- \h_0=- \sqrt {\a'} r\b (w'-\t w)$
so that
\eqn\non{  W (r, \a=0 ,\b , q_+=0) =
\sum_{w,w'=-\infty}^{\infty}
 \exp [{- I_0(r)  }]\  \exp[ -{\pi  \h _0^2 \ov 2 \t_2}]
\ {  \h_0 \theta'_1(0|\t )\ov \theta_1(\h_0|\tau )} \ , }
   \eqn\zeii{ \ I_0 (r) \equiv  \pi r^2  \t_2\inv { (w'-\t w)(w'-\bt w) } \ , }
\eqn\yuy {
 \ \ \h_0=  \sqrt{\a'}\b r(w'-\t w) \  . }
The representation  \wwwe\
is not the simplest one possible  for  $Z$ in the case  when $q_+=0$
 (the sums over $w,w'$ give $\delta$-functions,  and thus  the integrals over
$\h, \td \h$ in \wwwe\ can be computed explicitly, see below)
but its advantage is that it has a straightforward generalization to the case
of $q_+\not=0$.
Indeed,   according to \lagge\
to include the dependence on $q_+$ one is to make the transformation \xdx\ in
the action \laoy.  This transformation  can be represented as
$ x \to  \exp (iq_+ y_*) \hat x, \  \hat x = \exp (iq_1 u' + iq_2 v') x$,
 where $u',v'$ are single-valued parts of $u,v$. Then $\hat x$ is also
single-valued  and can be used as a new integration variable instead of $x$.
The $q_+$-dependent  analogues of \laoy, \aux,\ffa\
 are  thus obtained by the formal
substitution $ x \to  \exp (iq_+ y_*)  x$.
One finds that  \ffa,   and thus \ggg--\eee,    have the same form with
\eqn\repl{ \b C_0 \to \b C_0 +   q_+ \del y_* \ , \ \ \ \  \ \
\a \C_0 \to \a \C_0 +   q_+ \bd y_* \ , }
$$ \h \to \h    +  q_+ R(w'-\t w) \ , \ \  \
\ \ \ \td \h\to \td\h  +   q_+ R(w'-\bar \t w)    \ , $$
\eqn\replq{  \h
 \equiv  {1\ov \pi}  \beta C_0 +  q_+ R(w'-\t w)  , \   \ \td \h\equiv  {1\ov
\pi}  \a \C_0 +   q_+ R(w'-\bar \t w)  \ . }
Then the general expression for the partition function
is given by  \etr\ with \wwwe\ replaced by\foot{In considering formal singular
limits of the partition function (like
$R=\infty$, see below) it is more convenient to use the original integration
parameters $C_0, \ \C_0$  instead of  $\chi, \td\chi $ (cf. eq. \replq ).}
\eqn\wew{  W(r, \a,\b,q_+) \ =  \ r (\a'\a\b\t_2)\inv
\sum_{w,w'=-\infty}^{\infty}
 \int d\h d\td\h\  }
$$
\times \exp\big( -  \pi (\a' \a\b\t_2 )\inv
 [ \h \td \h   + \sqrt {\a'} r(q_+ + \b )  (w'-  \t w) \td \h
 + \sqrt {\a'} r (  q_+ -\a )      (w'- \bar \t w)\h $$
$$
+\  \a' r^2 q_+ (q_+ + \b -\a)  (w'-  \t w)(w'- \bar \t w)]\ \big) \ $$
$$ \times \
 \exp[{-{\pi  (\h - \td \h)^2 \ov 2 \t_2}}] \
 {  \h \td \h |\theta'_1(0|\t )|^2 \ov \theta_1(\h|\tau )
\theta_1 (\td\h |\bar \t ) } \ . $$
Like the measure in \etr\ this expression is $SL(2,Z)$ modular invariant
(to show this one needs to shift $w,w'$ and redefine $\h, \td \h$). The
expression in the approach where the integrals over the constant parts
$x_0,x_0^*$
are left until the very end is obtained by an obvious modification:
the $\h\td\h$ term in the exponential in \wew\   is  replaced by
$\h\td\h(1+\a\b x_0x_0^*)$. If one explicitly integrates over $x_0,x_0^*$,
then  the integrand in \wew\ is multiplied by the factor in \consm.
In all cases the  integrands of $Z$ are modular invariant.

Given that the model \lagg\ is $\vp$-dual to  the
 model \lag,   the two  should lead to the same partition function.
Starting with  \lag\ one should thus be able to reproduce the same expression
\etr,\wew\ in a simpler way.
Separating   $u$ and $v$ into  the ``winding"  and single-valued parts
($u=y_* + u', \  v=y_* + v'$)  and integrating over $u',v'$ in \lag \
one  again must introduce the  integral over
the auxiliary constant parameters  $C_0, \C_0$ in order to ``split" the
zero mode factor $\langle \del \td\vp \rangle \langle \bd \td \vp\rangle $.
Then \lag\  is transformed into
\eqn\kuk{  L=  C_0 \C_0  + \C_0   (\del y_*  +\a\del\tilde\vp )
 - C_0  (\bd y_*  + \b \bd \td\vp)  }
$$
 - q_+ \del\tilde\vp \bd y^*   + q_+ \bd\tilde\vp \del y^*  +
\del\rho\bd\rho
+ \rho^{-2 }\del\tilde\vp\bd \tilde\vp  +
\t_2 {\cal R}(\p_0 - \ha \ln \r^2)   \ . $$
Making $now$ the path integral  duality transformation  $\td \vp\to \vp $
one obtains the  same action as in \ffa,\repl\
  which is quadratic in $x, x^*$.
The resulting expression for $Z$ is thus equivalent to \etr,\wew.

To conclude, we have found the explicit representation for the partition
 function
in terms of the  two auxiliary {\it ordinary} integrals.
In general,  $Z (r, \a,\b,q_+)$  depends on four real dimensionless
parameters  ($R/\sqrt{\a'}$, \ $R\a,$  $ \ R\b, \ Rq_+$)  and  has several
symmetry properties which follow  from \wew.
 $Z$  is  symmetric under $\a \leftrightarrow - \b$ as well as under the
simultaneous  changing  of the signs of $\a$,  $\b$  and $q_+$
\eqn\syt{ Z(r,\a,\b,q_+) = Z(r,-\b,-\a ,q_+)=Z(r,-\a,-\b,-q_+)=
 Z(r,\b ,\a,-q_+)\ .  }
It  is also invariant under the  $y$-duality
which transforms the theory with  $y$-period  $2\pi R$ and  parameters
$a_+=q_+ -\a,\  c_+ =q_+ +\b, \ c_--a_- =\a-\b $ into the theory  with
$y$-period  $2\pi \a'/R$  and  parameters $-a_+,\  c_+, \ c_- - a_-$ or
parameters  $a_+, \ - c_+, \ c_- -
a_-$  (see \ldu,\ded,\dee).  This is seen explicitly from the representation
\wew:
doing the Poisson  resummation (see (5.55))
 in $w$ or in  $w'$  (which  is equivalent to
performing  the duality transformation in $y$) one obtains the same
expression for the exponential in \wew\
with  the parameters interchanged according to the above relations, namely,
 \eqn\duy{  Z(r,\a,\b,q_+) = Z(r\inv, q_+, \b -\a + q_+ ,\a)   =
Z(r\inv, \a -\b - q_+ , - q_+  ,-\b)  \ . }
Combining  \syt\ and \duy\ we also learn  that
\eqn\syte{ Z(r,\a,\b,q_+) = Z(r,-\b,-\a ,\a-\b-q_+) =  Z(r,\a,\b ,\a-\b-q_+)
\ .  }
 When  $\a=q_+$  ($a_+=0$) or $\b=-q_+$  ($c_+=0$)
the duality relations \duy\  retain their  standard  ``circle" form
 \eqn\duyf{Z(r,\a,\b,\a) = Z(r\inv, \a, \b , \a) \ ,  \ \ \
 Z(r,\a,\b,-\b) = Z(r\inv, \a, \b ,-\b)  \ . }
These duality  relations  will be  manifest  also in the representation  for
$Z$   derived in the operator approach  in  Section 5.3.

As follows from \wew,
the   expression for $Z$ \etr,\wew\    simplifies substantially
when  $\a\b q_+(q_+ + \b -\a)=0$, i.e. when
one of the   parameters  $\a, \b, q_+$ or $ q_+ +\b-\a $
 vanishes (i.e.   when  at least one of the  two
magnetic fields in \fies,\deey,\deeyy\  is uniform).
Then the integrals over $C_0, \C_0$ can be computed  explicitly
and one obtains a direct generalization of \non.
In view of the relations \syt,\duy,\syte\
  these four  cases are equivalent.
For example, when  $q_+=0$   the integrals  over $\h,\td\h$ in \wwwe\
 can be  easily computed  if one notes that the sums over $w,w'$ produce
$\delta$-functions when there is no  quadratic term in $w,w'$.
 The result, of course,  is
 the same as the one   found by starting directly with the
model  \lduu\ which is $y$-dual to the $q_i=0$ model \laggo.
The path integral for \lduu\ can be computed without need to introduce the
auxiliary fields $C,\C$ in \aux:  as in  the $\a=0, q_i=0$ model \ruts\
the integral over $\td v$ restricts $\td u$
to  the zero-mode value
$\td u_*= \td y_*= 2\pi \td R (w \s_1 + w ' \s_2), \ \ \td R = \a'/R , \  $
and one finishes with the  partition function  \etr\ with (cf.
\wwwe,\non,\duyf)\foot{As  before  in \wew, we
have projected out the contribution
$ \sim \t_2(\h_0 \td \h_0)\inv $ (which is present  in the winding $(w,w')$
sector) of the integral over $x_0,x^*_0$. }
\eqn\fulle{ W(r,\a,\b, q_+=0)\ = \ { r\inv }
\sum_{w,w'=-\infty}^{\infty}\
 \exp[{-I_0 ( r\inv ) }]\  }
$$ \times  \exp[{-{\pi  (\h _0- \td \h_0)^2 \ov 2 \t_2}}] \
{  \h_0 \td \h_0 |\theta'_1(0|\t )|^2 \ov
\theta_1(\h_0|\tau )
\theta_1 (\td\h_0 |\bar \t )\ } \ , $$
\eqn\rrf{   \h_0 =  \sqrt{\a'}\b  r\inv (w'-\t w) \  ,
\ \ \ \  \td \h_0 =  \sqrt{\a'}\a  r\inv  (w'-\bar \t w) \  .  }
Starting  directly with \wew\ or using  \fulle\ and
\duy\ (i.e.  $ Z(r,\a,\b,0) = Z(r\inv , 0 ,\b-\a ,\a )$)
we can  find  also  a simple form of
 $Z$ in the case   when $\a=0$ or when $\b=0$, i.e. when
$\a\b=0$   (cf. \fulle,\wew)
\eqn\rrtr{
 W(r, \a,\b , q_+)\vert_{\a\b=0}\   =  \
\ r  \sum_{w,w'=-\infty}^\infty    \exp[{-I_0 (r ) }] \ }
$$ \times  \exp[{-{\pi  (\h _0- \td \h_0)^2 \ov 2 \t_2}}] \
{  \h_0 \td \h_0 |\theta'_1(0|\t )|^2 \ov
\theta_1(\h_0|\tau )
\theta_1 (\td\h_0 |\bar \t )\ } \ , $$
$$
\h_0  = \sqrt {\a'} (q_+ + \b )  r(w'-\t w) \ , \ \ \ \
\td \h_0 =\sqrt {\a'} (q_+ - \a) r(w'-\bar \t w) \ , \ \  \ \a\b=0 \ .    $$
Another interesting special case is $a_+=c_+=0$ (i.e.  $\a=q_+=-\b$)   when
 the modular integrand $W$  \wew\  formally  factorises into  the
$r$-dependent   part
$W_0(r)$
and an   $\a$-dependent   part $W_1 (  \a)$.\foot{This factorization
is valid under certain  analytic continuation assumption
since the  contour of integration over $\h, \td\h$ depends on
$w,w'$ according to \replq.}
  As follows from \wew,
\eqn\trtr{ Z(r, \a,-\a , \a) =c_1  \int {[d^2\t]_1} \
W_0(r)\  W_1 (  \a)\ , }
\eqn\yty{W_0 = r \sum_{w,w'=-\infty}^\infty  \exp[{- I_0 (r ) }]  \ ,   }
\eqn\hyh{
W_1 = - (\a' \a^2\t_2 )\inv  \int d\h d\td\h\ \exp \big(  {\pi
\h \td \h \ov \a' \a^2\t_2 }\big) \
}
$$ \times  \exp[{-{\pi  (\h - \td \h)^2 \ov 2 \t_2}}] \
{  \h \td \h |\theta'_1(0|\t )|^2 \ov
\theta_1(\h|\tau )
\theta_1 (\td\h |\bar \t )\ } \  . $$
$W_0$  is the same  as the  partition function
of a free boson  on a circle.
Clearly,  in this case $Z(r,\a)=Z(r\inv, \a)$.

Finally,  let us note   that in the limit of  non-compact $y$-dimension  ($R\to
\infty$)
$Z$  \etr,\wew\
reduces  to the partition function of the free  bosonic closed string theory.
 A simple way to see this is to note that for the non-compact  $y$  the
parameters $q_i$ can
be set  equal to zero by a coordinate transformation  (equivalently, in the
case of $R=\infty$ the winding sector
becomes trivial ($w=w'=0$) and thus according to \repl\ $Z$ does not depend on
$q_+$).
Then taking   $R\to \infty $ in \fulle\ one finds that $Z$ takes  the
 flat  space expression.
To show this in general starting  directly with \etr,\wew\
one should first use \replq\ to return to the
integral over the variables $C_0, \C_0$,  rescale   the latter by $R$
and then take the limit $R=\infty$ (in \trtr\ this effectively corresponds to
shrinking the contour of integration  over $\h,\td\h$ to zero so that
$W_1\to 1$).

$Z(R\to \infty)= Z_{free}$   is also clear directly  from the form of the path
integral for the
$y$-dual to $q_i=0$ theory \lduu: in  the non-compact case the zero mode of
$\td u$ is constant  and  after integrating out $\td v$ one gets a free
$x,x^*$-theory.
This generalizes a similar observation for the $\a=q_+=0$ model \ruts.\foot{In
the limit $R=\infty$ the
$\a=q_+=0$ model \lgg\  is equivalent to the model of \napwi\
which has  trivial (free) partition function \kk.}

\newsec{Solution of the string model: canonical operator approach}
In this section we shall  first derive the  expression for the
solution of the classical
equations of motion  for  the general  values of parameters of our model \lagg\
in terms of constant zero-mode parameters and  free oscillators. We  shall then
canonically quantize the model using a light-cone type gauge  and derive
the quantum Hamiltonian   (which will be fourth order in oscillators but
diagonal in Fock space).
The possibility to choose the light-cone gauge combined with conformal
invariance guarantees  the unitarity of the model.
Finally,  we will  show that the operator approach leads to the same expression
for the partition function that was found above in the path integral approach.

\subsec{General solution  of the classical equations of motion
and light-cone gauge}
\def\lmo{F(\r)  \rho^2}
\def \b {\beta }
\def \be {\beta }
Let us now return to the discussion of the solution
of the classical equations of motion  corresponding to our  model  \laggd\
(we shall consider the flat cylindrical  Minkowski world sheet  with
$\s \in [0,\pi)$, \  $-\infty <\t< \infty$, \ $\s_\pm=\tau\pm\s $)
\eqn\genlag{
L=\del _+ u\del_- v + \del_+ \rho\del_-\rho
}
$$ + \ \lmo [\del_+ \vp+(\b +q_1)\del_+ u +q_2\del_+ v]
[\del _- \vp  + q_1\del_- u +(q_2-\a )\del_- v]
 \ , \ $$ $$
F\inv (\r)  = 1 + \a\b\r^2 \ .  $$
To solve the  corresponding  equations of motion
\eqn\geneq{
\del_+[\lmo \del_-(\vp+(q_2-\a )v+q_1u)]+
\del_-[\lmo\del_+(\vp+q_2 v+(q_1+\b )u)]=0\ ,
}
\eqn\gend{
\del_+\del_-\rho-{\rho F ^2  (\r) }\del_+[\vp+(\b +q_1)u+q_2 v]
\del_-[\vp+(q_2-\a)v+q_1 u]=0\ ,
}
\eqn\gent{
\del_+\del_-v=-\be\del_+[\lmo \del_-(\vp+q_1 u+(q_2-\a)v)]\ ,}
\eqn\gent{
\del_+\del_-u=\a \del_-[\lmo \del_+(\vp+(q_1+\be) u+q_2 v)]\ ,
}
we shall utilize  the  $\vp$-duality relation between \genlag\ and
the  model \lag\  or
\eqn\lag{{{\td L}}= \del _+( u + \a \tvp)\del_- (v +  \b \tvp)
+ \del_+ \r \del_- \r + \r^{-2} \del_+ \tvp \del_-\tvp } $$
+ \del_+ (q_1 u + q_2 v ) \del_- \tvp -
\del_- (q_1 u + q_2 v ) \del_+ \tvp \ ,   $$
which has the equations of motion
\eqn\dualeq{
\del_+\del_- (u+\a\tilde\vp )=0\ ,\ \  \ \ \del_+\del_- (v+\b \tilde\vp )=0 \ ,
}
\eqn\rhophi{
\del_+\del_-\rho+\rho^{-3}\del_+\tilde \vp \del_-\tilde\vp =0\ ,\ \  \ \
\del_+ (\rho^{-2}\del_-\tilde\vp )+\del_-(\rho^{-2}\del_+\tilde\vp)=0\ .
}
Eqs.
  \rhophi \  are   the equations of motion
for the ``dual 2-plane"  model  \eqn\dup{{{\td L}}_0=\del_+\rho\del_-\rho
+ \rho^{-2 }\del_+\tilde\vp\del_- \tilde\vp \ . }
Since the solutions of the equations of motion  for two dual \sms
are in general related (locally)  by  $ \ \ (G_{\m\n} + B_{\m\n}) \del_a x^\n =
\ep_{ab} \del^b \td x_\m$, \   we can express  the solution
of \rhophi\  in terms of the solution of  the free model
  dual to \dup\
\eqn\freee{ L_0=\del_+\rho\del_-\rho+ \rho^2 \del_+\hat \vp \del_-\hat \vp =
\del_+ X \del_- X^* \ ,
\ \ \  \ \ \  X\equiv  \r e^{i\hat \vp} \ .  }
We get\foot{For a discussion of a relation between
2-plane  and dual 2-plane models see \refs{\rocver,\dvv,\kirr}.}
\eqn\dualsol{ \r^2 = XX^*\ , \ \ \ \hat \vp = {1\over 2i}\ln{X\ov X^*}\ , \ \ \
 X=X_+  (\s_+) + X_- (\s_-) \ , }  $$
\del_\pm \tilde \vp =\mp \r^2 \del_\pm  \hat \vp=\pm {i\over 2}(X^*\del_\pm
X-X\del_\pm X ^*)\  , $$
and thus
\eqn\jjj{
\tilde\vp(\s,\t  )= 2\pi\a' [J_-(\s_-) - J_+(\s_+)]  +{i\ov 2} \big( X_+ X^*_-
-X_+^*  X_-\big)\ ,}
\eqn\cude{
J_\pm(\s_\pm)\equiv {i\ov 4\pi\a '}\int_0 ^{\s_\pm}d\s_\pm\big( X_\pm \del_\pm
X^*_\pm  -
X_\pm^* \del_\pm  X_\pm\big)\ . }
The solution of \dualeq\ is then
\eqn\dualsole{
u=U_+ +U_- -\a\tilde\varphi \ ,\ \  \ \ v=V_+ +V_- -\b \tilde\varphi\  ,
}
where $U_\pm $ and $  V_\pm $ are arbitrary functions of $\s _\pm $.
Returning now to  the  system \geneq--\gent\
we conclude  that since it is  $\vp$-dual to \dualeq--\rhophi, $\ u,v,\rho$
have the  same expressions \dualsol,\dualsole\
while $\vp$ is found to  be given by
\eqn\gensol{
\vp+ q_1 u+q_2 v=-\b U_++\a V_- +\hat \vp \  .
}
Thus
 \eqn\gensol{
x\equiv \r e^{i\vp} = \exp[-i(q_1 u +q_2 v )] \exp (i\a V_- -
i\b U_+    ) X \ ,  }
in agreement with our previous discussion  \soo,\xdx.

Let us now take into account  the boundary conditions.
The physical coordinate
$x=\rho e^{i\vp }$ is single-valued, i.e.
$\ x(\s + \pi, \t )=x(\s, \t )$. This    implies that the free field $X=X_+ +
X_-$ must satisfy  the   ``twisted"    condition as  in \ffff,\yplumin\ (see
also \ruts)
\eqn\bcfz{ X(\s + \pi, \tau)= e^{ i\g \pi } X(\s,\tau)\ , \ \ \  \
 X_\pm  = e^{\pm i\g \s_\pm } \X_\pm  \ ,  \ \ \  \X_\pm (\s_\pm \pm \pi)=
\X_\pm (\s_\pm )\ , }
where $\X_\pm = \X_\pm  (\s_\pm)$ are as defined in eq. \fourie.
Since the scalar field $y=\ha (u+v) $ is compactified on a circle of radius
$R$,
\eqn\bcfuv{
u(\s+\pi, \t )=u(\s,\t )+2\pi wR\ ,\ \  \ \ v(\s+\pi, \t )=v(\s,\t )+2\pi w R\
, }
where $w$ is an integer winding number. Let us now determine $\gamma$.
Eqs. \jjj,\cude\ and \bcfz\   imply
\eqn\bcfvp{
\tilde\vp (\s +\pi,\t )=\tilde \vp (\s,\t ) -2\pi\a ' J\ ,\ \ \
J= J_L + J_R\ ,\ \  \  J_L\equiv J_+(\pi ) \ ,\  \  J_R\equiv J_-(\pi )\ .
}
We shall  see below that after  the
 quantization  $J $ becomes   the total angular momentum
operator and has  integer eigenvalues. This is consistent with the fact that
$\td \vp$ has period $2\pi \a'$,  as implied by its duality to $\vp$ or $\hat
\vp$ which have periods $2\pi$.
As follows from \dualsole,\bcfvp\
the boundary conditions \bcfuv\  are satisfied by setting
\eqn\uuvv{
U_\pm =\s_\pm p_\pm ^u    + U_\pm '\ ,
\ \ \ \ V_\pm =\s_\pm p_\pm^v    + V_\pm'\ ,
}
\eqn\nombre{
p_\pm^u= \pm ( wR - {\a\a'}J)+p_u  \ ,\ \ \
p_\pm^v= \pm(wR -  {\b \a' }J)+p_v  \ ,
}
\eqn\sspp{
p_u\equiv \ha (s-p) \ ,\ \  \ \  \ \ p_v\equiv \ha (s+p)\ ,
}
where $U_\pm' $ and $V_\pm' $ are single-valued functions of $\s_\pm $
and $s$ and $p$   are arbitrary  parameters
(later they will be expressed in terms of  the   Kaluza-Klein
momentum and the  energy of the string).
Then  it follows from  \gensol\ that \bcfz\  is satisfied provided
$\gamma $ (which is defined modulo $2$)
 is given by
$$\gamma =2 [q_1+q_2+\ha (\be-\a )]wR+\b p_u+\a p_v
 $$
\eqn\gamsol{
= \ (c_+ +a_+)wR+ \ha (\be+\a)s +\ha (\a-\b )  p  \ ,  }
where  $a_+=q_+ -\a, \ c_+=q_+ + \b$ (see \dee).

Starting from the general expression for the  classical
stress-energy tensor of the
theory \genlag\ and evaluating it on the general solution
\dualsol, \gensol\
one finds that it takes the ``free-theory" form\foot{The dilaton term is
 ignored in this section since  its  role is  only to
maintain the conformal invariance of the quantum theory.
This term
cancels out anyway once one performs  the transformation to the free-theory
variables.}
\eqn\tplus{
T_{\pm\pm}= \del_\pm U_\pm \del_\pm V_\pm   + \del_\pm
X \del_\pm X^*\ . }
This of course is not surprising since the on-shell values
of the stress-energy tensors in  the two dual \sms\  should be the same
(\tplus\ is precisely  $T_{\pm\pm}$ corresponding to the theory \lag,  see
\dualsol, \dualsole).
  It is convenient to fix the light-cone gauge, using the residual symmetry
to gauge away,    e.g.,   $U'_\pm $.
Then the  classical constraints $T_{--}=T_{++}=0$ can be solved    as
usual and  determine the remaining oscillators of $V'_\pm $
  in terms of the free fields  $X_\pm$. It is also straightforward to quantize
the model in the covariant formalism (which is more suitable, e.g., for a study
of scattering amplitudes in the operator approach)  but in order to  determine
the physical
 spectrum the light-cone gauge is as usual more convenient.

After using \uuvv,\bcfz\  $T_{\pm\pm}$ takes the form
\eqn\stressol{
T_{\pm\pm}=p_\pm^u p_\pm^v + p_\pm^u\del_\pm V_\pm'\pm  i \gamma
(\X_\pm \del_\pm  \X^*_\pm  -\X^*_\pm \del_\pm  \X_\pm )+\gamma  ^2 \X_\pm
\X^*_\pm  +\del_\pm \X\del_\pm  \X^*\ ,
}
where $\X_\pm$ have the standard mode expansions \fourie.
The classical expressions for the Virasoro operators $L_0, \tilde L_0$
are obtained by integrating over $\s $
\eqn\vira{
L_0\equiv {1\ov 4\pi\a' } \int_0^\pi d\s\  T_{--} =   {p_-^up_- ^v \ov 4\a' }
 +\ha\sum_{n }  \big(n+\ha \gamma \big) ^2 a_n^{*} a_{n} ,
}
 \eqn\soro{
\tilde L_0\equiv { 1\ov 4\pi\a' } \int_0^\pi d\s\  T_{++} =   {p_+^u p_+^v\ov
4\a'} + \ha
\sum_{n}   \big( n-\ha  \gamma  \big) ^2      \td a_n^{*} \td a_{n} \ .
 }
 Hence
the Hamiltonian  is given by
\eqn\hamilto {
H=L_0+{\td L}_0= { \textstyle {1\ov 8\a'} } (4 {w^2R^2 }+ s^2 -  p^2 ) } $$  +
\ha\sum_{n }  \big(n+\ha \gamma \big) ^2 a_n^{*} a_{n}
+\ha \sum_{n}   \big( n-\ha  \gamma  \big) ^2      \td a_n^{*} \td a_{n}
-\ha  wR (\a +\b )J+\ha \a'\a\b J^2\ ,
$$
where we have used \nombre.  $J$ is the angular momentum defined in
\bcfvp,\cude\
 which has the following mode expansion:
\eqn\momop{ J=J_R+J_L\ , \ \ \
J_R=-\ha\sum _n (n+\ha\gamma ) a^*_na_n\ ,\ \ J_L=-\ha \sum_n (n-\ha\gamma )
\td a^*_n \td a_n \  . }

\subsec{Operator quantization }
We can now quantize the  theory using the light-cone  operator
approach by imposing  the canonical commutation relations, in particular,
\eqn\canon{
[P_x(\s,\t), x^*(\s',\t)]=[P_x^*(\s,\t ), x(\s',\t)]= -i\delta(\s -\s')\   , }
$$ [x^i(\s,\t),\del_\s x^j(\s',\t )]=0 \ ,
$$
where $P_x=\ha (P_1+iP_2), \  P_x^*=\ha(P_1-iP_2) $ are the canonical momenta
corresponding to $x$ and $x^*$ in \genlag.
As a result,  $s,\ p$ in \sspp\  and the Fourier modes  $a_n, \td a_n$
will become  operators acting in a  Hilbert  space.
Again,  the duality   between \genlag\ and \lag\ and between \dup\ and \freee\
implies that  imposing  \canon\ is equivalent
to demanding the canonical commutation relations
for the  fields $X, X^*$ of the free (but globally non-trivial,
cf. \bcfz) theory
\eqn\canz{
[P_X(\s, \t ), X^*(\s', \t )]=[P_X^*(\s, \t  ), X(\s', \t )]= -i\delta(\s
-\s')\   , } $$
 \ \ [X^i(\s, \t  ),\del_\s X^j(\s', \t )]=0\ ,
$$
where   $P_X(\s,\t )= {1\ov 4\pi \alpha'} \del_\t X $.   Using  \bcfz,
eq.\canz\
implies
\eqn\fff{
[ a_n, a_m^{*}] =    2  (n+ \ha \gamma  )\inv   \delta _{nm}\ , \ \ \
[ \td a_n, \td a_m^{*}] =    2  (n - \ha \gamma )\inv   \delta _{nm } \ .
}
One also finds that   $s$ and $p$ in \nombre\  and thus $\g$ in \gamsol\
commute with the mode operators. It is  then  easy to check  directly that
 \canon\ are indeed satisfied.

The  string energy and  the Kaluza-Klein linear momentum operators are given
by\foot{As usual, the eigenvalue of $p_y$ is quantized since $[p_y, y_0]=-i$,
where $y_0$ is the compact zero mode  of $y$.}
 \eqn\epjr{
E=  \int_0^\pi d\s P_t\ ,\ \ \  \ p_y= \int_0^\pi d\s P_y = { m\ov R}  \ ,\  \
m=0, \pm 1, \pm 2 , ... \ .
}
 $P_t, P_y$ are the
 canonical momenta which correspond to \genlag,
or, equivalently (on the solution of the equations of motion),  to \lag\
(we use again  $a_\pm, c_\pm$ defined in \dee)\foot{We assume that the free
$\del u \del v$ term in  the Lagrangian is taken in the ``symmetrized" form,
i.e. the  total derivative term $\ha (\del_- u \del_+ v - \del_+ u \del_- v)$
is added to \genlag,\lag. If one does not add such term the expression for  $E$
is shifted by $wR$-term. }
 \eqn\motos{
P_t={1\ov 2\pi \a'}(-\del_\tau t - a_-\del_+\td \vp
+   c_- \del_-\td \vp )\ ,\ \
\ \
 P_y={1\ov 2\pi \a'}(\del_\tau y - a_+ \del_+\td \vp
+ c_+ \del_-\td \vp )\ .  }
Using \jjj,\dualsole,\uuvv,\nombre\  and  integrating over $\s $ we get
\eqn\robo{
E=  - { {1\ov 2\a '}} [p-   \a'(c_- + a_-) {\hat J}]   \ ,\ \ \
\   p_y=  { {1\ov 2\a '}} [ s+ \a'   (c_+ +a_+) \hat J]   \ .  \
}
Here  $\hat J$ is the angular momentum operator  obtained by
symmetrizing the classical expression $J=J_R+J_L$ in \momop\
\eqn\momot{ \hat J=\hat J_R+\hat J_L=  -\four \sum _n (n+\ha\gamma ) (a^*_na_n
+a_na^*_n) -  \four \sum_n (n-\ha\gamma )
( \td a^*_n \td a_n  + \td a_n \td a_n^*)\  . }
Expressing  $s$ and $p$ in terms of $E, p_y$ and $\hat J$
\eqn\sssp{
p= - 2\a' [ E - \ha (c_- + a_-) {\hat J}] \ , \ \ \ \
s = 2\a' [ p_y -\ha  (c_+ +a_+) \hat J]   \ ,   }
we can  represent  $\gamma $ \gamsol\ in the form
\eqn\gamlim{
\gamma =(a_+ + c_+)wR +\a'[(c_+ - a_+)p_y+(a_- - c_-)E] }
$$+\   \ha \a '(a_+^2 -a_-^2 - c_+^2 + c_-^2  )\hat J\ ,
$$
which is consistent with the property  that  $\g$   commutes
with the mode operators in \fff.

The Virasoro operators $\hat L_0$ and $\hat {\td L_0}$ are obtained  by
symmetrizing
the mode operator products in \vira,\soro.  In particular, starting from
\hamilto\ and using \sssp\
we get  the quantum Hamiltonian
\eqn\hamiltof {
\hat H  = \hat L_0 + {\hat {{\td L}}}_0 =
 - \ha \a' [ E - \ha (c_- + a_-) {\hat J}]^2
 + \ha \a'  [ p_y -\ha  (c_+ +a_+) \hat J]^2 } $$  +
 \ha \a' [ {\a'}\inv wR - \ha (c_+ -a_+) \hat J]^2
-{\textstyle {1\ov 8}} \a' (c_- -a_-)^2 \hat J^2 $$
 $$  + \
\four \sum_{n }  \big(n+\ha \gamma \big) ^2 (a_n^{*} a_{n} + a_n a^{*}_{n})
+\four  \sum_{n}   \big( n-\ha  \gamma  \big) ^2     (
 \td a_n^{*} \td a_{n} + \td a_n \td a^{*}_{n})  \ .
$$
The  sectors of states  of the  model  can be  labeled  by  conserved
quantum numbers:  the   energy $E$, the angular momentum  $\hat J$ in the
$x_1,x_2$ plane, the
linear $p_y=m/R$ and winding  $wR$  Kaluza-Klein  momenta or ``charges"
(and also  by momenta in additional spatial dimensions).
The value of $\g$  \gamlim\ in  a given sector   depends  on $ E, \ \hat J, \
m, \ w$
as well as on the
parameters $a_\pm,\  c_\pm, \  R$
which determine the strength of the corresponding background fields.

In agreement with  the defining relations in \bcfz\  the
expressions for $\hat H$, $\hat J$ and the commutation relations \fff\
are invariant under $\g \to \g +2$ combined with the corresponding renaming
of the mode operators $a_n \to a_{n+1}, \ \td a_n \to \td a_{n-1}$.\foot{
Let us note that the region of $|\gamma | \approx 2 $  corresponds to
 values of   field strengths or quantum numbers which are of
  Planck order.
The fact that  the  mass spectrum  depends on $\gamma $ only modulo 2 suggests
that physics at  strong fields with $\gamma
=2+\epsilon $
is equivalent to the weak field regime with $\gamma =\epsilon $, i.e. implies
certain  periodicity in  dependence on the  field strengths.}

 The states belonging to the  $\g=0$  (in general,  $|\gamma| = 2n$, $
n=0,1,...,$)
 ``hyperplane" in the $(m,w,E,J)$ space
are special.
 For these states   the  translational invariance on the plane is restored: the
  zero-mode  oscillators $a_0, a_0^*,  \td a_0, \td a_0^*$  are  replaced by
the zero mode coordinate and conjugate  linear momentum.\foot{Strictly
speaking, this is true provided
$\g=0$ is satisfied with no constraint imposed on the orbital part of the
angular momentum $J$. Otherwise  one gets just one   continuous  (radial)
quantum number   in the ``plane" part of the spectrum (see Section 5.3).}

Restricting  for the moment the consideration to the sector of states with
 $ 2 >  \g  > 0 $,
where $\g $ is defined by eq. \gamlim ,
 one can   introduce  (as in   \ruts) the  normalized creation and annihilation
operators
 which will be used to define the Fock space of our model
(the subindices $\pm$ correspond  to the components with spin ``up" and
``down" respectively)
\eqn\ope{
[ b_{n\pm}, b_{m\pm }^{\dagger}] = \delta _{n m} \ ,\ \
 [\td b_{n\pm}, \td b_{m\pm }^{\dagger}] = \delta _{n m} \ ,\ \
[b_0,b_0^{\dagger}]=1 \ , \ \ [\td b_0,{\td b}_0^{\dagger}]=1 \ , }
\eqn\rightope{
 b_{n+}^{\dagger}= a_{-n} \omega_- \ ,\ \  b_{n+}= a_{-n}^* \omega_-\ ,\ \
 b_{n-}^{\dagger}= a_{n}^* \omega_+ \ ,\ \  b_{n-}= a_{n} \omega_+\ ,
}
\eqn\leftope{
 \td b_{n+}^{\dagger}= \td a_{-n} \omega_+ \ ,\ \  \td b_{n+}=\td a_{-n}^*
\omega_+\ ,\ \
\td b_{n-}^{\dagger}=\td a_{n}^* \omega_- \ ,\ \ \td b_{n-}=\td a_{n} \omega_-\
,\
}
 \eqn\ttt{
b_0^{\dagger}=\ha \sqrt{\gamma } a_0^*  ,\  \ b_0=\ha \sqrt{\gamma }a_0
 , \  \ {\td b}_0^{\dagger}=\ha \sqrt{\gamma } \td a_0  ,\ \ \td b_0=\ha
\sqrt{\gamma }\td a_0^*
\  ,  }
\eqn\ooom{
\omega_\pm \equiv \sqrt {  \ha \big( n \pm \ha {\gamma } \big) }, \
 \ \ \  n=1,2,..., \ \   \ \ \  0<  \g  < 2 \ . }
Then the  angular momenta operators \momot\ become (after normal ordering)
\eqn\angulr{ \hat J= {\hat J}_R+\hat J_L = \tilde b^{\dagger}_0 \tilde b_0
-b^{\dagger}_0 b_0 + S_R + S_L =J \ , }
\eqn\angull{ {\hat J}_R= - b^{\dagger}_0 b_0  - \ha  +\sum_{n=1}^\infty \big(
b^{\dagger
}_{n+}b_{n+} - b^{\dagger}_{n-}  b_{n-} \big)\equiv J_R-\ha
 \ ,}
$$
{\hat J}_L= \tilde b^{\dagger}_0 \tilde b_0  + \ha  +\sum_{n=1}^\infty \big(
\tilde b^{\dagger
}_{n+}\tilde b_{n+} - \tilde b^{\dagger}_{n-} \tilde b_{n-} \big)\equiv J_L +
\ha
 \ . $$
 The operators $L_0$ and $\td L_0$ will be normal ordered
with the
ordering constant being fixed by  the Virasoro algebra.
The Virasoro operators $L_n, \td L_n, \  n\neq 0$ can be  obtained in the
standard way as the Fourier components of the stress-energy tensor,  e.g.,
$L_n=(4\pi\a')^{-1}\int d\s e^{-i2\s n} T_{--}= p_- ^u V_n + {\cal L}_n.$
 In the light-cone gauge $L_n$  are required to vanish so
that their ``transverse" parts ${\cal L}_n$  are proportional to the modes
 of the operator $V'_\pm $,
\eqn\vfourie{ V_+'  =  2i  \a' \sum_{n\neq 0} {1\ov n} \tilde V_n
\exp (-2in \s_+)
  \  , \ \ \ \
V_- ' =  2i \a' \sum_{n\neq 0} {1\ov  n} V_n
\exp (-2in \s_-)
  \ . }
One  finds for the ``transverse" parts
of $L_n$
\eqn\viraa{
{ \cal L}_n \equiv -p_-^uV_n=\ha \sum_k (k+\ha\g)(k-n+\ha\g) a_k a_{k-n}^*
}
$$=
\sum _{k=1}^\infty [(k-\ha\g)(n+k-\ha\g )]^{1/2}
b^\dagger_{k+} b_{(n+k)+}
+\sum_{k=0}^\infty [(k+\ha \g)(n+k+\ha \g)]^{1/2}b^{\dagger}_{k-}b_{(n+k)-}
$$
$$
-\sum_{k=0}^{n-1} [(k+\ha \g)(n-k-\ha \g)]^{1/2}b_{k-}b_{(n-k)+}\ ,
$$
$$
{\cal L}_{-n}\equiv -p_-^u V_{-n}={\cal L}_n^{\dagger}\ ,\ \ \ n=1,2,3,...\ \ ,
$$
and similarly for $\td {\cal  L}_n$. It is an easy exercise to verify
that after  adding the contribution of the remaining 22 free scalar fields, the
operators ${\cal L}_n$ satisfy the  Virasoro algebra with  the central
charge equal to 24 and a trivial cocycle which shifts
 the free-theory normal ordering constant in  $L_0$  from 1 to
$1-\fourth \gamma (1-\ha\gamma) $ (or to
$1-\fourth \gamma' (1-\ha\gamma') $,   where $\g' = \g - 2k$
and $k$ is integer,
in the case when $2k < \g< 2k + 2$).
This corresponds to regularizing the infinite sums arising in the normal
ordering  process by the using   the generalised $\zeta$-function
regularisation \zere.
 Similar result is found in the
 open
string theory in  a constant magnetic field  \abo\ and is
typical to the case of   a free scalar field with twisted boundary conditions
(which appear also in orbifold or ``cone"  models, see, e.g.,
\refs{\orb,\dabh,\stro}).
 We will see that this shift is also  consistent with the modular
invariant  path integral
approach discussed in Section 4.

Inserting \gamlim,\robo \ into \vira--\hamilto\ and replacing the  zero mode
operators $E$ and $p_y$ by their eigen-values  we
obtain
the  expressions for  the Virasoro operators  and
the Hamiltonian
  \eqn\lzero{
{\hat L}_0=\four \a'  \big(-E^2+ p^2_a  + Q_-^2 \big)
 +\ha {\cal H} + N -c_0 \   ,
}
\eqn\lzerob{
{\hat {{\td L}}}_0=\four \a' \big(-E^2+p^2_a + Q_+^2\big) +\ha {\cal
H}+{\tilde N}-c_0\   ,
}
\eqn\hamiltoni{
\hat H= \ha \a' \big( -E^2 +  p_a^2 + \ha Q_+^2 +
\ha Q_-^2  \big) +{\cal H}+ N+  {\td N}-2c_0
\ , }
\eqn\jeisq{
{\cal H}\equiv  -
\a'   [(c_+Q_+-c_-E) J_R+(a_+ Q_- - a_- E) { J}_L] }
$$  +
 \ha \a' \big [  (c_+^2-c_-^2 )  J_R^2 +(a_+^2-a_-^2 ) J_L^2
+  (a_+^2+c_+^2-c_-^2 -a_-^2)  J_R  J_L  \big] \ .  $$
Here $Q_\pm$ are the left and right combinations of the Kaluza-Klein  linear
and winding momenta (which play the role of charges in the present context)
\eqn\defi{  Q_\pm \equiv  {1\ov \sqrt {\a'}}  ({m\ov r} \pm  wr)= p_y \pm
{wR\ov\a '} \  , \ \ \  \ \ \
   \  r\equiv {R\ov \sqrt{\a' }}\ ,
}
\eqn\czero{
c_0 \equiv 1- \four \gamma(1-\ha \gamma) ,
}
and  $N$ and $ {\td N}$ are the  standard
operators\foot{We have  introduced extra  22 spatial
free dimensions with momenta  $p_a$ and oscillators $a_{na}, \td a_{na}$
to ensure that the total space-time dimension is 26. }
\eqn\nnn{
 N= \sum_{n=1}^\infty n ( b^{\dagger}_{n+}b_{n+}+ b^{\dagger}_{n-}b_{n-}
+ a^{\dagger}_{na} a_{na} ) \ ,\ }
\eqn\nn{
 {\td N}= \sum_{n=1}^\infty n ( \td b^{\dagger}_{n+}\td b_{n+}+ \td b^{\dagger
}_{n-}\td b_{n-}+\td a^{\dagger}_{na} \td a_{na}   )  \ . }
 The Hamiltonian for the case of $\g=0$ is obtained by adding
$\ha\a'(p_1^2+p_2^2)$ and replacing
$-b_0^{\dagger}b_0-\ha $ and $\td
b_0^{\dagger}\td b_0 + \ha $   in  $\hat J_R$ and
$\hat J_L$ in \angulr,\angull\ by one half  of the  center of mass orbital
momentum $(x_1p_2-x_2p_1)$.

As explained above, the mass formula for states with $2k<\gamma <2k +2,\
k=$integer,  can be found in a similar way   by  renaming the creation and
annihilation operators.
The result is the same as  in  \lzero-\hamiltoni\ with the replacement
$ \g \to \g '= \g-2k $ in $c_0$.

Note that  the continuous
momenta $p_{1,2}$  corresponding to the zero modes of the
coordinates $x_{1,2}$  of the plane are  effectively replaced in the $0<  \g <
2  $   sector by the integer eigenvalues
$l_R, l_L =0,1,2,...$ of the zero-mode parts
$b_0^{\dagger}b_0$ and $\td
b_0^{\dagger}\td b_0$   of $\hat J_R$ and
$\hat J_L$. Thus  the ``2-plane"  part of  the spectrum is  discrete
(but, as mentioned above,  it  is continuous  when  $\g=0$).
 For example, in the Melvin model case (most of) the string states are
``trapped"  by the flux tube (see also Section 5.3).
This result is  consistent with   a picture of a closed string moving
 in a   magnetic field
orthogonal to the plane (see \ruts\ and below).



The Hamiltonian  \hamiltoni\    is  non-trivial:
because of the   angular momenta squared terms
it is, in general,    of $fourth$
order in creation/annihilation operators.
It is clear from our construction that \hamiltoni\ is, at the same time, also
the Hamiltonian for the  $\vp$-dual theory \lag, i.e.
it can be derived by starting directly with \lag\  (the origin of the quartic
terms in $\hat H$ can be  traced, in particular,
 to the presence of the $\a\b \del \td \vp \bd \td \vp $ term in \lag).\foot{In
the  $\vp$-dual theory the expression \gamsol\ for $\g$ in \bcfz\
follows from the condition that the zero mode of the momentum conjugate to $\td
\vp$ should take integer eigenvalues to make $\exp (i\td \vp p_{\td \vp})$
single-valued.}
The
two  $\vp$-dual
\sms\ \genlag\ and \lag\  thus represent the same CFT.
The quartic terms are absent only when $a_+=\pm a_-, \ c_+=\pm c_-$ (i.e. when
$\a\b q_+ (q_+ + \b -\a) =0$).
 In the case of $a_+=a_-=0, \ c_+=-c_-=\b$
corresponding to the model \lgg\  $\hat H$  in \hamiltoni\
 reduces to  the  quadratic  Hamiltonian  found in \ruts.
As for general values of the parameters, the spectrum can still be computed
explicitly, just as in the special
case considered in \ruts,
 since the
Hamiltonian \hamiltoni\ is in a {\it diagonal}  form (note that $N, \td N, J_L$
and
$J_R$ are diagonal in Fock space).

  It is easy to  see that (in agreement with a  discussion in Section 2)  the
coordinate-invariant  physical quantities  depend only on the $three$
parameters $\a,\b,q_+$ (as well as on $r$) even though $\hat H$ contains also
$q_-$.
The dependence of  on $q_-$ reflects
 a  choice of the frame that was made, and translates into the expected frame
dependence of the energy eigenvalues.

The Hamiltonian \hamiltoni\   is invariant under
the following transformations
\eqn\ddud{
r\to r\inv  \ ,\ \ \  m\leftrightarrow  w\ ,\ \ \ \ \ a_+\to -a_+ \ , }
\eqn\drdu{
r\to r\inv  \ ,\ \ \  m\leftrightarrow  -w\ ,\ \ \ c_+\to -c_+ \ . }
 This is consistent with the observation made in Section 2
that the $y$-dual of our \sm \lagg\ is  an
equivalent \sm \ldu\ with either  $y\to \td y$ and  $a_+\to -a_+$ or $y\to -
\td y$  and $c_+\to -c_+$  \ded. As was already  mentioned after \ded,
an interesting feature of our model is that
a nontrivial structure of the Lagrangian \laggd\ or \lag\
involving a  coupling   between  the two angular coordinates $y$ and $\vp$
implies that the $y$-duality  transformation ($r\to r\inv$),  in general,
must  be accompanied  by a  change in  one extra parameter.

The presence of the $O(\g^2)$ normal ordering term in \hamiltoni,\czero\
implies (see \gamlim) that the quantum Hamiltonian contains  $O(\a'^2)$ terms
of one  order higher in $\a'$. This is a consequence of the regularisation
(normal ordering) prescription we used  which is consistent with the
reparametrisation invariance of the theory (in particular, the  Virasoro
algebra and  modular invariance of the partition function).
Such higher order term is also consistent with
current algebra  approaches.
Indeed, there are two special cases when  our model becomes equivalent to  a
special
WZW or coset model:

 (i)   the non-compact ($R=\infty$) limit of the
constant magnetic field model  \lgg\ is equivalent (see \ruts) to
the $E^c_2$ WZW model \napwi\  for which the quantum stress tensor
contains one order $1/k \sim \a' $ correction term \refs{\kk,\sfees} (which is,
indeed, equivalent to the term  appearing in \hamiltoni\ in this limit);

(ii) the non-compact limit of the
Melvin model \lggm\  (when all
 coordinates are formally taken to be non-compact) can be related \tsem\  to  a
 special limit of the $SL(2,R)\times R/R$
gauged WZW model, or,  equivalently,  to the    $E^c_2/U(1)$ coset  theory
\sft,  the quantum Hamiltonian of which (obtained, e.g.,
 by a taking a limit in the
standard semisimple coset model Hamiltonian)
 also contains  a $1/k$ correction term \sft.

As follows from \lzero,\lzerob,  the Virasoro conditions $\hat L_0=
{\hat {{\td L}}}_0 =0$ become
\eqn\rotat{ \hat H\equiv H -2 +  \ha \g - \four \g^2 =0 \ , \ \ \ \
\  N- {\td N}=mw
  \ . }
The analysis of the spectrum in the special case $q_1=q_2=0$, $\a =0$,
was performed   in \ruts. The spectrum for the general class of models
considered here displays similar qualitative features, in particular,
 tachyonic instabilities. We shall discuss some of its
properties  in Section 6.

\subsec {Point-particle limit: zero level  scalar (tachyon) spectrum in the
Melvin model}
\def\T{\tilde T}
To illustrate  our  construction of the string  Hamiltonian  and  the
solvability of the model it is useful
to discuss the point-particle limit. Let us consider, for example, the case of
the Melvin model \lggm\
 ($a_+=0, \ c_+ = 2 \b , \ a_-=c_- =0$).\foot{The case of the constant magnetic
field model \lgg\ was discussed in detail in \ruts.
Other cases can be considered in a similar way (with special care taken
in comparing the  definitions of the
string and particle momenta because of the presence of the antisymmetric
tensor.}
The  point-particle limit of its  action is
(we omit the trivial $x_3$ direction)
\eqn\mepp{ S = {1\ov 4\a'} \int d\t \big[- \dot t^2  + \dot \r^2  + {F(\r)
\r^2}
\dot \vp (\dot \vp +  2\b \dot y)  + \dot y^2  \big] \  ,  \ \ \ F\inv =
1+ \b^2 \r^2 \ ,   }
with the   Hamiltonian being
\eqn\melh{H= \a'\big[ - p_t^2 +
p^2_\r + \r^{-2} F^{-2}(\r) p_\vp^2 + F\inv (\r)p_y^2  -2 \b F\inv (\r) p_\vp
p_y  \big] \  }
$$= \a'\big[ - p_t^2  + (p_y-\b p_\vp)^2 +
\b^2 p_\vp^2 + p^2_\r  + \r^{-2} p_\vp^2
+ \b^2 (p_y - \b p_\vp)^2 \r^2 \big] \ . $$
The scalar product
is defined with respect to the measure $\sqrt {-G} e^{-2\p} = \r$.
The Hamiltonian commutes with the Kaluza-Klein charge operator ($p_y$) and
the angular momentum operator  ($p_\vp$) and the   quantum mechanical problem
reduces to that of a
 free two-dimensional  oscillator.

 The corresponding Klein-Gordon  operator
is the Laplacian which appears in the (zero winding sector) tachyon equation
\eqn\kg{ \Delta = - {1\ov{\sqrt{ -G} e^{-2\p} }} \del_\m ({\sqrt{ -G}  e^{-2\p}
}G^{\m\n} \del_\n) \ , \ \ \  \ \
\a' (\Delta + O(\a'))  T = 4 T  \ ,  }
where  the  Melvin model ``$D=5$   metric" (with $x_3$ direction omitted)
 and the dilaton  are
\eqn\melme{ ds^2 = - d t^2  + d \r^2  + {F(\r) \r^2}
d \vp (d\vp + 2\b dy)  + d y^2 } $$ =  - d t^2  + d \r^2  + {F(\r) \r^2}
 (d\vp + 2\b dy)^2  + F(\r) d y^2   \  ,    \ \ \
 e^{2(\p-\p_0)}=F(\r) = (1+ \b^2\r^2)\inv \ . $$
Let us first ignore   possible higher order  $\a'$-terms
in the tachyon equation. Then
we get
   \eqn\mees{
\big[ - \del_t^2  + \r\inv \del_\r(\r\del_\r)
+ (\r^{-2} + 2\b^2  + \b^4 \r^2) \del_\vp^2 }  $$
\  + \   (1 + \b^2\r^2) \del_y^2
-2\b (1+ \b^2 \r^2) \del_\vp\del_y \big] T=
-4{\a'}\inv  T  \ . $$
 If we consider a particle state with a given energy $E=p_t$,
charge $p_y=m/R$ ($m=0,\pm 1, ...$) and orbital momentum
$p_\vp= l\ $ ($l=0, \pm 1, ... $)
 $$T= \exp (iEt + ip_y y + il\vp ) \ \T(\r) \ , $$
then
\eqn\schr{
\big[-{\r}\inv  \del_\r (\r\del_\r )
+ l^2\r^{-2}+   \g_0^2 \r^2 \big] \T= \m^2 \T \ , }  $$
\g_0 = \b (p_y -\b l)\ , \ \
\ \ \  \m^2 \equiv  E^2 + 4{\a'}\inv    - (p_y -\b l)^2  - \b^2l^2
\ .  $$
This  is  the Schr\"odinger equation
for a two-dimensional  oscillator
with the  frequency $\g_0 $.
In what follows we assume that $\g_0 >0$
(the spectrum for $\g_0<0$ is the same since the Hamiltonian
depends  on  $\g_0^2$).
Provided $\g_0 \not=0$ the spectrum is discrete, i.e. \schr\ has  normalizable
solutions
only if
\eqn\spee{  \m^2 =  2 \g_0 (2k + |l| +1) \ , \ \
\ \ \ k=0,1,2,... \ . }
To make the  analogy with  the solution  of   the string problem more explicit
let us  derive \spee\ by introducing the  creation/annihilation operators
\eqn\ppw{ C  \equiv  -i(\del_x^* + \ha \g_0  x)\  , \ \ \  \ \
 \td C  \equiv  -i(\del_x + \ha \g_0  x^*)\  , \  \ \
 \ \ }
$$ C ^\dagger\equiv  -i(\del_x - \ha \g_0 x^*)\ ,  \ \
\  \ \ \td C ^\dagger\equiv  -i(\del_x^*- \ha \g_0 x)\ ,  \  \ \ \
 x= \r e^{i\vp} \ , \  \ x^*= \r e^{-i\vp} \ ,$$
$$
\ \  [C , C ^\dagger]= \g_0  \ ,  \ \   \ \
[\td C , \td  C ^\dagger]= \g_0 \ , \ \ \ [C , \td C ] = [C , \td C ^\dagger]=0
\ ,   $$
\eqn\ops{ p^2_\r + \r^{-2} p_\vp^2 = - 4 \del_x\del^*_x =  (C ^\dagger + \td C
) ( \td C ^\dagger + C )  \ , \ \ \
\g_0^2\r^2 = (C -\td C ^\dagger)(C ^\dagger -\td C )\ , $$
 $$  \ \ \ p_\vp =-i\del_\vp= x\del_x - x^* \del^*_x=
\g_0\inv (\td C ^\dagger \td C   - C ^\dagger C   )\ . }
As a result, \schr\ is equivalent to
\eqn\scre{ 2(C C^\dagger + \td C^\dagger \td C) \T=
  2\g_0(b^\dagger b + \td b^\dagger \td b + 1 )T=  \m^2 \T \ , \ \  }$$  b =
\g_0^{-1/2} C \ , \ \  \ \
 \td b = \g_0^{-1/2} \td C  \ , \ \ \ \
[b, b^\dagger]= 1  \ ,  \ \  \
[\td b, \td  b^\dagger]= 1 \ . $$
The eigen-functions are thus given by
\eqn\tacf{
 T_{E,p_y, l_R,l_L}   =  e^{iEt + ip_y y } (b^\dagger)^{l_R}(\td
b^\dagger)^{l_L}  \exp (- \ha\g_0  xx^*) \ , }
   with the eigen-values
\eqn\eig{\m^2= 2\g_0 (l_R + l_L + 1)\ ,
\ \ \ \ \ \
l_{R}, l_{L} =0, 1,2, ... \ . }
This is the same condition as in \spee\ since according to \ops\
the orbital momentum eigen-value is
\eqn\speee{ l=  l_L-l_R\ , \ \ \  \ \
l_R + l_L = 2k + |l| \ . }
The resulting tachyon spectrum is the same as the semiclassical (leading order
in $\a'$) spectrum that
 follows from the string  constraints \rotat\ with
the Hamiltonian \hamiltoni\ depending on  the ``angular momenta"  operators
\angulr,\angull.
At the zero string excitation
level $S_L=S_R=N=\td N=0$
and the eigen-values of $\hat J_R$ and $\hat J_L$
are $-l_R -\ha $ and $l_L + \ha$ so that $\hat H=0$ reduces to
($p_a, \ a=3, ..., \ $ are momenta of additional dimensions)
\eqn\spece{ M^2\equiv E^2- p_a^2=
 - 4{\a'}\inv   + p_y^2 }
$$  -4\b p_y \hat J_R  + 4\b^2  (\hat J_L + \hat J_R)\hat J_R
- 2\a' \b^2(p_y - \b \hat J)^2
$$
$$ = - 4{\a'}\inv    + p_y^2  + 2\b p_y (2l_R + 1)
- 2\b^2 (l_L-l_R)(2l_R + 1) - 2\a' \b^2[p_y - \b (l_L-l_R)]^2
 \ ,  $$
where  it is assumed that the eigen-value $2\a' \b  [p_y - \b (l_L-l_R)]$ of
the operator $\g$ in \gamlim\
 is positive,
i.e $\g= 2 \a' \g_0 >0 $.  The $O(\a')$ correction  comes from the
$\g^2$ term  in $c_0$ in \hamiltoni.
It is easy to see that the  two expressions \eig\
and \spece\
for the point-like tachyon spectrum   indeed  agree up to the  $O(\a')$-term in
\spece .
A similar correspondence between the point-particle spectrum and the  zero
level string spectrum
can be established also for other
choices of background parameters (the solution
of the tachyon equation in the
case of the constant magnetic field model \hts, \lgg ,
 which is similar to the Landau spectrum, was already given in \ruts).

We see that for $\g=2\a'\g_0=2\a'\b[p_y -\b (l_L-l_R)]\not=0$
the scalar particles are ``localized"
near the core of  the flux tube, i.e. oscillate near $\r=0$ (they can, of
course, move freely  along the flux tube direction $x_3$).
Note that   even
the neutral tachyon states   with nonzero orbital momentum are trapped
by the flux tube.
Similar  ``bound state" interpretation should apply for higher  excitations in
the string
spectrum.
 Another interesting feature is that,
in contrast to the   Landau spectrum or the spectrum of the  standard 2d
oscillator,  here there is no usual degeneracy in the energy
 since the frequency  $\g_0$ itself   depends on $l$ (there is still
 smaller
degeneracy $\b(p_y -\b l) \to -\b( p_y -\b l)$).
When $\g_0=0$, i.e. $ \ p_y=\b  l \not=0, $ the  spectrum becomes continuous,
i.e. $\m^2$ in \schr\ can take arbitrary positive values
(yet the translational invariance on the plane is not fully restored since the
eigen-value of the orbital momentum is subject to the constraint
$l=p_y/\b$).
 This is possible only  for special values of the magnetic field strength,  $\b
R =m/l$. The spectrum is also continuous for $p_y=0, \ l=0$
 when the solution  of \schr\ which    decays at infinity and is bounded at
zero is  given by the zeroth Bessel function $\td T = J_0(\m \r)$
 with arbitrary $\m$.\foot{The solution of  a similar (uncharged) Klein-Gordon
equation
in the $a=0$ Melvin background was discussed in \gibwi, where
it was also  found   that the
spectrum contains both discrete ($l\not=0$) and continuous ($l=0$) branches.}

In general, the tachyon equation \kg\ contains  higher order  $O(\a'^2)$
terms.
Such terms
  are scheme-dependent and may be non-vanishing in the scheme in
which
the exact expressions for the
\sm couplings (metric, dilaton and antisymmetric tensor)
do not depend on $\a'$ (for a discussion of $\a'$-corrections to the
tachyon equation see \refs{\sftse, \sfew}\  and refs. there).
The exact form of the tachyon equation which is usually  hard to determine
at the $\s$-model level is  determined  by the underlying CFT.
  In the present model it can be
fixed by using directly the Hamiltonian  \hamiltoni\ or
 the relation to the coset model (for a similar discussion
in the case  of the constant magnetic field model see \ruts).
The appearence  of the $\a'$-correction term in the point-particle limit of the
string Hamiltonian  \hamiltoni\ or in  \spece\
is consistent with the presence of the ``quantum"
 $1/k$  correction  term in the Hamiltonian
of the special limit of the $SL(2,R)\times R/R$ WZW model \tsem\ or
in the Hamiltonian of  the
  $E^c_2/U(1)$ coset model \sft\ which, as was mentioned above,
 is related to the ``non-compact" limit of the Melvin model.\foot{As usual  in
the  gauged WZW  models, there
are two possible interpretations of the $\a'$-correction terms in the tachyon
spectrum:
in the ``CFT scheme" in which the tachyon equation retains its  leading-order
Klein-Gordon form they come from the $\a'$-corrections in the background \sm
fields \refs{\dvv,\bsf}; in the ``leading-order scheme"
in which the \sm fields have semiclassical values
they originate  from the corrections  to the tachyon equation \sftse.}
To find the $\a'$-correction term in the tachyon  equation
which produces the  $O(\a')$-term in \spece\
one may start, e.g.,
 with the quantum action of the $SL(2,R)\times R/R$ gauged  WZW model
with the full $1/k$-dependence included (see \sfew)
and take the special  limit $1/k=\a' \ep^2, \ q_0 = -1 +\b^{-2} \ep^2, \
r= 2\ep \r, \ \theta= \vp + \b y, \ \td \theta = \b\inv \ep^2 y, \ \ep \to 0$
($q_0$ is the parameter of embedding of the subgroup, $r,\theta, \td \theta$
are
$SL(2,R)$ coordinates). As a result, one finds the exact ($\a'$-corrected)
 form of the Melvin model action from which one reads off  the following
metric  and dilaton (cf. \melme)
\eqn\melmed{ ds^2 = - d t^2  + d \r^2  + {F(\r) \r^2}
 (d\vp + 2\b dy)^2  + F'(\r) d y^2   \  ,   }
$$  e^{2(\p-\p_0)}=[F(\r) F'(\r)]^{1/2}= \r\inv \sqrt {- G} \  ,
\ \ \ \  \  F'= (1+ \b^2\r^2 -2\a'\b^2)\inv \ . $$
These expressions correspond to the ``CFT scheme"
where the exact tachyon equation is the Klein-Gordon one for the exact metric
and dilaton.
As a result, the differential operator in \mees\  gets an  extra  term $
-2\a'\b^2(\del_y - \b \del_\vp)^2$
 so that $\m^2$ in \schr\ is shifted by $2\a'\b^2(p_y -\b l)^2$
and the spectrum \spee\ becomes exactly equivalent to \spece.

\subsec { Partition function from the  operator  quantization}
In  the operator formalism,  the one-loop  partition function
is obtained by using the Hamiltonian to propagate the  states  along  the
cylinder and taking the trace to
identify its ends (and imposing the Virasoro constraint with
 a Lagrange multiplier). Then
\eqn\zoper{
Z=\int {d^2\t\ov \t_2} \int dE\prod_{a=1}^{22} dp_a
\sum_{m,w=-\infty}^\infty  \Tr \exp \big[ 2\pi i( \t {\hat L_0} - \bar \t {\hat
{\td L}_0} )\big]\ ,
}
where  $\hat L_0$ and $\hat {\td L}_0$ are the Virasoro operators
\lzero\ and \lzerob\ constructed above.
Our aim is to compute $Z$  defined by \zoper\ and  to show its equivalence to
the expression  obtained  in the path integral approach \etr,\wew.
In order to integrate over $E$ and perform
the Poisson resummation, it is convenient
to represent  \lzero\ and \lzerob\  in the form (cf. \hamiltoni)
\eqn\complesq{ {\hat L}_0 =  \four  w^2 r^2   -   \ha mw  +N-1 - \four \a'(
E-c_-\hat J_R-a_-\hat J_L) ^2 }
$$  +  \four  [{m r\inv } - \sqrt {\a'} (c_+\hat J_R+ a_+ \hat J_L)]^2
-\ha  wr\sqrt {\a'} (c_+  \hat J_R-a_+ {\hat J}_L)
   + \a'\a\b \hat J_R\hat J_L  -\eit \g ^2 ,
 $$
 \eqn\kmk{
  {\hat{ \td  L}}_0 =  {\hat L}_0  + mw  + {  {\td N}} -  N\ . }
It is convenient to express the part of the exponential factor in \zoper\
containing $\gamma^2 $
in the following way:
$$
\exp{(\ha  \pi \t_2 \gamma^2})=\sqrt{\t_2}\int d\n \  \exp ( -\ha \pi \t_2
\n^2-\pi
\t_2 \gamma \n) \ ,
$$
where $\n$ is an auxiliary variable.
The  term  $\pi \t_2\gamma \n$    can be absorbed into
a redefinition of $\hat J_R\to \hat J_R'\equiv \hat J_R-\ha \n $,
$\hat J_L\to \hat J_L'\equiv \hat J_L+\ha \n $
as can be easily verified.

The  gaussian integrals over $E, p_a $ give an extra factor of  $\t_2^{-23/2}$.
 By using the Poisson resummation formula, one can trade the sum over the
discrete loop momentum $m$
for the sum over the conjugate winding number $w'$:
\eqn\poiw{
  \sum_{m =-\infty}^{\infty}{\cal  F}(m) = \sum_{w'=-\infty}^{\infty} \int d\m\
 e^{2\pi i \m  w'} {\cal  F} (\m)
\  ,  } $$
 {\cal F} (m) \equiv  \exp \big(  -\pi \t _2 [{m r\inv }-\sqrt {\a'} (c_+\hat
J_R'+ a_+ \hat J_L')] ^2-
2\pi i mw\t_1 \big) \  .
  $$
Integrating over $\m $ we get
$$
Z= \int d^2\t \ \t_2^{-13} \  \sqrt{\t_2}\int dx \  \exp ( -\ha \pi \t_2 \nu ^2
)  \ \Tr \ {\cal Z}\ ,
$$
\eqn\posson{
{\cal Z} = {r }
\sum_{w,w' =-\infty}^{\infty} \exp[  -{\pi r^2 \t\inv_2}(w'-\t w)(w'-\bar\t w)]
\   \exp \big[2\pi i(\t  N -\bar \t  {\td N})\big] }
$$ \times \exp[{2\pi i(w'-\t w) r \sqrt{\a'} c_+ \hat J_R'}]\
\exp[{2\pi i(w'-\bar \t w) r\sqrt{\a'}a_+ \hat J_L'}]  $$ $$
\times \   \exp( -4\pi \t_2 \a'\a\b  \hat J_R'\hat J_L')
\ .
$$
In order to compute the trace in \zoper\
and relate the result  to  the path integral
expression \zee\
it is convenient to ``split" the $\hat J_R\hat J_L$-term by introducing the
auxiliary variables, inserting   the following  identity:
\eqn\unos{
1=   {4 \t\inv _2 }
\int d\l d\bar \l } $$ \times \exp\big( -4\pi \t_2 ^{-1} [\l -\ha  r (w'-\t
w)-i\t_2\sqrt{\a'}\a \hat J_L']
[\bar \l +\ha   r (w'-\bar \t w)- i \t_2\sqrt{\a'}\b \hat J_R']   \big) \ .
$$
Then the first and the last exponential factors in \posson\
are cancelled out  and we are left with the following expression
\eqn\zzzzz{
 Z=c_1 \int {d^2\t\ \tau_2^{-14} } e^{4\pi \t_2} \sqrt{\t_2}\int d\n \  \exp (
-\ha \pi \t_2 \n^2 )
 \sum_{w,w'=-\infty}^\infty   \int d\l  d\bar \l  }
$$ \times \exp \big({- 4\pi \t_2 ^{-1}
 \big[ \l \bar \l     -  \ha  r (w'- \t w) \bar \l  +   \ha  r (w'-\bar \t w)
\l  \big] \big) } \  {\cal W}(\n, \l,\bar \l,  w,w';\t,\bar \t  ) \ , $$
where
\eqn\oooo{
{\cal W} \equiv \Tr  \exp  \big[{2\pi i(\t N - \h \hat J_R'})\big] \
\Tr   \exp \big[-2\pi i( \bar \t \td N   + \td \h \hat J_L')\big] \  , }
\eqn\ooo{
 \h\equiv - \sqrt{\a'}[ 2  \b \l + q_+  r (w'-\t w) ] \ ,\ \  \ \
\td \h \equiv - \sqrt{\a'}[ 2 \a\bar  \l   + q_+  r (w'-\bar\t w)] \ . }
The traces can now be easily computed  by using that
\eqn\ree{
\Tr \exp \big[2\pi i (\t N  - \h \hat J_R')\big]
 = {1\ov  \sqrt{\t_2 }  }
{\pi \h \ov \sin \pi \h }  \exp(\pi i  \chi \n) } $$ \times
\prod _{n=1}^\infty  (1-e^{2\pi i n\t })^{-22}\big[1- e^{2\pi i( n\t +
\h)}\big]\inv
\big[1- e^{2\pi i( n \t-\h)}\big]\inv \ , $$
and a similar expression for the ``left"  part.
The    factor $\pi \h /\sqrt {\t_2 }$ comes  from the normalization of the zero
mode  so that \ree\ has a regular $\h \to 0$ limit (see also  \ruts).
 As a result,
$\cal W $ \oooo\ can be expressed in terms of the Jacobi $\theta_1 $-function
of the torus (cf. \yy, \wew). After integrating over $\n$ we obtain
\eqn\atrazas{
{\cal W}={ \t_2}\inv  |f_0(e^{2\pi i\t })|^{-48} \
\exp\big[-{\pi(\h-\td\h)^2\ov 2\t_2}\big]\   { \h \td \h |\theta'_1(0|\t
)|^2
\ov \theta_1(\h|\tau )\theta_1 (\td\h |\bar \t ) }
\ .
}
The final expression for $Z$ in \zzzzz,\atrazas\  becomes    the same as
found in the path integral approach  \wew\
if  we  transform the integration variables from $\l, \bar \l$ to $\h, \td \h$
(or, equivalently, identify the auxiliary variables $C_0$ and $\bar C_0$ in
\ffa,\repl\
with
$-2\pi \sqrt {\a'} \l$ and $-2\pi \sqrt {\a'} \bar \l$ in \zzzzz).
The zero-mode normalization  used in \ree\ directly corresponds to the
prescription of projecting out the constant mode factor \consm\
we used in \wew\ (if one does not use this normalization and directly computes
the
trace in \ree\ one finds  the expression for ${\cal W}$ \atrazas\ without
the factor $\h\td \h/\t_2$  which is singular in the free-theory limit).

The operator formalism  representation  for $Z$ \zoper, \complesq\ makes its
 duality invariance properties manifest.
 It is also clear  why, e.g., the case of
$\a\b=0$ is special: here there is no $\hat J_R\hat J_L$ term  in  \kmk\
and thus the traces can be computed
without introducing the auxiliary integrals  \unos.
Then one finds again \rrtr.
 To reproduce  the  expression  for $Z$ in the $q_+=0$
 case in the operator approach
one is to  do the  Poisson resummation not in $m$ but in the winding  number
$w$. Then the $\hat J_L \hat J_R$ term  disappears   again and we  obtain
\fulle.
Equivalently, one may  note that  when $q_+=0$
the sums over $w', w $ in \zzzzz\   give  $\delta$-functions
and the integrals over $\l,\bar \l$ are easily computable
(one may  still apply
\poiw\  to  arrive at \fulle).

\newsec{Some physical properties of particular models}
Having obtained  a  diagonal  string Hamiltonian \hamiltoni\
 and constraints \rotat\ it is  straightforward
 to determine the spectrum of this   string  model.
This was  already done in the special case  of the
 uniform magnetic field model  ($\a\b =0$, $q_i=0$)
in ref. \ruts. Its space-time background
\hts\ is represented by a homogeneous space metric and
non-trivial antisymmetric tensor.
 Below,  in Section 6.1, we  shall examine the spectrum
in  a  complementary
 special case,   $\a=\b$ (or $c_-=a_-$), which includes, in particular, the
Melvin model ($\a=\b=q_+$).
The corresponding non-singular ($\a\b >0$)  space-time backgrounds
are   no longer homogeneous but  have  vanishing  antisymmetric tensor (see
\mmeh\ and Section 3.3).

The general property  of the spectrum observed in \ruts \ was  the appearance
of
tachyonic instabilities, typically associated with states with
angular momentum aligned along  the magnetic field.
Similar  instabilities are   present in  point-particle field
theories
in  external magnetic fields (and may  lead to a  phase transition with
restoration of  some symmetries, see  \nielsen).
In the context of
open string theory they were   observed in ref. \abo\
and further investigated in ref. \ferrara.
The new feature of
 the   closed string theory   \refs{\ruts , \thermal} is the existence
    of states
 with arbitrarily large charges.
Since the critical
magnetic field at which a given state of a charge $Q$ becomes tachyonic
is of order of $(\a' Q)\inv $, the generic pattern is that there is  an
infinite number  of  tachyonic
instabilities
for any given finite value of the magnetic field.\foot{One should keep in mind
that these are
tree-level results.
 At  the one-loop level the mass of a given mode
with charge $Q$ is expected to receive corrections of  order $O(Q^2g^2)$ \
($g$ is the string coupling) which can be neglected for $Q\ll  1/g$.
This suggests that the minimum magnetic field at which tachyonic instabilities
first appear is ${\bf B}_{\rm cr} \sim  (\a'Q)\inv  \sim O(g/\a' )$
which is  essentially of Planck order for  a reasonable value of $g$.}
As we shall discuss below,
in the  general case of $\a\b q_+(q_+ + \b -\a)\not=0$  there exist
also other types of tachyonic instabilities
 associated with the presence of  non-trivial geometrical background.

 Below we shall consider only  the sector with $0 <  \g <2 $, where $\g $ is
defined in  \gamlim . Other sectors can be analysed in a similar way.

\subsec{Spectrum  of  models with vanishing antisymmetric tensor ($\a=\b$)}
It is easy  to show that  the term in $\hat H$ \hamiltoni\ which is linear in
the energy  of a string state $E$   is directly related with the
presence of  an    antisymmetric tensor background. When
 $\a=\b $ the
antisymmetric tensor is absent and this term   can be eliminated
with  choosing  the  frame with  $a_-=0$, i.e.  $q_- = -\a $.
Then the corresponding background  gauge fields \vee\
have no electric components,
so  that the models  with $\a=\b $ can be
 characterized by the two  magnetic field strength parameters
${{{\bf B}_{\rm v}}_0}=a_++c_+ = 2q_+$ and ${{\bf B}_{\rm a}}_0=a_+-c_+ =-
2\b$. The particular
(self-dual) case  of $a_+=0$ corresponds to the dilatonic
Melvin model, while the case  of $a_+=c_+$ (i.e. $\a=\b=0$)
corresponds to the ``Kaluza-Klein"  ($a=\sqrt 3$) Melvin model \melvi.
 The spectrum is determined by  the conditions  \rotat, i.e. $\hat H=0$ and
$N-\tilde N=mw $, where
 (see  \hamiltoni,\angulr,\angull)
\eqn\hamelv{
  \hat H =  \ha   \big( - \a' M^2   + w^2r^2 +{m^2\ov r^2} \big)
+  N+  {\td N} -2 -\four \g^2 }
$$- \
\a'   \big(c_+  Q_+ \hat J_R +a_+Q_-\hat J_L \big) +
\ha \a'   ( \hat J_R + \hat J_L)\big(c_+^2 \hat J_R+a_+^2 \hat J_L\big)   \  ,
\ \ \ M^2=E^2-p_a^2 \ .
$$
 $M^2$ represents the mass,   invariant
with respect to the residual Lorentz group in the hyperplane  orthogonal to
the plane $x_1, x_2$ ($p_a$ are momenta in  possible additional spatial
dimensions).
 We recall that in presence of a non-trivial background including the magnetic
field the
momenta  $p_{1,2}$ of the transverse coordinates $x_{1,2}$ of  a
 closed string
 are traded  for  the
two (right and left)   Landau-type quantum numbers   $l_R, l_L =0,1,2,..., $
associated with the
zero-mode operators $b_0^{\dagger}, b_0$ and $\td b_0^{\dagger},\td b_0$.
The center of mass orbital momentum is $l=l_L-l_R=0, \pm 1, \pm 2 , ...$ (see,
e.g., \speee).
The Hamiltonian \hamelv\  defines the spectrum of states with  quantum numbers
such that
$0< \g <2  $ where $\g$  is given by (see \gamlim)
\eqn\cooo{\g=(a_+ + c_+) wR +\a'[(c_+ - a_+)mR\inv +(a_- - c_-)E] }
$$
+ \ \ha \a '(a_+^2 -a_-^2 - c_+^2 + c_-^2  )(l +S_R +S_L) \ .
$$
As discussed  in Sections 5.2, 5.3, for the special  states with  $\g=0$
(${\rm mod}\ 2$)  the   mass formula
takes the form
\eqn\gamcero{
 M^2= E^2-p_a^2-p_1^2-p_2^2={\a'}\inv\big( -4 -\ha \g^2 +2N+  2{\td N}+  w^2r^2
+{m^2\ov r^2}\big)
}
$$
-  2   \big[c_+ Q_+ (\ha l + S_R)  + a_+Q_- (\ha l + S_L) \big] +
   ( l + S_R + S_L)\big[c_+^2 (\ha l + S_R) +a_+^2(\ha l + S_L)\big]   \  ,
$$
where $l$ is the  eigen-value of the center of mass  orbital momentum
$(x_1p_2-x_2p_1)$  and the quantum numbers are subject to the constraint $\g=0$
(see \cooo).

 A novel feature of the $\a=\b$  model as compared to
the special $\a\b=0$  model studied
in \ruts\  is the presence  in $\hat H$ of  the additional
 terms {\it quartic} in the  oscillators.
Let us  consider the  spectrum of lowest-level neutral
states  ($N=\tilde N=0$,  $m=w=0$).
For these states $\g $ \cooo\ takes the form:
$\g =\ha \a'(c_+^2-a_+^2) (l_R-l_L) , \ \ l_{L,R}=0,1,2,...$.
Then the  generalization (to the case of $a_+\not=0$) of  the expression
\spece\
for the tachyonic spectrum for the Melvin model is
\eqn\tachyo{
M^2=-4{\a'}^{-1}+\ha   ( l_R -l_L) [c_+^2(2l_R+1)- a_+^2(2l_L+1)]
-{\eit}\a' (a_+^2 -c_+^2)^2l^2\ . }
 Let us take  for definiteness  $c_+^2 > a_+^2$.  A closer  inspection of
eq.\tachyo\ shows that, in the range $0<\g<2$, the new contribution
modifying the free-theory  value $\a'M^2=-4$ is positive definite,
  and that the usual tachyon disappears for  sufficiently large  values of
the magnetic fields $a_+, c_+$. This can be attributed to the effect of
curvature of the corresponding background which produces a mass gap in the
Laplace operator.
Similar conclusion is true for the ``massless" level ($N=\td N=1$)   neutral
$(m=w=0)$
states  with $0< \g <2$: they receive positive corrections to their
$M^2$.\foot{For the massless level ($N=\td N=1$) perturbations this
implies, in particular, the stability of the corresponding field-theoretic
Melvin solutions
(see \refs{\gibma,\gibwi} and refs. there).}


The situation is different in the charged sector.
Let us recall that in  the case of the  constant  magnetic field model
($\a=q_i=0$) it was shown  in  \ruts\
that   the term  in $H$
proportional to $c_+Q_+\hat J_R$ gives
rise  to  the  tachyonic instabilities   which are  similar  to magnetic
instabilities in gauge theories.  For a fixed value of the  parameter
 $c_+$   one finds an infinite number of tachyonic states in the spectrum.

 Instabilities caused by the linear in $\hat J_{L,R}$ terms
in $H$ are present also in the $\a=\b$ models  (in particular, in the Melvin
model).
There  are,  in fact,  infinitely many tachyonic charged states at higher
levels.
 Let us first consider the  level one   state
with $w=0$, $m >0 $, $\  N=\td N=1$,  $l_R=l_L=0, \  S_R=1, S_L=- 1$ (i.e.  $
\hat J_{R,L}=\pm \ha $)
which  corresponds to  a   ``massless"  scalar
state with  a  Kaluza-Klein charge.
We may assume without  loss of generality that $R >  \sqrt{\a'}$ (if $R<\sqrt
{\a'}$ a similar discussion applies  with  $m$ replaced by  $w$).
The expression for the mass that follows from \hamelv\ is ($2\b= c_+-a_+$)
\eqn\grave{
 M^2=  p_y(p_y -2 \b-2\a'\beta^2 p_y )    \  , \ \ \ \  p_y =  mR\inv  \ .
}
For $R>>1$, $M^2$ becomes negative when $\b >\b_{\rm cr}\cong \ha p_y$.
For these states   $\g =2\a' \b p_y  $ and thus $\g <2$ if
 $\b >\b_{\rm cr}$ and  $\a' p^2_y <2$ (i.e. $R^2 > \ha  \a' m^2$).
The critical value of the magnetic field goes to zero as $R\to\infty $.
In the noncompact  $R=\infty $ theory  $p_y$ becomes a continuous parameter
representing the momentum of the  ``massless" state in the $y$-direction.
Thus the ``massless" state
 with  an infinitesimal momentum $p_y$ becomes tachyonic for an infinitesimal
value of $\b $
(corresponding  to
an infinitesimal deformation of the metric and antisymmetric  tensor
backgrounds).

To illustrate the appearance of an infinite number of tachyons, let us
consider the $a=1$ Melvin model with $r=1$ where
\eqn\masmelv{
\a'M^2=-4+4\td N+\a' (Q_+ - c_+J_R)^2+\a ' c_+^2J_RJ_L + \eta\ ,\ \ \ \
\eta\equiv \g(1- \ha \g)\leq \ha \ . }
Consider states with $\td N=J_L=0,\ m=w$. Then a given state becomes tachyonic
for
\eqn\criticc
{
(\sqrt{\a' } Q_+ - \sqrt{4-\eta })/J_R<\sqrt{\a' }c_+<(\sqrt{\a' }Q_+ +
\sqrt{4-\eta })/J_R \ ,
}
with
$
\g=-2J_R+1+\sqrt{(2J_R-1)^2+2\a'  Q _+^2-8}  .
$
All states with $J_R\cong Q_+/c_+=(2/c_+)\sqrt{N} ,\ N>>1$, will become
massless in the infinitesimal interval $c_+=Q_+/J_R -\epsilon$ and
$c_+=Q_+/J_R $.
The emergence of an infinite number of massless particles suggests an
enhancement of  gauge symmetries of the theory (for a further discussion see
\thermal ).

Next, let us consider the  $a=\sqrt {3}$ (``Kaluza-Klein") Melvin  model
where  the expression for $M^2$ in \hamelv\ (for $0< \g <2$)  and  for  $\g$ in
\cooo\  are given by ($a_+=c_+=q_+, \ a_-=c_-=0$)
\eqn\kkme{ M^2 =2{\a'}\inv (-2 + N+\td N)   +   (p_y -  q_+  \hat J)^2
 +   [{\a'}\inv wR - q_+ (\hat J_R -\hat J_L)]^2 } $$
- \  q_+^2    (\hat J_R -\hat J_L)^2  -2{\a'}\inv q_+^2w^2R^2
\ ,  \ \  \   \ \ \g = 2 q_+ w R  \ . $$
It follows from \gamcero\  that  in the non-winding (i.e. $\g=0$) sector  this
model is stable, i.e. has  no new  instabilities (except the usual flat space
tachyon).  As for the winding sector, there exists a range of parameters $q_+,\
R$
for which we find again the same linear instability as in the $a=1$ Melvin
model \grave.


\subsec{Gyromagnetic factors }
 In the case  when $B_{\m\n}=0$,
the standard definition for the gyromagnetic factor in terms of the
non-relativistic expansion gives a frame independent result. In the weak field
limit the mass formula takes the form (see \hamiltoni,\hamelv;  $S_{L,R}$ are
the spin parts of the left and right angular momenta  \angulr,\angull)
\eqn\weakl{
 M^2= M^2_0-2(c_+Q_+S_R+a_+Q_-S_L) } $$ + \
[(2l_R+1)c_+Q_+-(2l_L+1)a_+Q_-]+O(c_+^2,a_+^2)
\ ,  $$
$$
\a'M_0^2=-4+2N+2\td N+\ha Q_+^2+\ha Q_-^2 \ .
$$
Note that in the  case of the self-dual model   $a_+=0$ (or $c_+=0$)
we get the familiar field-theory  expression  in the non-winding sector
($w=0$)
\eqn\fieii{
M^2=M_0^2-2e{\cal B} S_R+e{\cal B} (2l_R+1) +O(R^2{\cal B}^2)\ , } $$
\  e=Q_+R\ ,\ \  \ \ {\cal B} =c_+R\inv  \ , \  \ \ \ l_R=0,1,...,
\ .  $$
 In the models under consideration we have two $U(1)$ gauge fields with the
magnetic
strengths
determined by
$c_+=\ha ({{\bf B}_{\rm v}}_0- {{\bf B}_{\rm a}}_0 ) $ and
$a_+=\ha ({{\bf B}_{\rm v}}_0+ {{\bf B}_{\rm a}}_0 )$.
The associated (tree-level) gyromagnetic factors  can be easily read  off from
the Hamiltonian. In the  $B_{\m\n}=0$  case, i.e.  when  $a_-=c_-=0$,
  the $g$-factors  reduce to the form  derived   in ref. \rs\ (see \weakl)
\eqn\gyros{
g_R={2S_R\ov S}\ ,\ \ \ \ \ \ \ g_L={2 S_L\ov S} \ ,  \ \ \ \ S=S_R + S_L
\ . }
As was  found in \ruts, in the case of the
 uniform magnetic field model (which has $B_{\m\n}\neq 0$) there  exists
another contribution  to the $g$-factor coming from the term in $H$ which is
linear in   energy.
Since the magnetic dipole moment  is expected to  be
a background-independent property of the state, this  seems  puzzling.

 The  resolution of this puzzle is the following.
In closed string theory
a magnetic field  background necessarily induces other fields (curved metric,
dilaton, antisymmetric tensor, etc.) and their  consistent
configuration is not unique.
  If one insists on having   strictly
 uniform magnetic field  background, like the one  with $c_+=-c_-=\b $
 considered in \ruts,  this  implies  the presence of an antisymmetric tensor.
 Then  the term  linear in $E$   in  the Hamiltonian  ($\a' c_- E\hat J_R $)
produces    an additional
contribution to the gyromagnetic coupling ($-  \mu { \b} $) and
thus
the $g$-factor gets  an extra  term proportional to the energy.
Note, however, that in order to determine  the magnetic dipole moment
 it is sufficient to consider  a  magnetic
field which is approximately constant in some  finite region.
For generic values of  our parameters  the  magnetic fields   in \fies\
 are  indeed  constant in the region near   $\r=0$  (and  are equal  to $a_+\pm
c_+$, i.e.  do not depend on
 $c_-$ and $a_-$).   That  means
that,  in general (even when  $B_{\mu\nu}\neq 0 $),
the linear  term in $H$
does not, in fact,  produce a contribution to the gyromagnetic ratio.
This suggests that $g_{R,L}$ in \gyros\  are the correct,
 background independent values of the gyromagnetic factors, in accordance with
the suggestion  of ref. \rs \ (see also \seng).  These values
are not inconsistent with the   values
 characteristic to    charged rotating  black holes (see \refs{ \rs , \seng}),
i.e. do not contradict the conjecture \susskind\
about the  correspondence between string states and black holes.

Another interesting special case   (including, in particular, the Kaluza-Klein
Melvin solution \melvi) is when $a_+=c_+$. As follows from \hamelv,   the
magnetic dipole moment  is  then given by
\eqn\maal{
\mu= {\a' \ov 2M} [ r\inv m S + w r (S_R-S_L)]  \ . \ \ }
As a consequence, the  ordinary  Kaluza-Klein states  with $w=0$
(i.e. with charges of
 ``non-winding" origin)  have the standard Kaluza-Klein  field-theory  value of
the gyromagnetic factor,  $g =1$  (see also \rs).

\subsec{  Singular  backgrounds}
Given a  solvable model  one is  able, in principle,
to address the  important question of  how space-time singularities of a  given
background
 are reflected in the properties of the  string solution (CFT) it  represents,
for example, if there is  something
 special happening   with  the spectrum
of  physical  string states  and  the  partition function
for the values of parameters when curvature may have singularities.

As  was mentioned  in Section 3 (see also (A.1)), the  singularities
appear  in the cases when
\eqn\eqy{ \a\b<0  \ , \ \ \ \  {\rm or} \ \ \ \ \
q_+(q_+ +\b-\a )<0 \ . }
These conditions imply that $\b \neq \a$, i.e. require the presence of an
antisymmetric tensor background.
There is an additional instability  in the spectrum which  occurs
only for singular geometries  and leads to a  tachyonic mass
even for the usual (transverse) graviton state.
It is related to the presence of the $O(\hat J^2)$ terms in the Hamiltonian
and is present  even  in the
absence of a magnetic field background.\foot{The
 quartic terms in the  Hamiltonian \hamiltoni\
(which can be eliminated by a proper choice of $q_-$   when $\a\b q_+(q_+
+\b-\a)=0$), are not necessarily associated with singularities. Indeed, there
are  models  with singular geometry  as, e.g., the one with  $q_+=0$,
$\a\b<0$, where the quartic term is absent
 (conversely, there are regular geometries with a
quartic term in $H$ which were discussed  in  Section 6.1).}
Consider,  for example,  the  model with
  $a_+=c_+=0$ or $\a=-\b=q_+$ (see \fies).
 The corresponding singular  background \qeqqp\ was mentioned in Section 3.2.
In this case  the energy formula for the graviton state, $N=\td N=1, m=w=0$,
is given by   (see \hamiltoni;   we choose the frame  with $p_a=0$ and
$c_-=-a_-$)
\eqn\nogauge{
 [E- c_-(\hat J_R-  \hat J_L)]^2 +  2\a'c_-^2E^2 = - 4c_- ^2\hat J_R \hat J_L\
{}.
}
Thus   all the states with $\hat J_R\hat J_L  >0$  have  complex energy
for infinitesimal values of the parameter $\b = \ha (a_- - c_-)$.\foot{A
pathology
of such states is reflected also in the fact that
the value of $\gamma$ in \cooo\
$\g= \a'(a_--c_-)[E - \ha (a_-+ c_-) \hat J]$ which depends on $E$ also
becomes  complex.  }
When the energy gets an imaginary part,   the partition function   develops
a new divergence  coming from the  modular region  $\t_2 \to \infty$
(see \zoper\ and ref. \ruts).

  The  backgrounds  with \eqy\
 generically  have
singularities  which may be related to
naked and black string  type  ones
(in what follows we  drop the trivial $x_3$-direction).
  Consider,  for example,  the case of $a_+=c_+$ or  $\a=-\b $ in the
frame with $c_-=-a_-$. The corresponding metric is (see  \qqqs)
\eqn\blacko{
ds^2=-{dt^2 \ov 1-c_-^2\rho^2 }  +d\rho^2+{\rho^2\ov 1+(c_+^2-c_-^2) \rho
^2}d\vp ^2 \ .
}
For  $c_+=0$ this metric is related by a coordinate transformation to \qeqqp.
Changing
$\rho =1/\td \r $  and replacing $\vp$ by a noncompact coordinate
 one can see that near the  singularity  \blacko\  coincides with the metric of
 the
extreme $D=3$  black string
 in  \hoho.
For $c_+=a_+\not=0$  let us make
 the analytic continuation
 $c_\pm\to i c_\pm $   and at the same time replace
$\vp\to i\vp $ ($\vp $ will still be an angular variable). Then  the background
fields and  the
corresponding \sm   remain real.
Introducing  the coordinate $
x=-{1/\r^2} +c_-^2\  $
one can represent the metric  \blacko\ in the form
\eqn\blackth{
ds^2=-(1-{\m\ov x})dt^2+{dx^2  \ov 4   (x-\m)^3}
+{d\vp ^2 \ov c_+^2-x }
\ ,\ \  \ \  \m\equiv c_-^2 \ .
}
The corresponding  geometry has two singularities, at $x=0$ and $x=c_+^2$
(this is clear, e.g. from
the expression for the curvature scalar in  (A.1)).
 For $x \to\infty $ the curvature approaches
a constant value. If $c_+^2\gg c_-^2=\m$  one can consider the region $x\ll
c_+^2$ where
the metric takes  the  form
\eqn\blackf{
ds^2=-(1-{\m\ov x})dt^2+{d x^2  \ov 4   (x-\m)^3}
 + c_+^{-2} d\vp ^2
\  .
}
This  is recognized as the metric of a (2+1)--dimensional black hole
 with an event horizon
 at $x=\m$.

\newsec{Concluding remarks}
 In this paper we have presented an explicit solution of a class of non-trivial
string models describing curved space-time backgrounds with uniform magnetic
fields.
The corresponding CFT's  are simpler than  generic coset models (having ``free"
central charge and just one $\a'$-correction term in the Hamiltonian which is
diagonal in the free-field Fock space),  are unitary (as follows, e.g.,  from
the possibility to choose a light-cone gauge)
and thus  (along with the model of ref.\napwi\ studied
in \kk\ and its ``compact" generalisation discussed in \ruts)
provide  first examples of  consistent  solvable  conformal string models with
explicit   $D=4$  curved space-time interpretation.

The method of constructing exact string  solutions
and solving the corresponding 2-dimensional conformal  theories
 used in this paper  can be  applied   also to a number of related
models.
 Solvable cases  include, e.g.,
various analytic continuations of  \lag, \lagg\ and their duality transforms,
a generalization \laggde\  from  Section 2.4,
as well as
superstring   and heterotic  string generalizations.
In addition, there are  extensions of our  class of backgrounds (mentioned in
Sections 2.4 and 3.7) which
are  also exact string solutions, but
 they correspond to more complicated CFT's which
cannot  be analysed  using the methods of  Sections 4 and 5.

There are also   possible
interesting applications.
In particular,
the family of string  backgrounds  described  above
 provides a laboratory to study the issue of tachyonic instabilities in string
theories.
We have seen that some of the   members  of
 this family
 are, in  a certain  sense,  ``more unstable" than  others.
 For example,   the $a=\sqrt{3}$ Kaluza-Klein theory
does not have the infinite number of instabilities associated with  large
charge states   discussed in Section 6.1.
 It may happen that some  related  (superstring) models
 may   be stable
 for some  special values of   parameters.
This might   suggest a   mechanism    to
select a stable  string vacuum.

\newsec{Acknowledgements}
We would like to thank  E. Kiritsis, C. Kounnas,  R. Metsaev, K. Sfetsos and E.
Verlinde  for
useful discussions of related questions.
A.A.T. is grateful to the CERN Theory Division
for   hospitality while this work was in progress.
He  also acknowledges the  support
of PPARC and of NATO grant CRG 940870.

\appendix {A}{Curvature scalar for the metric \qqqs}
The curvature scalar corresponding to the metric \qqqs \ is given by
\eqn\curvsc{
R=\ha(A_0+A_2\r^2+A_4\r^4+A_6\r^6)F^2\F^2 \ ,
}
where
\eqn\ssw{F=(1+\a\b \r^2)^{-1}\ ,\ \ \ \F=[1+q_+(q_+ +\b-\a )\r^2 ]^{-1}\ ,
}
and
\eqn\ssew{
A_0=\a ^2 +10\a\b+\b ^2 +12 q_+(q_++\b-\a )\ ,\ \  $$ $$   A_2=\a\b (\a-\b
)^2+q_+(q_++\b-\a ) [36\a\b +(\a+\b )^2 ]\ ,
}
$$
A_4=\a\b q_+ (q_++\b-\a ) [4q_+(q_++\b-\a )+(\a +\b )^2]\  , $$ $$   \ A_6=-8
[\a\b q_+ (q_++\b-\a )]^2\ .
$$
For $\a\b<0$ or $q_+(q_+ +\b-\a )<0$ it is singular at the points where $F$ or
$\F $ diverge, and generically it goes to zero as $\r \to\infty $
(special cases are e.g. $\a=0$, $q_+=0$ where $R$=constant).
At $\r =0 $ it has a finite value given by $R=\ha A_0$.

\vfill\eject

\listrefs

\end

\\
Title:  Exactly solvable string models of curved space-time backgrounds
Authors: J.G. Russo   and   A.A. Tseytlin
Comments:  52 pages,  harvmac
(Revised version: important clarifications
added, the discussion of the spectrum is modified, a new section  on  solution
of tachyon  equation  in  Melvin  background added)
Report-no: CERN-TH/95-20,  Imperial/TP/94-95/17
\\
We consider a new 3-parameter class of exact 4-dimensional solutions
in closed string theory and solve the corresponding string model,
determining the physical spectrum and  partition function.
The background fields (4-metric, antisymmetric tensor, two
Kaluza-Klein vector fields, dilaton and modulus) generically
describe axially symmetric stationary rotating (electro)magnetic
flux-tube type universes. Backgrounds of this class include
both the dilatonic  (a=1 and a=\sqrt 3) Melvin solutions
and the uniform magnetic field solution, as well as some
singular space-times. Solvability of the string $\s $ model
is related to its connection via duality to a simpler model
which is a ``twisted" product of a flat 2-space and a space
dual to 2-plane. We discuss some  physical properties of this
model as well as a number  of generalizations leading to other
exact 4-dimensional string solutions.
\\

There was a suggestion \refs{\gibb,\gibwi,\gibma}
to interpret  the Melvin-type solutions of the  higher dimensional
 (dilaton) Einstein-Maxwell  theory
as alternatives to the standard Kaluza-Klein compactification on compact
spaces.
The idea was to consider  the $(\r,\vp)$ part of the Melvin space as an
internal one. Though this 2-space is non-compact, it is nearly
closed and   the corresponding
 scalar Laplacian    has  discrete branch in the spectrum (see also Section
5.3).  The string Melvin model \tsem\
 studied  in this paper  may be considered as a  string-theory
implementation of the idea of using  a non-compact space
as an internal one
(in our model the internal space  is actually  3-dimensional
$(\r, \vp, y)$).
Since the spectrum of the string mass operator for the Melvin model
is explicitly computable, this makes possible to determine the
corresponding masses of particles  moving in flat spatial  directions.
Unfortunately, as in the case of the particle theory limit \gibwi,
this idea does not actually work in  the case of the Melvin model:
though most of the states in the spectrum belong to its discrete  branch,
there  are also  special ``zero mode" states (e.g.,
scalar state with zero charge and orbital momentum in
\spece)  which  have continuous mass parameter.
It may happen, however, that there are related string models
which (like  modifications of the Melvin solution  by adding a cosmological
term discussed  in \gibwi) may not have this deficiency and yet  be explicitly
solvable.